\documentclass[11pt,a4paper]{article}
\pdfoutput=1
\usepackage{jheppub,hyperref}

\usepackage{multirow, graphicx,amssymb,url,mathrsfs,amsmath}
\usepackage{wrapfig,boxedminipage,setspace,subfigure,epsfig}
\usepackage{amsxtra,amstext,latexsym,dsfont,amsfonts}
\usepackage{color,eucal}

       \def\e  {\epsilon}
\def\s  {\sigma}       

\renewcommand{\a}{\alpha}      \renewcommand{\b}{\beta}

\newcommand{\be}{\begin{equation}}
\newcommand{\ee}{\end{equation}}
\newcommand{\ba}{\begin{array}}
\newcommand{\ea}{\end{array}}

\newcommand{\dd}{\mathrm{d}}

\newcommand{\abs}[1]{\left| #1 \right|}

\title{Thermoelectric Conductivities at Finite Magnetic Field and the Nernst Effect}

\author[a]{Keun-Young Kim,}
\author[a]{Kyung Kiu Kim,}
\author[b]{Yunseok Seo,}
\author[b]{and Sang-Jin Sin}

\emailAdd{fortoe@gist.ac.kr}
\emailAdd{kimkyungkiu@gmail.com}
\emailAdd{yseo@hanyang.ac.kr}
\emailAdd{sjsin@hanyang.ac.kr}

\affiliation[a]{ School of Physics and Chemistry, Gwangju Institute of Science and Technology,
Gwangju 500-712, Korea
}

\affiliation[b]{ Department of Physics, Hanyang University, Seoul 133-791, Korea }

\abstract{
We study the thermoelectric conductivities of a strongly correlated system in the presence of a magnetic field by the gauge/gravity duality. We consider a  class of  Einstein-Maxwell-Dilaton theories with axion fields imposing momentum relaxation.  General analytic formulas for the direct current(DC) conductivities and the Nernst signal are derived in terms of the black hole horizon data.  
For an explicit model study, we analyse in detail the dyonic black hole modified by momentum relaxation. 
In this model, for small momentum relaxation, the Nernst signal shows a bell-shaped dependence on the magnetic field, which is a feature of the normal phase of cuprates.  
We compute all alternating current(AC) electric, thermoelectric, and thermal
 conductivities by numerical analysis and confirm that their zero frequency limits precisely reproduce our analytic DC formulas, which is a non-trivial consistency check of our methods. We discuss the momentum relaxation effects on the conductivities including cyclotron resonance poles. 
  }


\begin{document}

\maketitle

\section{Introduction}

Strongly coupled electron systems show many interesting phases such as non-Fermi liquid, high $T_c$ superconductor and pseudo gap phase. Some of the most important and basic experimental observables in investigating those systems are the conductivities: electric($\sigma$), thermoelectric($\alpha,\bar{\alpha}$), and thermal($\bar{\kappa}$) conductivity.  Therefore, it is essential to develop a theoretical method to compute conductivities to explain and guide experiments. However, due to strong coupling,   the perturbative analysis of quantum field theory does not work   and  we  don't have a reliable systematic tool to compute them. 

Gauge/gravity duality is an approach for such strong coupling problems,  and it has been developed as a method  for conductivity~\cite{Hartnoll:2009sz,Herzog:2009xv}. Some early works treated systems which have translation invariance.  However, at  finite charge density the direct current(DC) conductivities in such systems are infinite. To solve this problem, it is essential to introduce momentum relaxation. 
For this, several ideas have been proposed.

The most straightforward way is to impose inhomogeneous boundary conditions of 
the bulk fields to break translation invariance~\cite{Horowitz:2012ky, Horowitz:2012gs,Horowitz:2013jaa,Ling:2013nxa,Chesler:2013qla,Donos:2014yya}. Massive gravity models studied in~\cite{Vegh:2013sk,Davison:2013jba,Blake:2013bqa,Blake:2013owa,Amoretti:2014ola} give mass terms to gravitons, which break the spatial diffeomorphisms (not the radial and temporal ones), and consequently break momentum conservation of the boundary field theory with translational invariance unbroken~\cite{Blake:2013bqa}.
Holographic Q-lattice models and  those with massless linear dilaton/axion fields ~\cite{Donos:2013eha,Donos:2014uba,Andrade:2013gsa,Gouteraux:2014hca, Taylor:2014tka, Kim:2014bza,  Bardoux:2012aw,Iizuka:2012wt, Cheng:2014qia,Blake:2014yla} take the advantage of a continuous global symmetry of the bulk theory.
Some other models utilise a Bianchi VII$_0$ symmetry to construct black holes dual to helical lattices~\cite{Donos:2012js, Donos:2014oha,Donos:2014gya,Erdmenger:2015qqa}.  All these models yield finite DC conductivities as desired. 

However, all models with momentum relaxation, except \cite{Blake:2014yla},  did not include the magnetic field. Since the transport properties at finite magnetic field such as the quantum Hall effect, the Nernst effect, and the Hall angle, are also important basic probes for strongly correlated electron system, it is timely and essential to develop the methods  for them in the presence of momentum relaxation. 
Indeed, the holographic analysis on conductivities at finite magnetic field was one of the pioneering themes opening up the AdS/CMT(condensed matter theory)~\cite{Hartnoll:2007ai, Hartnoll:2007ih, Hartnoll:2007ip}. 
The purpose of our paper is to extend them by implementing  momenturm relaxation holographically\footnote{In \cite{Hartnoll:2008hs}, momentum relaxation is introduced perturbatively in the hydrodynamic limit.}.  This paper is also a companion of  \cite{Kim:2014bza,Kim:2015dna,Kim:2015sma}, where  all  DC/AC electric($\sigma$), thermoelectric($\alpha,\bar{\alpha}$), and thermal($\bar{\kappa}$) conductivities are analysed thoroughly in the absence of magnetic field.
 
We consider a general class of  Einstein-Maxwell-dilaton theories with axion fields imposing momentum relaxation.  
First, we derive the analytic formulas for DC electric($\sigma$), thermoelectric($\alpha,\bar{\alpha}$), and thermal($\bar{\kappa}$) conductivity in terms of black hole horizon data following the method developed in ~\cite{Donos:2014cya}. Based on these formulas we discuss the model independent features of the Nernst signal.  
Notice that the Nernst signal \eqref{eN0}
\be
e_N = -{(\sigma^{-1}\cdot \alpha)_y}^{x}\,,  \nonumber
\ee
is zero in the holographic model without momentum relaxation, since the electric conductivity is infinite. 
Thus, momentum relaxation is essential for the Nernst effect. 
The Nernst signal has interesting properties which could support the existence of the quantum critical point(QCP).  
As we approach to the QCP or the superconducting domain,  the strength of the Nernst signal becomes stronger and shows a non-linear dependence on the magnetic field, which is different from the expectation based on the Fermi liquid theory \cite{Wang:2006fk} 
\footnote{These anomalous behavior may be described by a liquid of quantized vortices and anti-vortices in non-superconducting phase. See \cite{Anderson:2007aa} for a speculative point of view.}. See \cite{Hartnoll:2007ih, Hartnoll:2008hs}, for pioneering works on the Nernst effect by the holographic approach and the magnetohydrodynamics with a small impurity effect. 
We deal with similar topics by means of a general class of holographic models encoding momentum relaxation, where we {\it assume} that momentum relaxation is related to finite impurity density, which could be large. Note, however, that this relation is not proven yet.

After discussions on a class of models,  we study in detail the dyonic black hole background~\cite{Hartnoll:2007ai, Hartnoll:2007ih, Hartnoll:2007ip}, modified by the specific axion fields introduced in \cite{Andrade:2013gsa}.  We numerically compute AC electric($\sigma$), thermoelectric($\alpha,\bar{\alpha}$), and thermal($\bar{\kappa}$) conductivity and confirm their zero frequency limits agree to the DC formulas that we have derived analytically.  It recovers the results in \cite{Hartnoll:2007ai} if momentum relaxation vanishes. We discuss the momentum relaxation effect on the conductivities including  the cyclotron resonance poles, which was first observed in \cite{Hartnoll:2007ip}. 

This paper is organized as follows. 
In section \ref{dchorizon}, we consider a general class of  Einstein-Maxwell-dilaton theories with axion fields and derive general formulas for the DC electric, thermoelectric, and thermal  conductivity at finite magnetic field as well as the Nernst signal. 
In section \ref{sec3}, as an explicit example, we analyse the dyonic black brane with the axion hair and discuss the Hall angle and the Nernst effect. 
In section \ref{DyonicBH}, we continue our analysis on the model introduced in section \ref{sec3}.  We compute the AC electric, thermoelectric, and thermal conductivity numerically.  The momentum relaxation effect on AC conductivities and the cyclotron resonance poles are discussed. We compare the zero frequency limit of our numerical AC conductivities with the DC analytic formulas derived in section \ref{sec3}. In section \ref{seccon} we conclude.  

{\bf{Note added:}}  While this work was near completion, we noticed the appearance of \cite{Amoretti:2015gna, Blake:2015ina,Lucas:2015pxa} which have some overlap with ours. \cite{Amoretti:2015gna} deals with a massive gravity model at finite magnetic field. \cite{Blake:2015ina} considers the same class of models as ours. \cite{Lucas:2015pxa} obtains general expressions for conductivities at finite magnetic field using the memory matrix formalism.

\section{General analytic DC conductivities at finite magnetic field}\label{dchorizon}

In this section, we will derive analytic formulas for the DC conductivities($\sigma, \alpha, \bar{\alpha}, \bar{\kappa}$)  in the presence of a magnetic field, from a general class of  Einstein-Maxwell-Dilaton theories with axion fields($\chi_1, \chi_2$)
\be \label{action01}
S=\int \dd^4 x \sqrt{-g}\left[ R -\frac{1}{2}\left[(\partial \phi)^2 +\Phi_1 (\phi) (\partial \chi_1)^2 +\Phi_2 (\phi)(\partial \chi_2)^2 \right] -V(\phi)-\frac{Z(\phi)}{4} F^2 \right] \,,
\ee
where $F = \dd A$ and  all $\Phi_1(\phi)$, $\Phi_2(\phi)$, and $Z(\phi)$ should be non-negative for the positive energy condition. Without the magnetic field, analytic formulas for conductivities($\sigma, \alpha, \bar{\alpha}, \bar{\kappa}$) were computed in \cite{Donos:2014cya} while, with the magnetic field, the electric conductivity $\sigma$ was presented in \cite{Blake:2014yla}. Here we compute all the other conductivities as well.  We employ the method developed in \cite{Donos:2014cya,Donos:2014yya}, but a finite magnetic field poses some technical subtlety. We will explain how to treat it. 
The action (\ref{action01}) yields equations of motion:
\begin{align}
R_{MN} -\frac{1}{2} g_{MN} {\cal L}-\frac{1}{2}(\partial_{M}\phi)(\partial_{N}\phi)-\frac{1}{2}\sum_{i=1,2} \Phi_i (\phi)(\partial_{M} \chi_i)(\partial_{N} \chi_i )-\frac{Z(\phi)}{2}F^{P}_{M} F_{PN} =&0 \,,  \label{eom01} \\ 
\nabla_{M}\left(Z(\phi) F^{MN}\right) =&0  \,,  \label{eom02} \\ 
\nabla_{M}\left(\Phi_i (\phi) \nabla^{M} \chi_i \right) =& 0 \,, \label{eom03}  \\ 
\nabla^2 \phi -\frac{1}{2}\sum_{i=1,2} \frac{\partial \Phi_i}{\partial \phi} (\partial \chi_i)^2 -\frac{\partial{V(\phi)}}{\partial \phi} -\frac{1}{4} F^2 \frac{\partial Z(\phi)}{\partial \phi} =& 0 \,, \label{eom04}
\end{align}
where, $\cal L$ is the Lagrangian density of the action \eqref{action01}.

To study the system at finite chemical potential with a background magnetic field($B$), we take the gauge potential as 
\be\label{ansatz2}
A = A_M \dd x^M = a(r) \dd t +\frac{B}{2}(x \dd y -y \dd x) \,.
\ee
We choose the axion fields
\begin{equation} \label{axion1}
\chi_1 = k_1 x\,, \qquad \chi_2 = k_2  y \,,
\end{equation}
which break translational invariance and can give rise to momentum relaxation~ \cite{Donos:2013eha, Donos:2014uba}.
A metric anasatz consistent with the choice \eqref{ansatz2} and \eqref{axion1} is 
\begin{equation}\label{ansatz}
\dd s^2 =  G_{MN} \dd x^M \dd x^N = -U(r) \dd t^2 +\frac{1}{U(r)} \dd r^2  +e^{v_1(r)} \dd x^2 +e^{v_2(r)} \dd y^2 \,, \\
\end{equation}
with $\phi = \phi(r)$. We consider the case in which the black hole solution exists and its horizon  is located at $r=r_h$, i.e.  $U(r_h) = 0$.  We further assume the ansatz
\be\label{const01}
\Phi_1(\phi) =\Phi_2(\phi) =:\Phi(\phi)\,, \qquad v_1 (r) =v_2(r)=:v(r)\,, \qquad k_1 = k_2 =:\beta \,.
\ee

First, the Maxwell equation (\ref{eom02}) yields a conserved charge $\rho$ in the $r$-direction
\be
\rho =\sqrt{-g} Z(\phi) F^{tr} = Z(\phi) e^{v(r)} a'(r) \,,
\ee
which is identified with the number density in the boundary field theory.  The axion equation \eqref{eom03} is trivially satisfied. The Einstein equation \eqref{eom01} and scalar equation \eqref{eom04} become 
\begin{align}
v''(r) =& -\frac{1}{2}(v'(r)^2 +\phi'(r)^2)  \,, \label{bg1} \\
U'(r) =&-\frac{e^{-2 v(r)}}{2 Z(\phi) v'(r)} \Bigg[\rho^2 +2 e^{2 v(r)} V(\phi) Z(\phi) +B^2 Z(\phi)^2 +e^{v(r)} Z(\phi)(2 \beta^2 \Phi(\phi) \nonumber  \\
&+e^{v(r)} U(r) (v'(r)^2 -\phi'(r)^2) \Bigg]  \,,  \label{bg2} \\
U''(r)=& \frac{e^{-2v(r)}\rho^2}{Z(\phi)}+ B^2 e^{-2v(r)} Z(\phi)+e^{-v(r)}\beta^2 \Phi(\phi) +\frac{U(r)}{2}\left(v'(r)^2 -\phi'(r)^2\right) \,,  \label{bg3}
\end{align}
by which we can obtain the background solutions for given $Z(\phi)$ and $V(\phi)$.
For example, if we choose
\be
\phi(r)= {\rm constant}\,, \qquad \Phi(\phi)=1\,, \qquad Z(\phi)=1\,, \qquad V(\phi)=-6/L^2 \,,
\ee
  the background becomes the AdS-dynonic black hole geometry  with the momentum relaxation. We will  discuss it in section \ref{sec3} in detail.  

To compute the conductivities for the general background, we consider small perturbations  around the background obtained by  \eqref{bg1}-\eqref{bg3}
\begin{align}
&\delta A_{i} = t \, \delta f_{i}^{(1)}(r) +\delta a_{i} (r)  \,,  \label{fluc1} \\
&\delta G_{t i} = t \, \delta f_{i}^{(2)}(r) +\delta g_{t i} (r) \,,  \label{fluc2} \\
&\delta G_{r i} = e^{v(r)} \delta g_{r i}(r) \,,  \label{fluc3}  \\
&\delta \chi_i = \delta \chi_i (r)\,, \label{fluc4} 
\end{align}
where $i=1,2$ and $t$ denotes the time coordinate. $\delta f_{i}^{(1)}(r)$ and $\delta f_{i}^{(2)}(r)$ are chosen 
\begin{align}
& \delta f_{i}^{(1)}(r) = - E_{i}  +\zeta_{i} a(r) \label{a11} \,, \\
 &\delta f_{i}^{(2)}(r) = -\zeta_{i} U(r), \label{a22}
\end{align}
where $E_i$ is the electric field and $-\zeta_i$ is identified with the temperature gradient~\cite{Donos:2014cya}. 
In spite of the explicit $t$ dependence  in \eqref{fluc1} and \eqref{fluc2}  all equations of motion of fluctuations turn out to be time-independent, which is the reason to introduce the specific forms of \eqref{a11} and \eqref{a22}.   Furthermore, the electric current  and heat current can be computed as the boundary($r \rightarrow \infty$) values of  $J^i(r)$ and $Q^i(r)$, 
\begin{align}
J^i(r) & =  Z(\phi)\sqrt{-g} F^{ir}(r)  \label{Jii}  \\ 
& = -Z(\phi)\left\{U(r)(-\e^{ij}B \delta g_{r j}(r) +\delta^{ij} \, \delta a_{j}'(r))+a'(r)\delta^{ij}\, \delta g_{t j}\right\} \,, \\
Q^i(r) & = U^2(r)\, \delta^{ij} \partial_r \left(\frac{\delta g_{t j} (r)}{U(r)}\right)-a(r) J^{i}(r)\,, \label{Qii}
\end{align}
which were identified in~\cite{Donos:2014cya} at $B=0$ and they are still valid at finite $B$.

Our task is to plug the solutions of the fluctuation equations into \eqref{Jii} and \eqref{Qii} and read off the coefficients of $E_i$ and $\zeta_i$.
First, the Maxwell equation yields
\begin{align}
0 &= \partial_{M} \left(Z(\phi) \sqrt{-g}F^{i M}\right) \nonumber \\
&= \partial_{r} \left(Z(\phi) \sqrt{-g}F^{i r}\right)+\partial_{t} \left(Z(\phi) \sqrt{-g}F^{i t}\right)   \label{Max1} \\
&= \partial_{r}J^{i} - B \e^{ij} e^{-v(r)} \zeta_{j} Z(\phi) \, .\nonumber
\end{align}
Therefore, the current at the boundary is given by
\begin{align}\label{Jx2}
J^{i}(\infty) &= J^{i}(r_h)+ B \e^{ij}  \zeta_{j}\int_{r_h}^{\infty} dr' e^{-v(r')} Z(\phi(r'))\cr
&\equiv J^{i} (r_h) + B \e^{ij}  \zeta_{j} \Sigma_1.
\end{align}
 Next, let us turn to the heat current, $Q^i$, \eqref{Qii}. It is convenient to start with the derivative of $Q^i$. 
\begin{align}
\partial_r Q^{i} &= 2 U(r)  \delta^{ij} \partial_r \left(\frac{\delta g_{t j} (r)}{U(r)}\right)+U^2(r)\delta^{ij} \partial^2_r \left(\frac{\delta g_{t j} (r)}{U(r)}\right) -a'(r) J^{i}-a(r) {J^{i}}' \cr
&= B \e^{ij} E_{j} e^{-v(r)} - 2 B \e^{ij} \zeta_{j} a(r)e^{-v(r)} Z(\phi)\,.
\end{align}
After using the Einstein equations for fluctuations with the ansatz \eqref{fluc3}-\eqref{fluc4},
\begin{equation}\label{einstein02}
\begin{split}
&2 U(r) \delta g_{tx}''(r)=\Big\{ 2 B^2 Z(\phi) e^{-2v(r)}+2 \beta^2 \Phi(\phi)e^{-v(r)}+U(r) (v'^2(r)-\phi'^2(r))  \Big\}\delta g_{tx}(r)  \\
&\qquad \qquad  \qquad -2 \rho e^{-v(r)}U(r) \left\{ \delta a_{x}'(r)+ B \delta g_{ry}(r) \right\}+ 2 B Z(\phi) e^{-v(r)}\left\{ \zeta_{y} a(r) -E_y \right\} \,, \\
&2 U(r) \delta g_{ty}''(r)=\left\{ 2 B^2 Z(\phi) e^{-2v(r)}+2 \beta^2 \Phi(\phi)e^{-v(r)}+U(r) (v'^2(r)-\phi'^2(r))  \right\}\delta g_{ty}(r)  \\
&\qquad \qquad  \qquad -2 \rho e^{-v(r)}U(r) \left\{ \delta a_{y}'(r)- B \delta g_{rx}(r) \right\}- 2 B Z(\phi) e^{-v(r)}\left\{ \zeta_{x} a(r) -E_x \right\}\,, \\
&\delta g_{r x} = \frac{1}{U(r)\left( B^2 Z(\phi)+\beta^2 e^{v(r)} \Phi(\phi)\right)}\Big[B \rho e^{-v(r)} \delta g_{ty} + B Z(\phi) U(r) \delta a_{y}'(r)  \\
& \qquad   +k \Phi(\phi) U(r) e^{v(r)} \delta \chi_x'(r) -\rho E_x +\left(\rho a(r) - e^{v(r)} U'(r) + e^{v(r)} U(r) v'(r)\right)\zeta_x\Big] \,,  \\
&\delta g_{r y} = \frac{1}{U(r)\left( B^2 Z(\phi)+\beta^2 e^{v(r)} \Phi(\phi)\right)}\Big[-B \rho e^{-v(r)} \delta g_{tx} - B Z(\phi) U(r) \delta a_{x}'(r) \\
&  \qquad  +k \Phi(\phi) U(r) e^{v(r)} \delta \chi_y'(r) -\rho E_y +\left(\rho a(r) - e^{v(r)} U'(r) + e^{v(r)} U(r) v'(r)\right)\zeta_y\Big]\,,
\end{split}
\end{equation}
we end up with a relatively simple expression for the heat current;
\begin{align}\label{Qx2}
Q^{i} (\infty)&= Q^{i} (r_h) + B \e^{ij} E_{j} \int_{r_0}^{\infty}dr' e^{-v(r')} Z(\phi(r'))- 2 B \e^{ij} \zeta_{j} \int_{r_0}^{\infty} dr' a(r') e^{-v(r')}Z(\phi(r')) \nonumber \\
&\equiv  Q^{i} (r_h) + B\e^{ij} E_{j} \Sigma_1 + B\e^{ij} \zeta_{j} \Sigma_2 \,.
\end{align}

In summary, we have two boundary currents:
\begin{align}\
  J^{i}(\infty) &= J^{i} (r_h) + B \e^{ij}  \zeta_{j} \Sigma_1 \,, \\
  Q^{i} (\infty) &= Q^{i} (r_h) + B\e^{ij} E_{j} \Sigma_1 + B\e^{ij} \zeta_{j} \Sigma_2 \,, 
\end{align}
where $Q^{i}(r_h)$ and $J^{i}(r_h)$ are functions at horizon, which can be further simplified by the regularity condition at the black hole horizon \cite{Donos:2014cya,Donos:2014yya}
\begin{equation}\label{nearhorizon}
\begin{split}
\delta a_{i} (r) &\sim -\frac{ E_{i}}{4\pi T} \ln (r-r_h) + \cdots  \,, \\ 
\delta g_{t i} (r) &\sim \delta g_{t i}^{(h)} + \mathcal{O}((r- r_h)) + \cdots \,,   \\
\delta g_{r i} (r) &\sim e^{-v(r_h)} \frac{\delta g_{t i}^{(h)}}{U(r_h)} +\cdots \,, \\ 
\delta \chi_{i} (r) &\sim \chi_{i}^{(h)} + \mathcal{O}((r- r_h))  +\cdots \,.
\end{split}
\end{equation}
Thus, the boundary currents yield
\begin{align}\label{bdcurrent}
J^{i} (\infty) &= -(\rho\, \delta^{ij} e^{-v_h}+ B\e^{ij} e^{-v_h} Z_h )\delta g_{t j}^{(h)}+ \delta^{ij} E_{j} Z_h + B \e^{ij}  \zeta_{j} \Sigma_1 \,, \\
Q^{i} (\infty) &= -4\pi T \delta^{ij} \delta g_{t j}^{(h)} + B\e^{ij} E_{j} \Sigma_1 + B\e^{ij} \zeta_{j} \Sigma_2 \,,
\end{align}
where $v_h = v(r_h)$ and $Z_h =Z(\phi(r_h))$. $\delta g_{t i}^{(h)}$ also can be replaced by the horizon data using the equations of motion \eqref{einstein02}.
The near horizon expansion of the last two equations in (\ref{einstein02}) gives
\begin{align}
 \delta g_{tx}^{(h)} &= \frac{e^{v_h}}{ B^2 Z_h+\beta^2 e^{v_h} \Phi_h}\Big[B \rho e^{-v_h} \delta g_{ty}^{(h)} - B Z_h  E_y -\rho E_x -4\pi e^{v_h}  T\zeta_x\Big] \,, \\
\delta g_{ty}^{(h)} &= \frac{e^{v_h}}{ B^2 Z_h+\beta^2 e^{v_h} \Phi_h}\Big[-B \rho e^{-v_h} \delta g_{tx}^{(h)} - B Z_h  E_x -\rho E_y -4\pi e^{v_h}  T\zeta_y\Big] \,,
\end{align}
which, in turn, give us two algebraic equations for $\delta g_{t i}^{(h)}$ of which solutions are
\begin{equation}\label{dgrx0}
\begin{split}
\delta g_{ti}^{(h)} &= \frac{e^{v_h}}{B^2( \rho^2 + B^2 Z_h^2 +2 \beta^2 e^{v_h}  Z_h \Phi_h) + {\beta^4} e^{2 v_h} \Phi_h^2} \\
& \qquad \qquad \times \Big[ -\beta^2 \rho e^{v_h} \Phi_h \ E_i -B( \rho^2 +B^2 Z_h^2 + \beta^2 e^{v_h}  Z_h \Phi_h) \epsilon_{ij} \ E_j  \\
& \qquad \qquad  \qquad + 4\pi Te^{v_h}( B^2 Z_h + \beta^2 e^{v_h} \Phi_h )\ \zeta_i - 4\pi T e^{v_h} B \rho \epsilon_{ij}\ \zeta_j \Big] \,.
\end{split}
\end{equation}
where   $\Phi_h \equiv \Phi(\phi(r_h))$. 
Finally, the conductivities are obtained by differentiating the boundary currents ($J^{i}(\infty)$, $Q^{i}(\infty)$) with respect to the external electric field($E_{i}$) or the thermal gradient ($\zeta_{i}$):
\begin{align}
\hat{\sigma}^{ij} &= \frac{\partial J^{i}(\infty)}{\partial E_{j}} =  -(\rho\,\delta^{ik} e^{-v_h}+ B\e^{ik} e^{-v_h} Z_h )\frac{\partial \delta g_{t k}^{(h)}}{\partial E_{j}}+ Z_h \,\delta^{ij} \,, \label{DCtrans01} \\
\hat{\alpha}^{ij} &= \frac{1}{T} \frac{\partial J^{i}(\infty)}{\partial \zeta_{j}} = -(\rho \,\delta^{ik} e^{-v_h}+ B\e^{ik} e^{-v_h} Z_h )\frac{1}{T}\frac{\partial \delta g_{t k}^{(h)}}{\partial \zeta_{j}} + \e^{ij}\frac{B}{T} \Sigma_1 \,, \\
\hat{\bar{\alpha}}^{ij} &= \frac{1}{T} \frac{\partial Q^{i}(\infty)}{\partial E_{j}} =-4\pi \delta^{ik} \frac{\delta g_{t k}^{(h)}}{\partial E_{j}}  + \e^{ij}\frac{B}{T} \Sigma_1 \,, \\
\hat{\bar{\kappa}}^{ij} &= \frac{1}{T} \frac{\partial Q^{i}(\infty)}{\partial \zeta_{j}} =-4\pi \delta^{ik} \frac{\delta g_{t k}^{(h)}}{\partial \zeta_{j}} + \epsilon^{ij}\frac{B}{T} \Sigma_2 \,, \label{DCtrans04}
\end{align}
where we put hats on the conductivities to distinguish them from the ones where  magnetization current  are taken out~\cite{Hartnoll:2007ih}.
More explicitly, with \eqref{dgrx0}, the general DC conductivity formulas are given as follows. \\
(i) The electric conductivity and Hall conductivity:
\begin{align}\label{Sxy}
\hat{\s}^{xx} &= \hat{\s}^{yy} =\frac{e^{v_h} \beta^2 \Phi_h \left(\rho^2 + B^2 Z_h^2 +e^{v_h}\beta^2 Z_h \Phi_h \right)}{B^2 \rho^2 +\left(B^2 Z_h+e^{v_h} \beta^2 \Phi_h \right)^2} \,, \\
\hat{\s}^{xy} &= -\hat{\s}^{yx} =\frac{B \rho \left(\rho^2 + B^2 Z_h^2 +2 e^{v_h}\beta^2 Z_h \Phi_h \right)}{B^2 \rho^2 +\left(B^2 Z_h +e^{v_h} \beta^2 \Phi_h \right)^2} \,.
\end{align}
(ii) The thermoelectric conductivity:
\begin{align}
\hat{\a}^{xx} &= \hat{\a}^{yy} = \frac{4 \pi e^{2 v_h} \beta^2 \rho \Phi_h}{B^2 \rho^2 +\left(B^2 Z_h +e^{v_h} \beta^2 \Phi_h \right)^2} \,, \\
\hat{\a}^{xy} &= -\hat{\a}^{yx} =\frac{4 \pi e^{v_h} B \left(\rho^2 + B^2 Z_h^2 +e^{v_h} \beta^2 Z_h \Phi_h \right)}{B^2 \rho^2 +\left(B^2 Z_h +e^{v_h} \beta^2 \Phi_h \right)^2} +\frac{B}{T} \Sigma_1 \,,  \label{Axy} \\
\hat{\bar{\a}}^{ij}  &=\hat{\a}^{ij} \,,
\end{align}
(iii) The thermal conductivity:
\begin{align}
\hat{\bar{\kappa}}^{xx} &=\hat{\bar{\kappa}}^{yy}= \frac{16 \pi^2 T e^{2 v_h} \left( B^2 Z_h + e^{v_h} \beta^2 \Phi_h \right)}{B^2 \rho^2 +\left(B^2 Z_h +e^{v_h} \beta^2 \Phi_h \right)^2} \,, \\
\hat{\bar{\kappa}}^{xy} &= -\hat{\bar{\kappa}}^{yx} =\frac{16 \pi^2 T e^{2 v_h} B \rho}{B^2 \rho^2 +\left(B^2 Z_h +e^{v_h} \beta^2 \Phi_h \right)^2}+ \frac{B}{T} \Sigma_2 \,, \label{Kxy}
\end{align}
where
\be \label{sss1}
\Sigma_1 =\int_{r_h}^{\infty} dr' e^{-v(r')} Z(\phi(r')) \,, 
\ee
and 
\be \label{sss2}
\Sigma_2 = -2  \int_{r_h}^{\infty} dr' a(r') e^{-v(r')} Z(\phi(r')).
\ee
The $\frac{B}{T}\Sigma_{i}$ terms in (\ref{Axy}) and (\ref{Kxy}) come from the contributions of the magnetization current and the energy magnetization current, which should be subtracted~\cite{Hartnoll:2007ih}. In particular, in the case of the dyonic black hole  in section \ref{22}\footnote{The magnetization and energy magnetization current in a general setup were computed in \cite{Blake:2015ina}},  $\Sigma_1 = \frac{{\cal M}}{B} $ and $\Sigma_2 =  \frac{2({\cal M}^E - \mu {\cal M})}{B}$, where $\cal M$ is the magnetization and ${\cal M}^E$ is the energy magnetization.  The relation between $\Sigma_i$ and the magnetizations are confirmed by \eqref{MandME}.
Therefore, the DC conductivities($\s^{ij}, \alpha^{ij}, \bar{\kappa}^{ij}$) read 
\begin{align}
\s^{ij} &= \hat{\s}^{ij} \,, \label{DCtrans1} \\
\alpha^{ij} &= \hat{\alpha}^{ij} -\frac{B}{T} \Sigma_{1} \epsilon^{ij}\,, \label{DCtrans2} \\
\bar{\kappa}^{ij} &= \hat{\bar{\kappa}}^{ij} -\frac{B}{T} \Sigma_{2} \epsilon^{ij} \,,  \label{DCtrans3}
\end{align}
which are expressed in terms of the black hole horizon data.

\subsection{Nernst effect}\label{sec21}
The thermoelectric conductivites play an important role in understanding high $T_c$ superconductors. In the presence of a magnetic field,  a transverse electric field can be generated by a transverse  or longitudinal thermal gradient. The former is called the `Seebeck' effect and the latter is  called the `Nernst' effect.

The electric current, $\vec{J}$, can be written in terms of the external electric field and the thermal gradient as follows;
\be
\vec{J} = \sigma \vec{E} -\alpha \nabla \vec{T},
\ee
where $\sigma$ and $\alpha$ are $2\times 2$ matrices. In the absence of the electric current, 
\be
E_{i} = {(\sigma^{-1}\alpha)_i}^{j} (\nabla \vec{T})_j.
\ee
Based on the definition of the Nernst effect, the Nernst signal ($e_N$) is defined as
\be\label{eN0}
e_N = -{(\sigma^{-1}\cdot \alpha)_y}^{x}.
\ee

The Nernst signal in cuprates shows different features from conventional metals, so it is one of the important observables in understanding  high $T_c$ superconductors. 
For example, in conventional metals the Nernst signal is linear in $B$, while in the normal state of a cuprate it  is  bell-shaped as a function of $B$.  See, for example,   Figure 12  in \cite{Wang:2006fk}.
At a fixed $B$, the Nernst signal increases as temperature decreases in the normal state of a cuprate and, in turn, near the superconducting phase transition the Nernst signal becomes much stronger than conventional metals as shown in Figure 20 in \cite{Wang:2006fk}.

Now we have the general formulas for the DC transport coefficients, (\ref{DCtrans1}) and (\ref{DCtrans2}),  we can compute a general Nernst signal  (\ref{eN0})
\be\label{eN1}
e_N =  \frac{4 \pi  e^{2 v_h} \beta^2  Z_h^2 \Phi_h B }{ \rho^4 + 2 e^{v_h} \beta^2 \rho^2 Z_h \Phi_h +Z_h^2 \left(B^2 \rho^2 + e^{2 v_h} \beta^4 \Phi_h^2 \right) } \,,
\ee
which is expressed in terms of the black hole horizon data.  By playing with the parameters, $Z_h, v_h$, and $\Phi_h$, we can simulate $e_N$ by the formula \eqref{eN1}. It may guide us in constructing more realistic models showing aforementioned cuprate-like properties and furthermore in understanding the physics of strongly correlated systems.

There are two comments on general features of the Nernst signal \eqref{eN1}. 
First, it is proportional to $\beta^2$ for small value of  $\beta$ and it goes as $1/\beta^2$ for large value of $\beta$. It has the maximum at $\beta=\beta_\mathrm{max}$
\be
\beta_{\mathrm{max}}^2 =\frac{e^{-v_h} \rho\sqrt{\rho^2 +B^2 Z_h^2}}{Z_h \Phi_h} \,.
\ee
Second, in the  limit, $\rho\rightarrow0$,  the  Nernst signal becomes
\be
e_N\Big|_{\rho=0} =\frac{4\pi B}{\beta^2 \Phi_h} 
\,.
\ee
%
%
%
In this regime, relevant to the quantum critical point, the Nernst signal is proportional to the inverse of $\Phi_h$ at fixed $B$ and $\beta$.
Because the Nernst signal increases as temperature decreases in the normal phase of cuprates~\cite{Wang:2006fk}, to  have a cuprate-like property, $\Phi_h$ should decrease as temperature decreases. 
This will be a restriction on $\Phi_h$ in model building.

\section{Example: dyonic black branes with momentum relaxation} \label{sec3}

\subsection{Model with massless axions}
As an explicit model,  we consider the Einstein-Maxwell system with massless axions.   
The action is  given by  (\ref{action01}) with the following choices 
\begin{align}\label{dyon01}
\phi =0\,,\quad \Phi_1 =\Phi_2 =1\,,\quad V(\phi) = - 6/L^2\,,\quad Z(\phi) =1\,,
\end{align}
where $L$ is the $AdS$ radius which  will be set to be $1$ from now on. Adding the Gibbons-Hawking term, we start with
\begin{align}
S_0 = \frac{1}{16\pi G}\int_{M}  \dd^{4}x \sqrt{-g} \left[   R +6 -\frac{1}{4}F^2  - \frac{1}{2}\sum_{I=1}^{2} (\partial\chi_I)^2   \right]
- \frac{1}{8\pi G} \int_{\partial M} \dd^3 x   \sqrt{-\gamma} K \,,
\end{align}
where $\gamma$ is the determinant of the induced metric on the boundary. $K$ is the trace of the extrinsic curvature tensor $K_{MN}$\footnote{Where $M,N=0, 1, 2, r $ are  the indices for bulk and  $\mu,\nu=0, 1, 2$  are the indices for the boundary coordinates.} defined by $-\gamma^P_M \gamma^Q_N\nabla_{(P} n_{Q)}$, where $n$ is the outward-pointing normal vector\footnote{In our case,  $n^M = \left(  0 , 0 , 0 ,  1/\sqrt{U(r)} \right)$. See \eqref{dyon02}.} to the boundary($\partial M$) which is at $r= \Lambda$.  From here, for simplicity, we also take $16\pi G =1$.
For the holographic renormalization, we have to add a counter action
\begin{align}
S_\mathrm{c}=\int_{\partial  M}  \dd x^3 \sqrt{-\gamma} \left( - 4 -R[ \gamma]  + \frac{1}{2}  \sum_{I=1}^2 \gamma^{\mu\nu}  \partial_\mu \chi_I \partial_\nu\chi_I    \right) \,,
\end{align}
and the finite renormalized on-shell action is
\begin{align}\label{Sren}
S_{ren} = \lim_{\Lambda \to \infty} \left( S_0 + S_c  \right)_{\mathrm{on-shell}} \,.
\end{align}
Since the boundary terms  do not change the equations of motion, the equations \eqref{eom01}-\eqref{eom04} are valid and yield, with \eqref{dyon01},
\begin{align}
&R_{MN} = \frac{1}{2}g_{MN} \left( R + 6 -\frac{1}{4}F^2  - \frac{1}{2}\sum_{I=1}^{2} (\partial\chi_I)^2 \right) +\frac{1}{2}  \sum_{I}  \partial_M \chi_I \partial_N \chi_I +\frac{1}{2} {F_M}^P F_{NP} \,,
\label{Rmn} \\
&\nabla_M F^{MN} =0 \,,  \qquad
\nabla^2 \chi_I =0 \,. \label{Epsi}
\end{align}

We want to find a solution of the equations of motion, describing a system at finite chemical potential($\mu$) and temperature($T$) in an external magnetic field($B$) with momentum relaxation.  
It turns out the dyonic black brane solution modified by the axion hair \eqref{axion1} does the job. i.e.
\begin{equation}\label{dyon02}
\begin{split}
\dd s^2  =& -U(r) \dd t^2 +\frac{1}{U(r)} \dd r^2  + r^2 \left(\dd x^2 +  \dd y^2\right) \,, \\
& U(r)= r^2 - \frac{ \beta^2}{ 2 } - \frac{m_0}{r} + \frac{ \mu^2 + q_m^2   }{4} \frac{r_h^{2}}{r^{2}} \,,  \\
&a(r) = \mu \left(  1- \frac{r_h}{r}   \right) \,, \qquad B = q_m r_h \,, \\ 
&\chi_1 = \beta x\,, \qquad \chi_2 =  \beta y \,,  
\end{split}
\end{equation}
where $r_h$ is the location of the  horizon and
\begin{align}
m_0 = r_h^3 \left(  1+\frac{\mu^2 + q_m^2}{4 r_h^2} - \frac{\beta^2}{2 r_h^2}     \right) \,.
\end{align}

\subsection{Thermodynamics} \label{22}

To obtain a thermodynamic potential for this black brane solution, we compute the on-shell Euclidean action($S^E$) by analytically continuing to the Euclidean time($\tau$) of which period is  the inverse temperature
\begin{align}
t = - i \tau \,, \quad  S^E = - i S_{ren} \,,
\end{align}
where $S^E$ is the Euclidean action. 
By a regularity condition at the black brane horizon the temperature of the system is given by the Hawking temperature,
\begin{align}\label{temperature}
T = \frac{U'(r_h)}{4\pi}=\frac{1}{4\pi} \left( 3r_h - \frac{\mu^2+q_m^2+ 2\beta^2}{4r_h}  \right) \,,
\end{align}
and the entropy density is given by the area of the horizon,
 \begin{align}
 s= 4\pi r_h^2 \,.
\end{align}

Plugging the solution (\ref{dyon02}) into the Euclidean renormalised action \eqref{Sren}, we have the thermodynamic potential($\Omega$) and its density($\mathcal W$):
\begin{align} 
S^E  = \frac{\mathcal{V}_2}{T} ( - m_0  -\beta^2 r_h  + q_m^2 r_h  ) \equiv  \frac{\mathcal{V}_2}{T} \mathcal W \equiv \frac{\Omega}{T} \,,
\end{align}
where $\mathcal{V}_2 = \int \dd x \dd y$. The potential density $\mathcal W$ can be expressed in terms of the thermodynamic variables as 
\begin{equation} \label{W1}
\begin{split}
\mathcal W =&  \frac{\Omega}{\mathcal{V}_2} = - r_h^3 -\frac{r_h}{4  }\left(\mu^2 + 2\beta^2 - 3 q_m^2 \right) \\
=&~\varepsilon -s T - \mu \rho \,, 
\end{split}
\end{equation}
where $\varepsilon$ and $\rho$ are the energy density and the charge density respectively. The second line is obtained by
using the relation 
\begin{equation} \label{onepfunction}
 \varepsilon = 2m_0 \,,\qquad \rho = \mu r_h \,.
\end{equation}
which is derived as follows. 
We want to compute one-point functions of the boundary energy-momentum tensor($T_{\mu\nu}$), current($J_\mu$) and scalar operators($\mathcal{O}_I$) dual to $\chi_I$. 
Our metric is of the following form
\begin{align}
ds^2 = \mathcal N^2 dr^2 + \gamma_{\mu\nu} (  dx^\mu + V^\mu dr  ) (  dx^\nu + V^\nu dr  )~~,
\end{align}
where $\mathcal N$ and $V_\mu$ are the lapse function and the shift vector. The extrinsic curvature tensor  has  non-vanishing  components, $K_{\mu\nu} = -\frac{1}{2\mathcal N} \left( \partial_r \gamma_{\mu\nu}  - D_\mu V_\nu - D_\nu V_\mu \right)$, where $D_\mu$ is a covariant derivative of the boundary metric $\gamma_{\mu\nu}$.  In terms of aforementioned variables, we define `conjugate momenta' of the fields as 
\begin{align}
\Pi _{\mu \nu } &\equiv \frac{\delta S_{ren} }{\delta  \gamma^{\mu \nu }}  =\sqrt{-\gamma}\left(  K_{\mu \nu }- K \gamma _{\mu \nu } - 2\gamma _{\mu \nu } + G_{\mu\nu}[\gamma]- \frac{1}{2} \partial _{\mu }\chi _I \partial _{\nu }\chi _I +\frac{1}{4} \gamma _{\mu \nu }\nabla \chi _I \cdot \nabla\chi_I   \right) \,,  \nonumber   \\
\Pi^{\mu } &\equiv  \frac{\delta S_{ren}}{\delta A_\mu} =-\mathcal N\sqrt{-\gamma} F^{ r\mu}  \,,  \qquad
\Pi_I \equiv  \frac{\delta S_{ren}}{\delta \chi_I}=\sqrt{-\gamma}\left(  -\mathcal N \nabla ^r \chi_I - ~ \square_\gamma \chi_I  \right)  \,.  
\end{align}
Thus, the one point functions are
\begin{align} \label{onep}
\left< T_{\mu\nu} \right> &= \lim_{r \to \infty}   \frac{2 r}{\sqrt{-\gamma}} \Pi_{\mu\nu} = \left( \begin{array}{ccc}
2m_0 & 0& 0\\
0&m_0&0\\
0&0&m_0
\end{array} \right)\,,~\nonumber\\
\left<J^\mu\right> &= \lim_{r \to \infty}  \frac{r^3}{\sqrt{-\gamma}} \Pi^\mu = \left( ~\mu r_h,~0,~0~\right)\,, \qquad \left<\mathcal O_I\right>= \lim_{r \to \infty} \frac{r^3}{\sqrt{-\gamma}}\Pi^I =0 \,,
\end{align}
which yield \eqref{onepfunction}.

Since the pressure $\mathcal{P} = -\mathcal{W}$, \eqref{W1} becomes  a Smarr-like relation
 \begin{align} \label{smarr1}
 \varepsilon +\mathcal P = sT + \mu \rho \,.
 \end{align}
Notice that the pressure is not equal to $\left< T_{xx}\right>$ since
\begin{align}
\mathcal P = \left< T_{xx}\right> + r_h \beta^2 - r_h q_m^2~~.
\end{align}
The magnetization($\mathcal M$) and the energy magnetization($\mathcal M^E$) are 
\begin{align} \label{MandME}
\mathcal M = -  \left( \frac{\delta  \Omega}{\delta B} \right)  = - q_m\,, \qquad \mathcal M^E = -  \left( \frac{\delta  \Omega}{\delta B^E} \right) = - \frac{1}{2}\mu  q_m\,,
\end{align}   
where $\delta B$ and $\delta B^E$ are defined by a linearized solution\footnote{For details we refer to Appendix C of \cite{Hartnoll:2007ih}.}
\begin{align} \nonumber
&\delta g_{ty} =  x U(r) \delta B_E ~,\quad \delta A_y =  x \left(  \delta B  + \mu\delta B_E  \frac{r_h}{r}  \right)~, \\
& \delta A_t =  - \frac{B}{2 r_h^2} \left(  1-  \frac{r_h^2}{r^2}  \right) \delta B_E   -  \frac{B}{\mu r_h^2} \left(  1-  \frac{r_h}{r}  \right) \delta B\nonumber~~~.
\end{align}
We find the first law of thermodynamics
\begin{align}
\delta \mathcal E = T\delta s + \mu \delta \rho - r_h \delta (\beta^2) -\mathcal M \delta B,
\end{align}
by combining the variation of the first line of \eqref{W1} with respect to $T, \mu,\beta^2$ and $B$:
 \begin{align}
 \delta \mathcal W = -\mathcal M \delta B -s \delta T - r_h \delta(\beta^2) - \rho \delta \mu \,,
 \end{align}
and the variation of \eqref{smarr1}.

\subsection{DC conductivities: Hall angle and Nernst effect}\label{sec:DCconduct}

In this section we study the DC conductivities of the dyonic black brane with momentum relaxation. 
Because we have derived the general formulas in section \ref{dchorizon},  we only need to 
plug model-dependent information (\ref{dyon01}) and (\ref{dyon02}) into \eqref{DCtrans1}-\eqref{DCtrans3}.

The electric conductivities yield
\begin{align}
\sigma^{xx} & =\sigma^{yy} =\b^2  r_h^2 \frac{ B^2 +r_h^2 \left( \mu^2 +\b^2 \right)}{r_h^2 \mu^2 B^2 + \left(B^2 +r_h^2 \b^2 \right)^2} \,, \label{dyonicSxx} \\
\sigma^{xy} & = - \sigma^{yx} = B r_h \mu   \frac{B^2 +r_h^2 \left( \mu^2 +2 \b^2 \right)}{r_h^2 \mu^2 B^2 + \left(B^2 +r_h^2 \b^2 \right)^2} \,. \label{dyonicSxy}
\end{align}
In the clean limit, $\beta \rightarrow 0$, these boil down to 
\be
\sigma^{xx} = \sigma^{yy} = 0 \,, \qquad  \sigma^{xy} = - \sigma^{yx} = \frac{r_h \mu}{B}=\frac{\rho}{B}, \nonumber
\ee
where temperature dependence drops out and we recover the results expected 
on general grounds from Lorentz invariance, agreeing with \cite{Hartnoll:2007ai}. 
In the limit $B \rightarrow 0$, the expressions become
\begin{equation}
\sigma^{xx} = \sigma^{yy} = 1+\frac{\mu^2}{\beta^2}\,, \qquad \sigma^{xy} =  \sigma^{yx} = 0 \,,  \nonumber
\end{equation}
which reproduces the result in \cite{Andrade:2013gsa,Kim:2014bza}.
The thermoelectric and thermal conductivities read
\begin{align}
\a^{xx} &= \a^{yy}  = \frac{4\pi r_h^5 \b^2 \mu}{r_h^2 \mu^2 B^2 + \left(B^2 +r_h^2 \b^2 \right)^2} \,, \label{t1} \\
\a^{xy} &= -\a^{yx} = 4\pi r_h^3 B  \frac{B^2 +r_h^2 \left( \mu^2 +\b^2 \right)}{r_h^2 \mu^2 B^2 + \left(B^2 +r_h^2 \b^2 \right)^2}\,,  \label{t2} \\
\bar{\kappa}^{xx} &= \bar{\kappa}^{yy} = 16 \pi r_h^4 T  \frac{B^2 +r_h^2 \beta^2}{r_h^2 \mu^2 B^2 + \left(B^2 +r_h^2 \b^2 \right)^2} \,,  \label{t3} \\
\bar{\kappa}^{xy} &= -\bar{\kappa}^{yx} \frac{16 \pi^2 r_h^5 \mu T B}{r_h^2 \mu^2 B^2 + \left(B^2 +r_h^2 \b^2 \right)^2} \,.\label{t4}
\end{align}

\begin{figure}[]
\centering
    \subfigure[$\sigma^{xx}$ ]
   {\includegraphics[width=4.8cm]{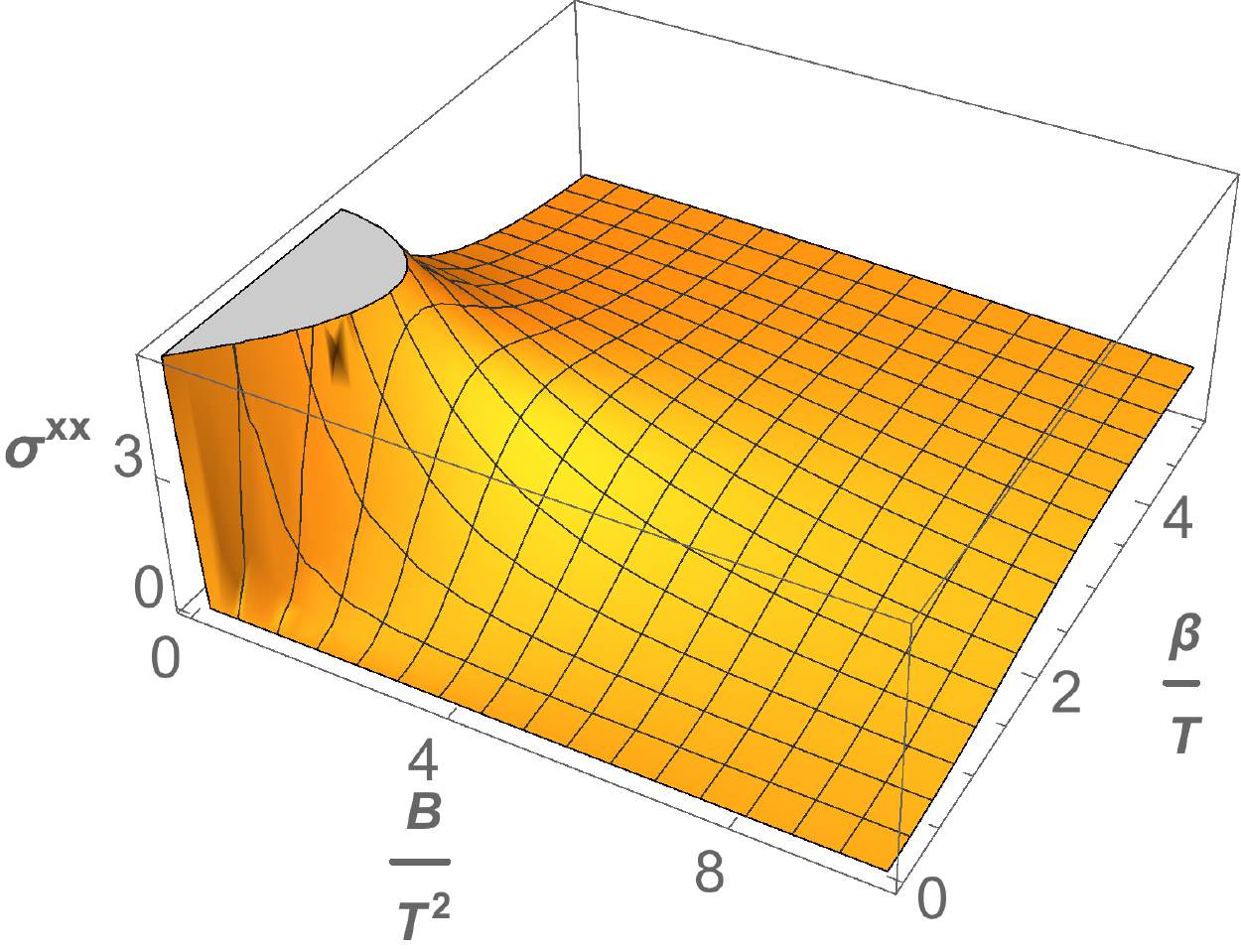} \label{}}
      \subfigure[$\alpha^{xx}$ ]
   {\includegraphics[width=4.8cm]{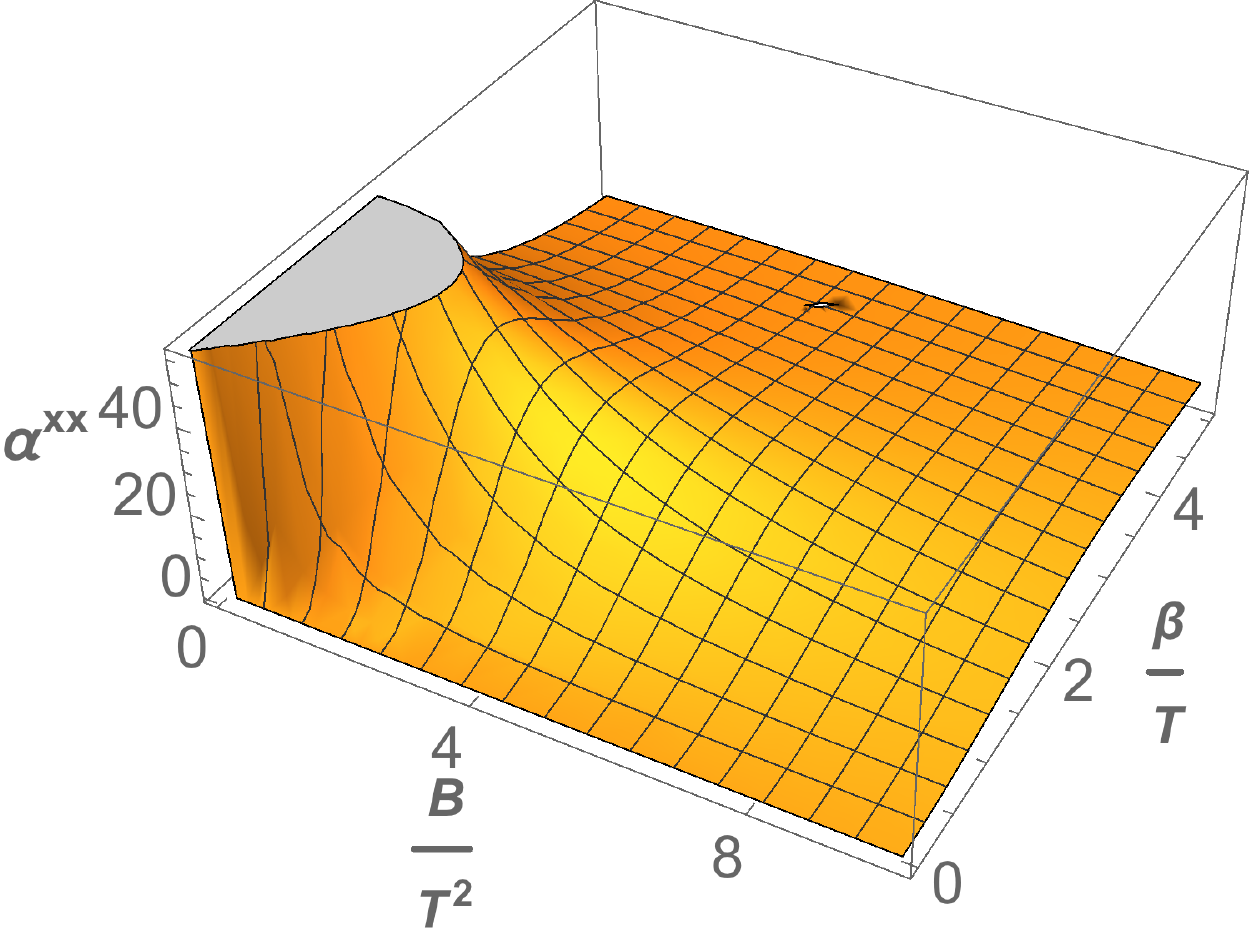} \label{}}
       \subfigure[$\bar{\kappa}^{xx}$]
   {\includegraphics[width=4.8cm]{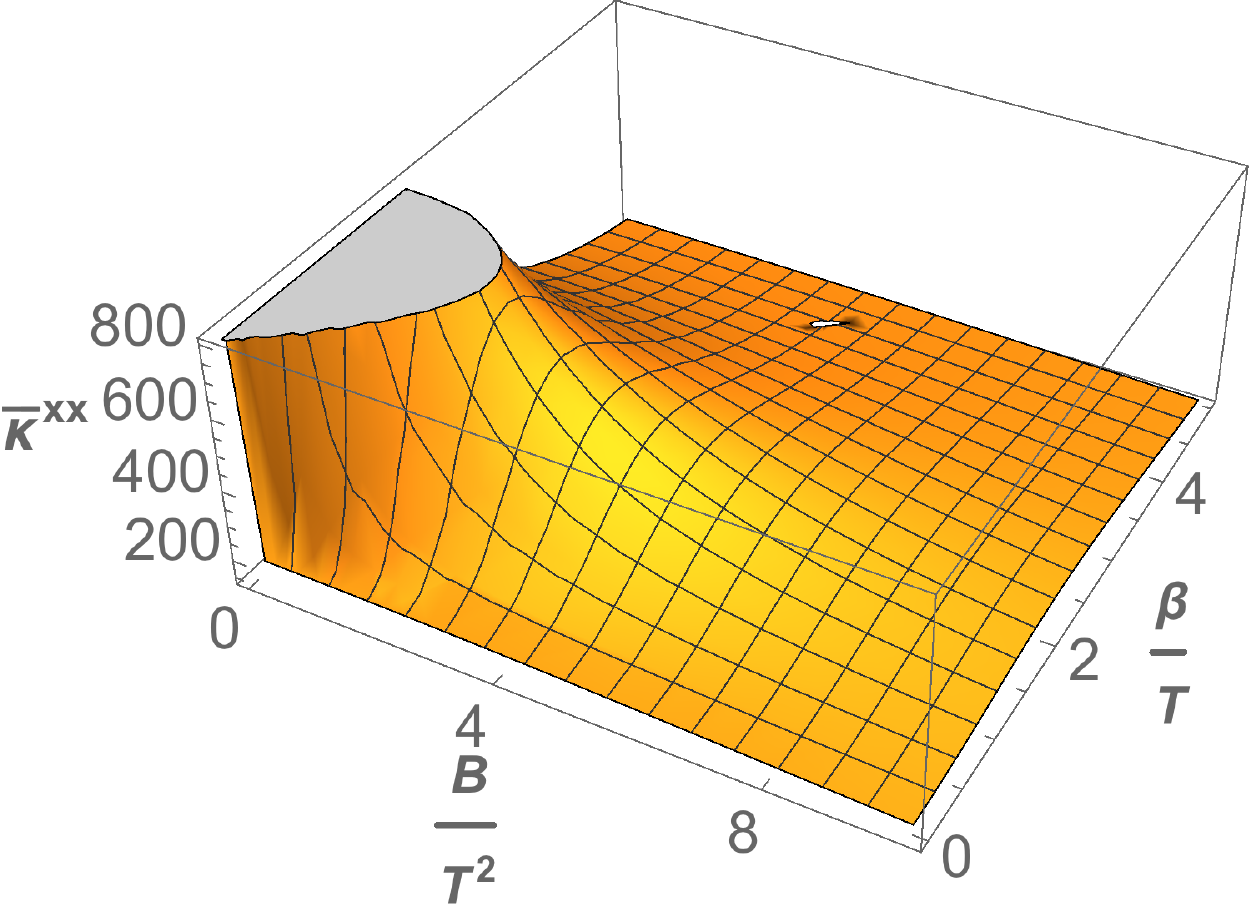} \label{}}
       \subfigure[$\sigma^{xy}$]
   {\includegraphics[width=4.8cm]{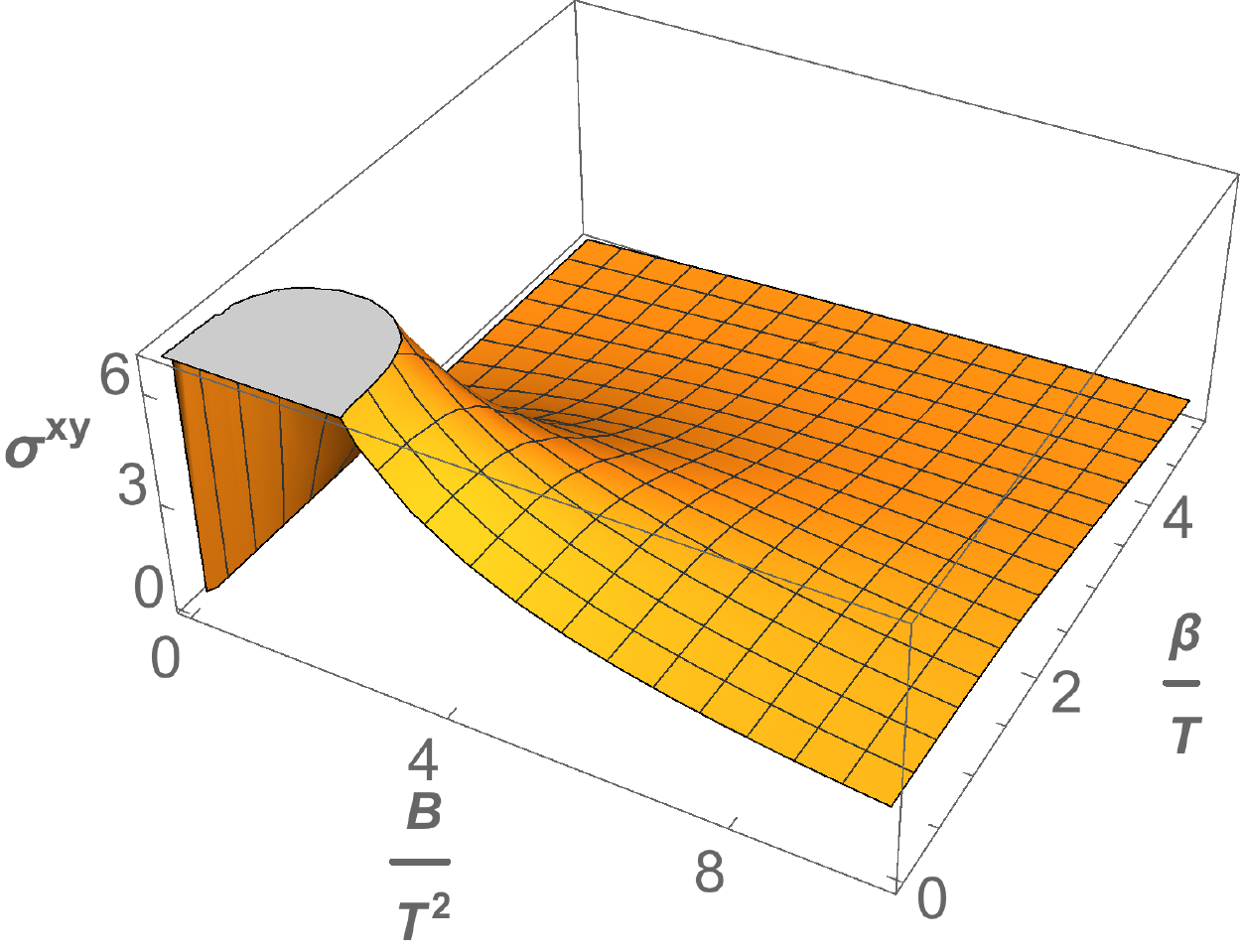} \label{}}
        \subfigure[$\alpha^{xy}$]
   {\includegraphics[width=4.8cm]{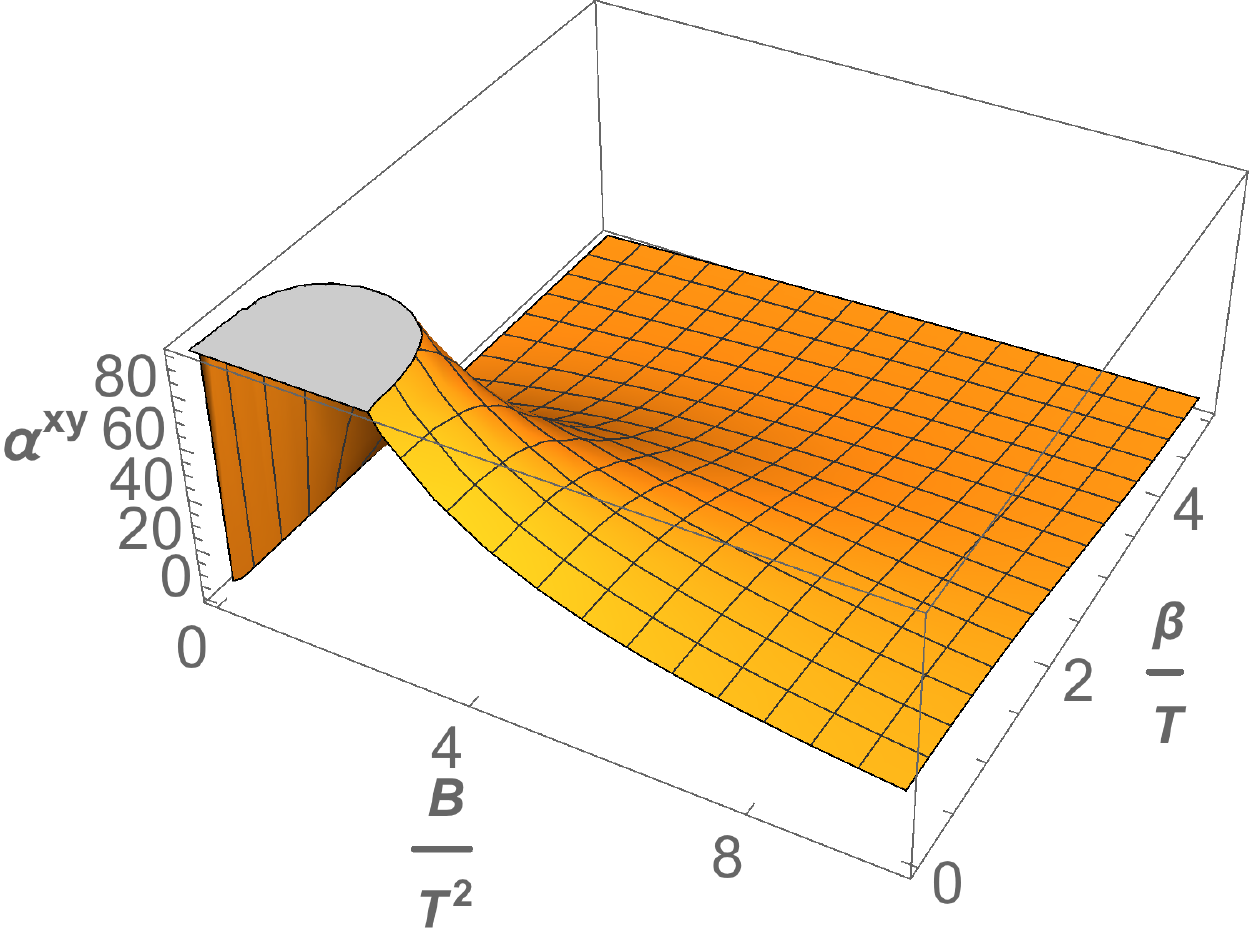} \label{} } 
     \subfigure[$\bar{\kappa}^{xy}$]
   {\includegraphics[width=4.8cm]{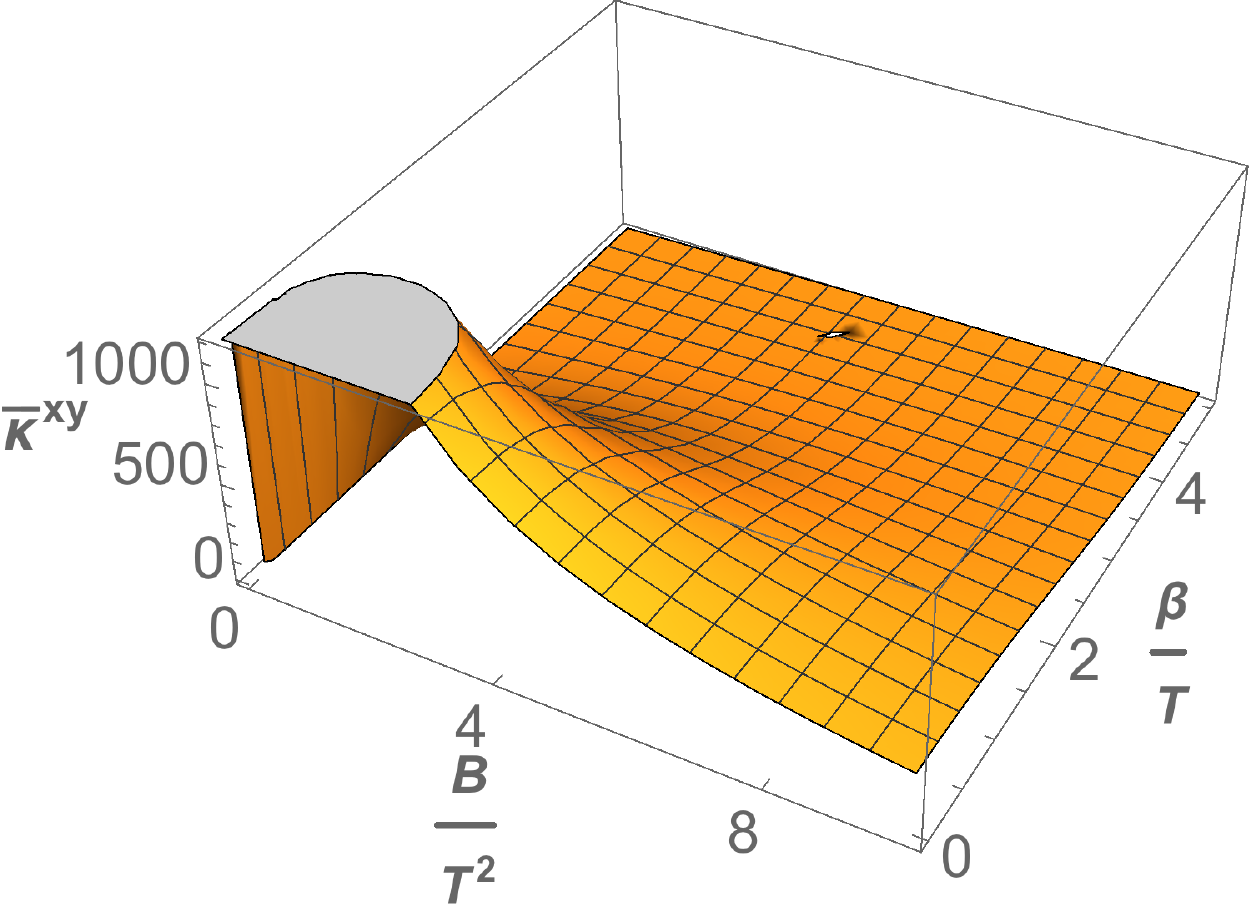} \label{} }
 \caption{$\beta/T$ and $B/T$ dependence of DC conductivities at fixed $\mu/T =4$
           } \label{fig:dcSxy}
\end{figure}

To see the effect of $\beta$ and $B$ on the conductivities, the formulas \eqref{dyonicSxx}-\eqref{t4} are not so convenient since $r_h$ is a complicated function of $T,B,\mu$, and $\beta$, as expressed in (\ref{temperature}).
Therefore, we make plots of conductivities in Figure \ref{fig:dcSxy}, where we scaled the variables by $T$ and fixed $\mu/T=4$. The $\sigma^{xx}, \alpha^{xx}$, and $\bar{\kappa}^{xx}$ are qualitatively similar; the $B$ dependence at fixed $\beta$ is monotonic, while the $\beta$ dependence at fixed $B$ is not.  The $\sigma^{xy}, \alpha^{xy}$, and $\bar{\kappa}^{xy}$ are similar;  the $\beta$ dependence at fixed $B$ is monotonic, while the $B$ dependence at fixed $\beta$ is not.
See Figure \ref{fig:numericalDC} for the cross sections at $B/T^2=1$.  As the chemical potential increases, the electric and thermoelectric conductivity generally increases while its overall 2-dimensional shape does not change qualitatively. However, the thermal conductivities behave differently; $\bar{\kappa}^{xx}$   broadens while $\bar{\kappa}^{xy}$   is sharpened.

\begin{figure}[]
\centering
      \subfigure[$B$-$\beta$ dependence at  $\mu/T =4$] 
   {\includegraphics[width=4.8cm]{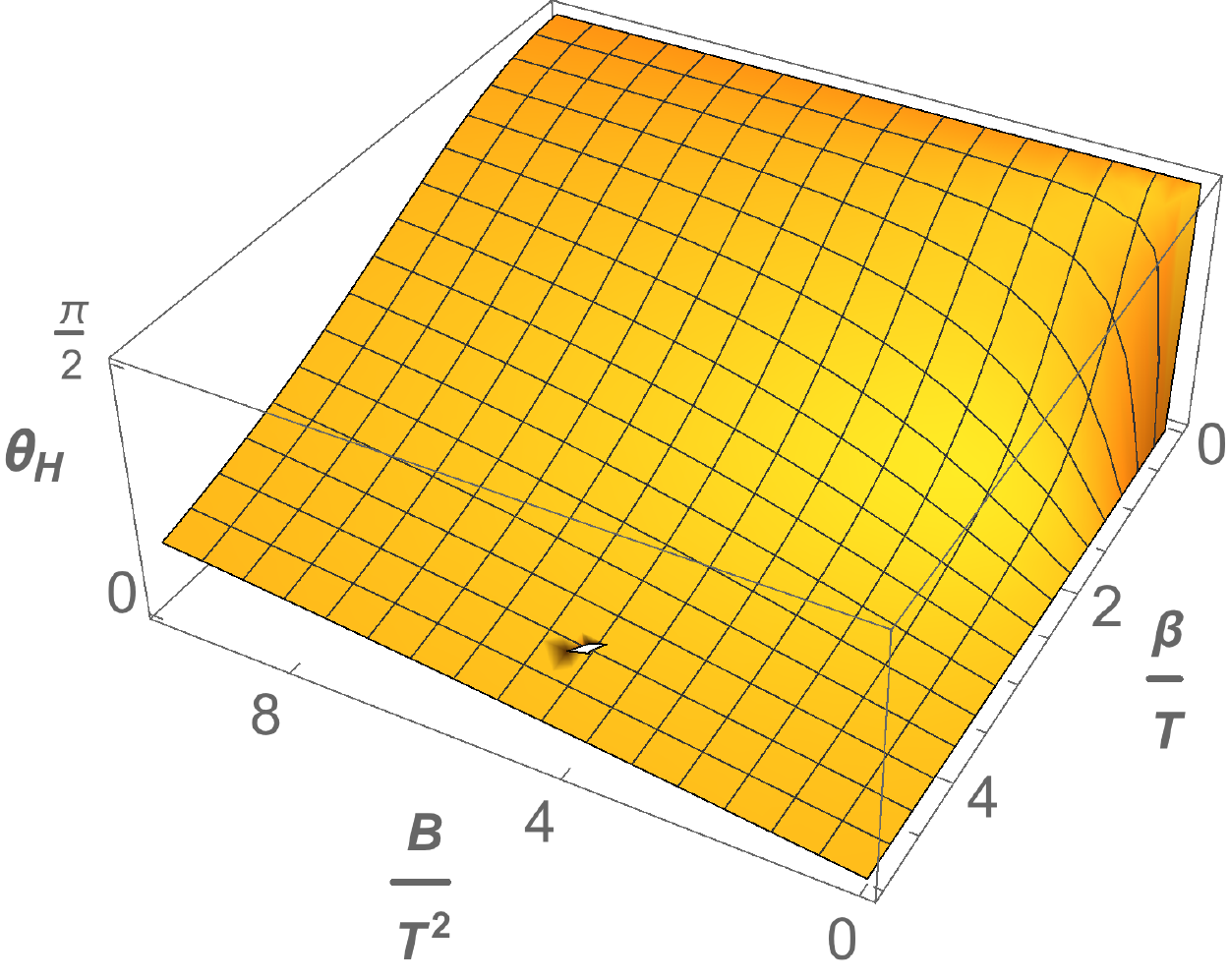} \ \  \label{}}
         \subfigure[$\beta$-$T$ dependence at  $B/\mu =1$ ]
   {\includegraphics[width=4.8cm]{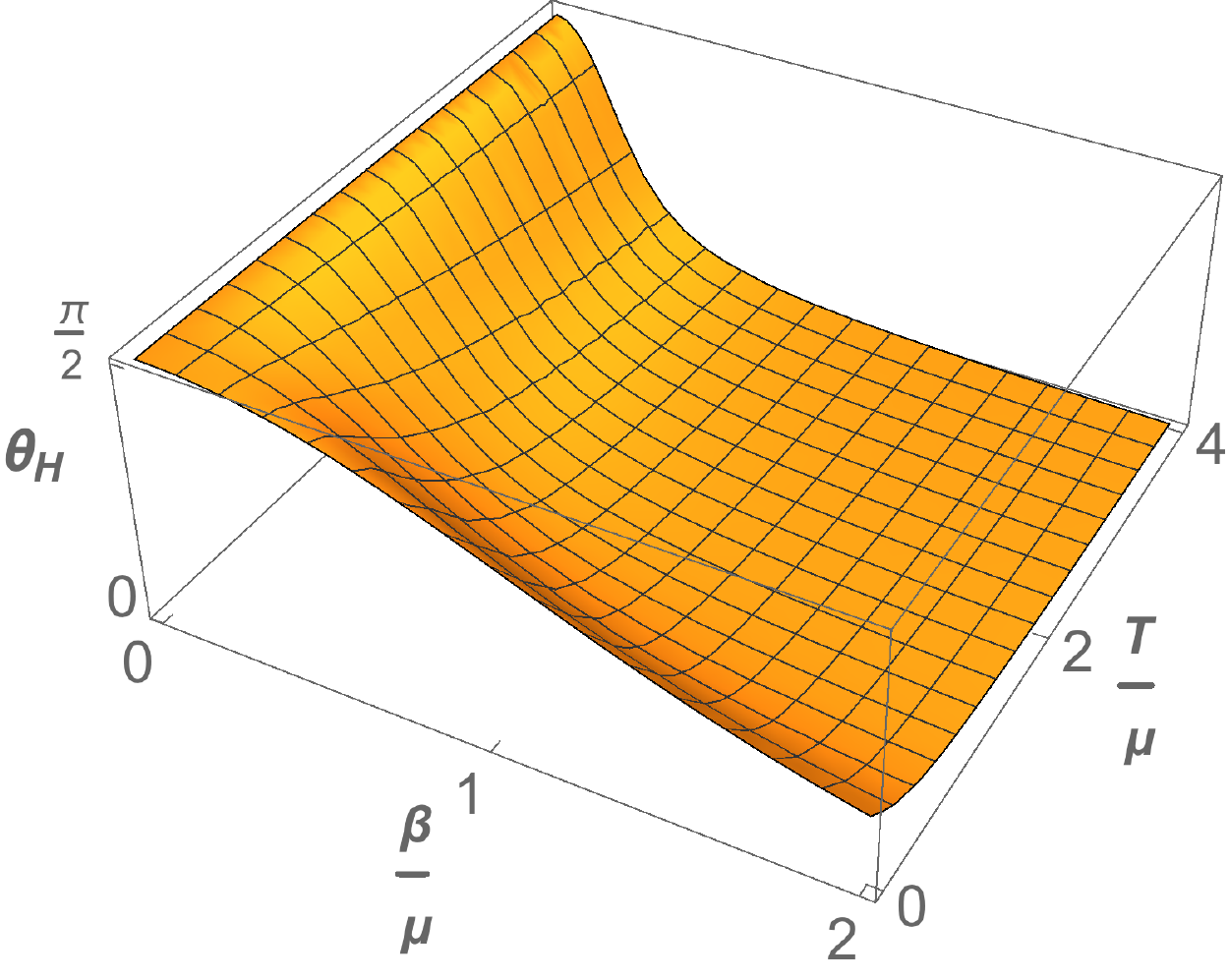} \label{}}
 \caption{Hall angle $\theta_H = \arctan(\sigma^{xy}/\sigma^{xx})$
           } \label{fig:Hall11}
\end{figure}

The Hall angle($\theta_H$) is defined by the ratio of the electric conductivities:
\be \label{Hall2}
\tan{\theta_H} \equiv \frac{\sigma^{xy}}{\sigma^{xx}} = \frac{\mu B}{r_h \beta^2} \frac{B^2 +r_h^2 (\mu^2 +2 \beta^2)}{B^2 +r_h^2 (\mu^2 + \beta^2)} \,,
\ee
which agrees to the result reported in \cite{Blake:2014yla}. As shown in Figure \ref{fig:Hall11},
the Hall angle $\theta_H$ ranges from $\pi/2$ to $0$. 
The angle increases as $B$ increases or $\beta$ decreases(Figure \ref{fig:Hall11}(a)). In the strange metal phase, we are interested in the temperature dependence of the Hall angle, which is proportional to $1/T^2$. In our case, we numerically found that the Hall angle ranges between $1/T^0$ and $1/T^1$(Figure \ref{fig:Hall11}(b)). In the large $T$ regime, the Hall angle always scales as $1/T$. It can be seen also from the formula \eqref{Hall2}, where if $T$ is large compared to the other scales, $r_H \sim T$ so $\tan \theta_H \sim 1/T$.

\begin{figure}[]
\centering
  \subfigure[ $\beta/T$ and $B/T^2$ dependence of the Nernst signal at $\mu/T=1$]
   {\includegraphics[width=4.8cm]{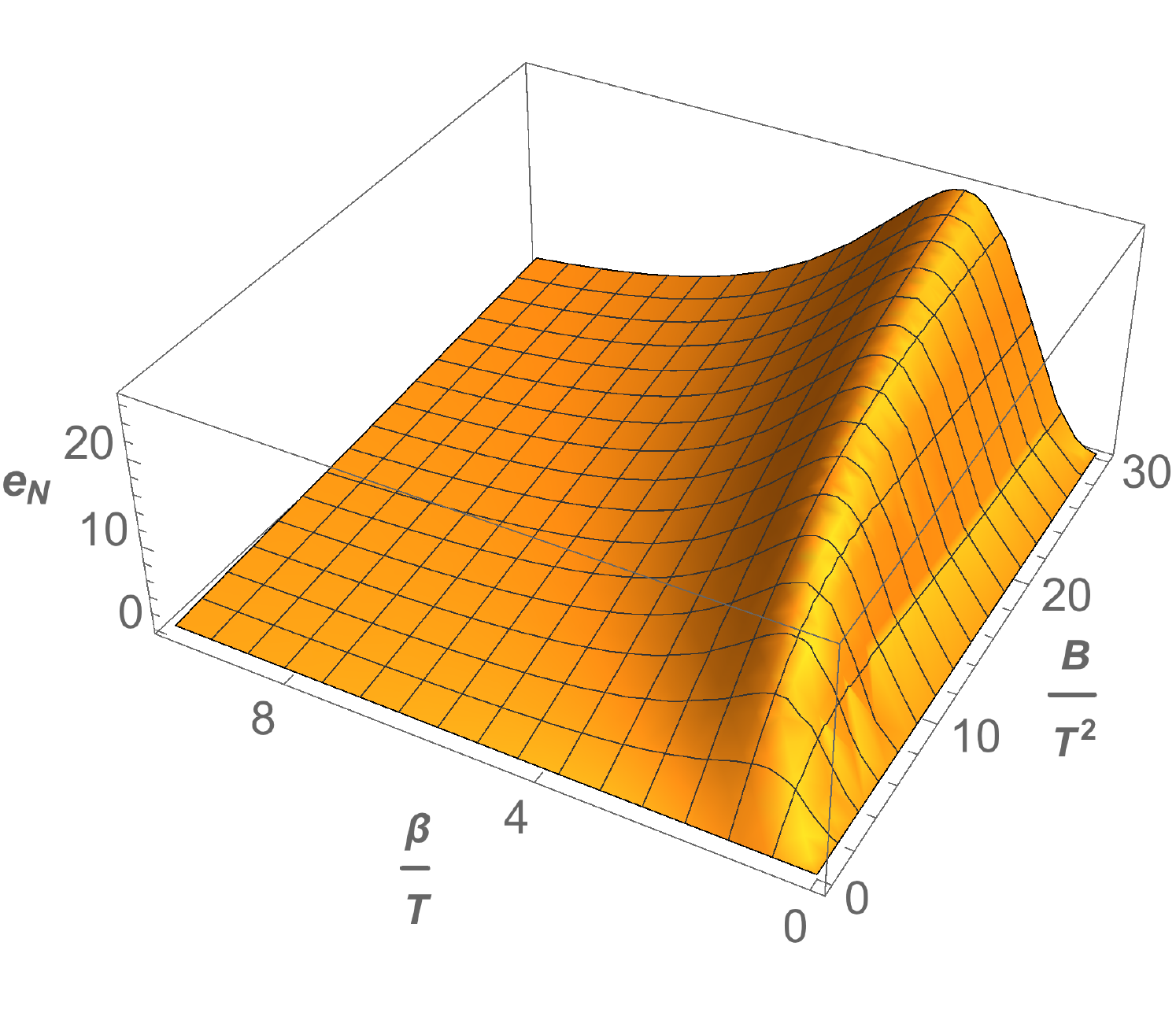} \label{}}
   \hspace{.5cm}
     \subfigure[The cross sections of (a) at $\beta/T=0.5, 1, 4$(red, green, blue)]
   {\includegraphics[width=5cm]{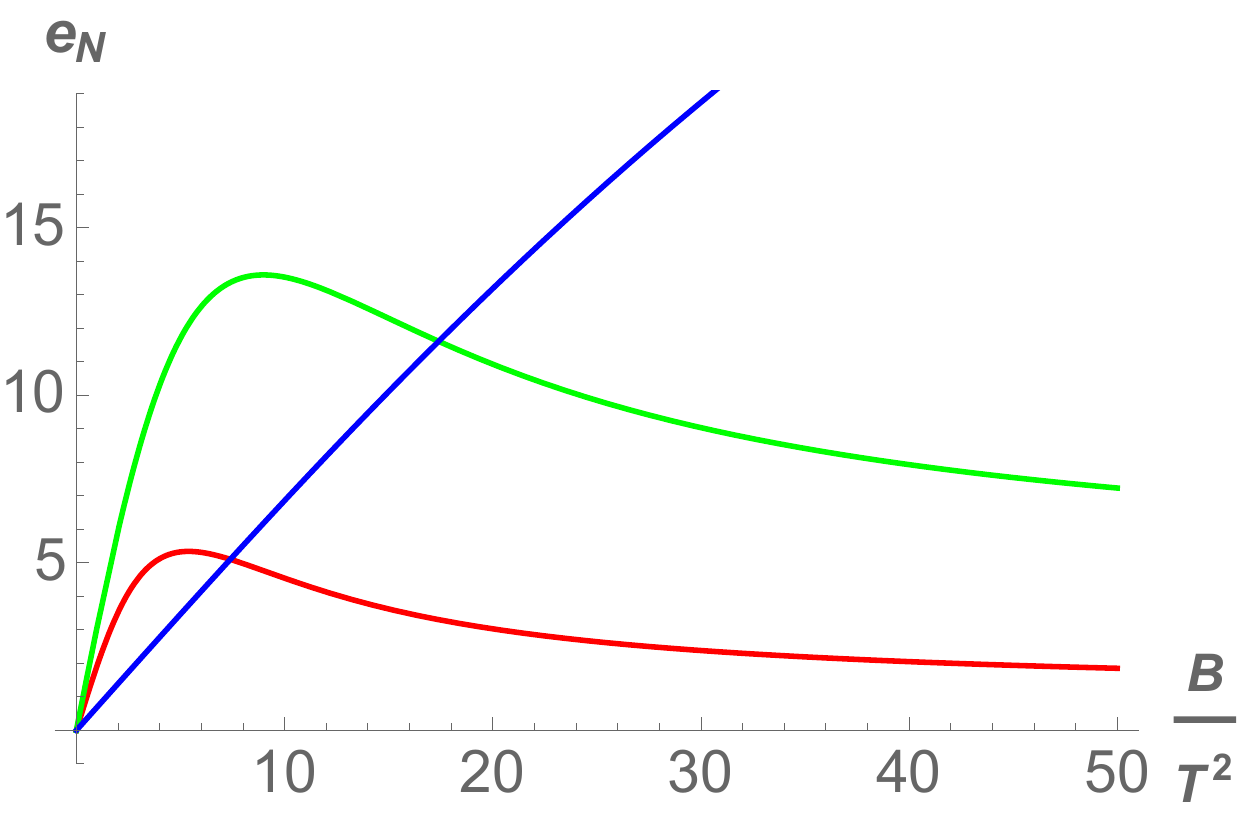} \label{} }
 \caption{Nernst signal 
           } \label{fig:NernstT}
\end{figure}

The Nernst signal \eqref{eN1} yields
\be\label{Nernst01}
e_N = \frac{4\pi r_h^2 \beta^2 B}{\mu^2 B^2 +r_h^2(\mu^2 +\beta^2)^2}.
\ee
As discussed in Section \ref{sec21}, the Nernst signal is linear in $B$ in conventional metals while it becomes bell-like in the normal state of cuprates~\cite{Wang:2006fk}. To see whether our model can capture this feature, we make a three dimensional plot of the Nernst signal as a function of $\beta/T$ and $B/T^2$ at fixed $\mu/T=1$ in Figure \ref{fig:NernstT}, where (b) is the cross section of (a) at fixed $\beta$.  The blue line is almost straight while the green and red ones are bell-like. Therefore, we find that our system shows
the transition from the normal metal(blue line) to cuprate-like state(green and red) as $\beta$
decreases, in the Nernst signal perspective. 
The green and red curves are similar to Figure 12  in \cite{Wang:2006fk} and it was proposed that the bell shape can be explained by the dynamics of the vortex liquid in non-superconducting phase~\cite{Anderson:2006, Anderson:2007aa}. It was also interpreted as an evidence of the pseudo-gap phase~\cite{Wang:2006fk}.  If we take this point of view, our model might be relevant to the pseudo-gap phase.

One may wonder if the blue line is qualitatively similar to the green and red ones for higher $B/T^2$. As  $B/T^2$ increases the blue line reaches the maximum and has almost plateaued out, which is not bell-shaped.
Furthermore, the numerical value of $B/T^2$ at the maximum $e_N(\sim 60)$ is about $1600$, which is too big to consider, compared to other scales.

\section{Numerical AC conductivities} \label{DyonicBH}

So far we have discussed the DC conductivities.  In this section,  we will consider the AC conductivities. To be concrete, we continue to investigate the model in the previous section, namely, the dyonic black brane with the momentum relaxation by the axion fields.

\subsection{Equations of motion and on-shell action} \label{sec41}

In section \ref{dchorizon} and \ref{sec3}, we have derived the DC conductivity in a gauge $g_{ri}  \ne 0$. In this section, to compute the AC conductivity, we will work in a gauge $\delta g_{ri} = 0$. The physical result should not be changed by this gauge choice\footnote{There is a subtle issue on the gauge choice and holographic renormalization. For more details we refer to \cite{Kim:2015sma}.}. Indeed, we will show the zero frequency limit of the AC conductivity (with $\delta g_{ri} = 0$) agrees to the DC conductivity (with $\delta g_{ri} \ne 0$) in section \ref{dchorizon}. It serves as a good consistency check of our numerical method.

To compute conductivities holographically it is consistent to turn on linear fluctuations of bulk fields, $\delta g_{ti}$, $\delta A_i$ and $\delta \chi_i$ at zero momentum in the $x$ and $y$ directions, where $i=1,2$(or $i=x,y$). The fields $\delta g_{ti}$ and $\delta A_i$ are related to the heat current and the electric current. They were introduced in \cite{Hartnoll:2007ai} and here we add $\delta \chi_i$ for momentum relaxation. 
The fields, $\delta g_{ti}$, $\delta A_i$ and $\delta \chi_i$, can be expressed in momentum space as
\begin{align} \label{flucA}
 \delta A_i(t,r) &= \int^{\infty}_{-\infty} \frac{d \omega}{2\pi}  e^{-i\omega t}  a_{i}(\omega,r)\,, \\
\delta g_{ti}(t,r) &=  \int^{\infty}_{-\infty} \frac{d \omega}{2\pi} e^{-i\omega t} r^2 h_{ti}(\omega,r)\,,  \label{flucg} \\ \label{flucPsi}
 \delta \chi_i(t,r) &= \int^{\infty}_{-\infty} \frac{d \omega}{2\pi} e^{-i\omega t}  \psi_{i}(\omega,r) \,,
\end{align}
where $r^2$ in the metric fluctuation \eqref{flucg} is introduced to make the asymptotic solution of $h_{ti}$ constant at the boundary($r \rightarrow \infty$), for the sake of convenience. 
From \eqref{Rmn} and \eqref{Epsi}, the linearized equations for the Fourier components are given as follows.\\
- Einstein equations:
\begin{align} \label{ee1}
\frac{q_m^2 h_{ti}}{r^4 U}+\epsilon_{ij}\frac{i \omega  q_m  a_j}{r^4 U}+\frac{\beta ^2  h_{ti}}{r^2 U}+\frac{i \beta  \omega   \psi _i}{r^2 U}-\frac{\mu   a_i'}{r^4}-\frac{4  h_{ti}'}{r}-h_{ti}''=&0 \,, \\
\epsilon_{ij}\frac{i U q_m  a_j'}{r^4 \omega }+\frac{i \beta  U  \psi _i'}{r^2 \omega }+\epsilon_{ij}\frac{i \mu  q_m  h_{{tj}}}{r^4 \omega }+\frac{\mu   a_i}{r^4}+ h_{ti}'=&0  \,, 
\end{align}
- Maxwell equations:
\begin{align} \label{me1}
\frac{U'  a_i'}{U}+\epsilon_{ij}\frac{i \omega  q_m  h_{{tj}}}{U^2}+\frac{\mu   h_{ti}'}{U}+\frac{\omega ^2  a_i}{U^2}+ a_i''=&0 \,, 
\end{align}
- Scalar equations:
\begin{align} \label{se1}
\frac{U'  \psi _i'}{U}-\frac{i \beta  \omega   h_{ti}}{U^2}+\frac{\omega ^2  \psi _i}{U^2}+\frac{2  \psi _i'}{r}+
\psi _x''=&0~.
\end{align}
Among these eight equations, only six are independent. 

Near the black hole horizon ($r \rightarrow 1$)\footnote{From here we set $r_h=1$ for convenience of numerical analysis. } the solutions are expanded as
\begin{equation} \label{nearh}
\begin{split}
&h_{ti} =  (r-1)^{\nu_\pm+1}(h_{ti}^{(I)} + h_{ti}^{(II)}(r-1) + \cdots),  \\
&a_i=  (r-1)^{\nu_\pm}(a_i^{(I)} + a_i^{(II)}(r-1) + \cdots )   ,\\
&\psi_i=  (r-1)^{\nu_\pm}(\psi_i^{(I)} + \psi_i^{(II)}(r-1) + \cdots ) \,,
\end{split}
\end{equation}
where $\nu_\pm = \pm i 4 \omega  /(-12 + q_m^2 + 2\beta^2 + \mu^2)$. In order to impose the incoming boundary condition relevant to the retarded Green's function~\cite{Son:2002sd}, we have to take $\nu = \nu_+$. It turns out that the 4 parameters $a_i^{(I)}$ and $\psi_i^{(I)}$ may be chosen to be independent  since  $h_{ti}^{(I)}$ and all higher power coefficients can be determined by them.

Near the boundary($r \rightarrow \infty$), the asymptotic solutions read
\begin{equation} \label{nearb}
\begin{split}
&h_{ti} =   h^{(0)}_{ti} + \frac{1}{r^2} h^{(2)}_{ti} + \frac{1}{r^3}h_{ti}^{(3)}+\cdots,  \\
&a_i=a_i^{(0)} + \frac{1}{r}a_i^{(1)}+ \cdots,\\
&\psi_i = \psi_i^{(0)} + \frac{1}{r^2} \psi^{(2)}_i + \frac{1}{r^3}\psi^{(3)}_i + \cdots~,
\end{split}
\end{equation}
where the leading terms $h^{(0)}_{ti}, a_i^{(0)}$, and $\psi_i^{(0)} $ are independent constants, which fix $ h^{(2)}_{ti} ,\psi^{(2)}_i$ completely by the equations of motion.
The nontrivial sub-leading terms, $h_{\mu\nu}^{(3)}, a_\mu^{(1)},$ and $\psi^{(3)}_I$ will be determined by the incoming boundary conditions at the horizon for the given leading terms, $h^{(0)}_{ti}, a_i^{(0)}$, and $\psi_i^{(0)}$.
The leading terms play the role of sources for the operators of which expectation values are related to $h_{\mu\nu}^{(3)}, a_\mu^{(1)},$ and $\psi^{(3)}_I$ respectively.

 Expanding the renormalized action (\ref{Sren}) around the dyonic black brane background and using the equations of motion, we obtain a quadratic on-shell action:
\begin{equation}
\begin{split}
S^{(2)}_{\mathrm{ren}}= \lim_{\Lambda \to \infty}\frac{1}{2}\int_{r=\Lambda} \dd^3 x
& \left [ \left(4 r^3 -\frac{ 4 r^4}{\sqrt{U(r)}}\right){\delta\tilde h}_{{ti}}^2-\frac{\beta  r^2  }{\sqrt{U(r)}}\delta\chi_i  \dot{{\delta\tilde h}}_{{ti}} \right. +\frac{r^2  }{\sqrt{U(r)}}\delta\chi_i  {\delta \ddot \chi }_i
\\-U(r) {\delta A}_i {\delta A}_i' & +r^4 {\delta\tilde h}_{{ti}} {\delta\tilde h}_{{ti}}'-r^2 U(r)  \delta\chi_i \delta\chi_i ' \left.-{\delta\tilde h}_{{ti}} \left(\mu  {\delta A}_i-\frac{\beta  r^2 \dot{\delta\chi_i }}{\sqrt{U(r)}}\right)~
 \right] \,,
\end{split}
\end{equation}
where $\delta \tilde h_{ti} \equiv r^{-2} \delta g_{ti}(t,r)$, `dot' denotes $\partial_t$ and `prime' denotes $\partial_r$.  We have dropped the  contributions from the horizon, which is the prescription for the retarded Green's function \cite{Son:2002sd}. In particular, with the spatially homogeneous ansatz \eqref{flucA}-\eqref{flucPsi}, the quadratic action in momentum space yields
\begin{equation} \label{onshell2}
S^{(2)}_{\mathrm{ren}} = \frac{\mathcal{V}_2 }{2} \int_0^{\infty} \frac{ \dd\omega}{2\pi}    \left(   -  \mu  \bar a_i^{(0)}  h_{ti}^{(0)}   -2m_0  \bar h_{ti}^{(0)}  h_{ti}^{(0)}  + \bar a_i^{(0)} a_i^{(1)} -3 \bar h_{ti}^{(0)} h_{ti}^{(3)}    + 3 {\bar \psi}^{(0)} \psi^{(3)} \right) \,,
\end{equation}
where $\mathcal{V}_2$ is the two dimensional spatial volume $\int \dd x \dd y $. The argument of the variables with the bar is $-\omega$. We dropped the range $(-\infty,0)$, which is the complex conjugate of \eqref{onshell2},  to obtain complex two point functions \cite{Son:2002sd}.

The on-shell action \eqref{onshell2} is nothing but the generating functional for the two-point Green's functions sourced by $a_i^{(0)}, h_{ti}^{(0)}$, and ${\psi}^{(0)}$. We may simply read off a part of the two point functions from the first two terms in \eqref{onshell2}.  The other three terms are nontrivial and we need to know the dependence of $\{a_i^{(1)}, h_{ti}^{(3)}, \psi^{(3)}\}$ on $\{a_i^{(0)}, h_{ti}^{(0)},{\psi}^{(0)}\}$.  However, the linearity of the equations (\ref{ee1})-(\ref{se1}) makes it easy to find out a linear relation between $\{a_i^{(1)}, h_{ti}^{(3)}, \psi^{(3)}\}$ and $\{a_i^{(0)}, h_{ti}^{(0)},{\psi}^{(0)}\}$. In the following subsection
we will explain how to find such a relationship numerically in a more general setup  and apply it to our case.

\subsection{Numerical method}

A systematic numerical method with multi fields and constraints was developed in \cite{Kim:2014bza} based on \cite{Amado:2009ts,Kaminski:2009dh}. We summarize it briefly for the present case and refer to \cite{Kim:2014bza,Kim:2015sma} for more details. Let us start with $N$ fields  $\Phi^a(x,r)$, ($a=1,2,\cdots, N$), which are fluctuations around a background.  Suppose that they  satisfy a set of coupled $N$ second order differential equations and the fluctuation fields depend on only $t$ and $r$: 
\begin{equation}
\Phi^a(t,r) = \int \frac{\dd \omega}{(2\pi)}  e^{-i\omega t}  r^p \Phi^a_\omega(r)\,, \label{newphi}
\end{equation}
where $r^p$ is multiplied such that the solution  $\Phi^a_\omega (r) $ goes to constant at the boundary($r \rightarrow \infty$). For example, $p=2$ in \eqref{flucg}.

Near horizon($r=1$), solutions can be expanded as
\begin{equation} \label{incoming}
\Phi^a(r) = (r-1)^{\nu_{a\pm}} \left( \varphi^{a} + \tilde{\varphi}^{a} (r-1) + \cdots \right) \,,
\end{equation}
where we omitted the subscript $\omega$ for simplicity and $\nu_{a+}(\nu_{a-})$ corresponds to the incoming(outgoing) boundary condition. In order to compute the retarded Green's function we choose the incoming boundary condition \cite{Son:2002sd}.  This choice reduces the number of independent parameter from $2N$ to $N$. There may be further reductions by  $N_\mathrm{c}$ if there are $N_\mathrm{c}$ constraint equations.  As a result, the number of independent parameter is $N-N_\mathrm{c}$  so we may choose $N-N_\mathrm{c}$ initial conditions, denoted by  ${\varphi^{a}}_{ \hat i }$($\hat i =1,2,\cdots, N-N_\mathrm{c}$):
\begin{equation} \label{init}
\begin{pmatrix}
    {\varphi^{a}}_{1} \ & {\varphi^{a}}_{2}\ & {\varphi^{a}}_{3}\ &  \ldots \ &  {\varphi^{a}}_{N-N_\mathrm{c}}
\end{pmatrix}
   =
\begin{pmatrix}
    1 & 1& 1&  \ldots & {} \\
    1 & -1& 1 & \ldots & {} \\
   1 & 1& -1 & \ldots & {} \\
    \vdots & \vdots & \vdots  & \ddots & {} \\
    1 & 1 & 1 & \ldots & {}
\end{pmatrix} \,.
\end{equation}
 Every  column vector ${{\varphi^a}_{\hat {i}}}$   yields a solution with the incoming boundary condition, denoted by ${{\Phi}^a}_{\hat i} (r)$, which is expanded as
\begin{equation}
{\Phi^a}_{\hat i}(r)  \rightarrow   \mathbb{S}_{{\hat i}}^{a}  + \cdots +  \frac{\mathbb{O}_{{\hat i}}^{a}}{r^{\delta_a}}  + \cdots   \qquad (\mathrm{near\ boundary})\,,
\end{equation}
where  ${\mathbb{S}^a}_{\hat i}$ are related to the {\it{sources}}, which are the leading terms of ${\hat i}$-th  solution,  and  ${\mathbb{O}^a}_{{\hat i}}$ are related to the expectation values of the {\it{operators}} corresponding to the sources($\delta_a \ge 1$). A general solution constructed by ${\Phi^a}_{\hat i}(r)$ is 
\begin{align} \label{GS}
\Phi_{\mathrm{in}}^a(r) \equiv \Phi_{\hat{i}}^{a}(r) c^{\hat{i}} \rightarrow   \mathbb{S}_{\hat{i}}^{a} c^{\hat{i}}  + \cdots  +  \frac{ \mathbb{O}_{\hat{i}}^a c^{\hat{i}} }{r^{\delta_a}}  + \cdots
 \qquad (\mathrm{near\ boundary})  \,,
\end{align}
with real constants $c^{\hat{i}}$. We want to identify $\mathbb{S}_{\hat{i}}^{a} c^{\hat{i}}$ with the independent sources $J^a$ but if there are constraints, it is not possible since $a>\hat{i}$. However, in this case,  there may be $N_\mathrm{c}$ other solutions corresponding to some residual gauge transformations~\cite{Kim:2014bza,Kim:2015sma}
\begin{equation}
{\Phi^a}_{\bar i}(r)  \rightarrow   \mathbb{S}_{{\bar i}}^{a}  + \cdots +  \frac{\mathbb{O}_{{\bar i}}^{a}}{r^{\delta_a}}  + \cdots   \qquad (\mathrm{near\ boundary})\,,
\end{equation}
where $\bar i$ runs from $N-N_\mathrm{c} +1$ to $N$.   This extra basis set, $\Phi^a_{\mathrm{c}}(r)  =   \Phi_{\bar i}^{a} c^{\bar i}$, generates a general solution together with the ingoing solutions.

In our case, $N=6$ and $N_\mathrm{c} = 2$, which corresponds to the constraints $g_{ri} = 0$.  
There are two sets of additional constant solutions of the equations of motion (\ref{ee1})-(\ref{se1})
\begin{equation}
h_{ti} = h_{ti}^0\,, \quad a_i = -\frac{i q_m \epsilon_{ij} h_{tj}^0}{\omega} \,, \quad \chi_i = \frac{i \beta h_{ti}^0}{\omega} \,,
\end{equation}
where $h_{ti}^0$ is arbitrary constant and $i, j =1,2$.
Therefore, the explicit expression for $\mathbb{S}^a_{\bar{i}}$ is 
 \begin{equation}
{\mathbb{S}^a}_{\bar i}=
\left(
\begin{array}{cc}
 {\mathbb{S}^a}_{5}   &
{\mathbb{S}^a}_{6}  
\end{array}
\right)
  =   \left(
\begin{array}{cc}
  0& -i \frac{q_m}{\omega} \\ 
   i \frac{q_m}{\omega}  &0   \\ 
   1&0 \\
    0& 1  \\ 
    i \frac{\beta}{\omega}&0  \\
    0& i \frac{\beta}{\omega}
\end{array}
\right) .  \nonumber 
\end{equation}
These can be understood as residual gauge transformations keeping $g_{ri} = 0$, which are generated by the vector fields, of which non-vanishing components are $\xi^x = \epsilon^x  e^{-i\omega t}$ and $\xi^y = \epsilon^y  e^{-i\omega t}$ ($\epsilon^i$ are constants). i.e. ${\cal L}_\xi A_i= - q_m \epsilon_{ij}  \xi^j$, ${\cal L}_\xi g_{ti}=-i\omega r^2 \delta_{ij} \xi^j$ and  ${\cal L}_\xi \chi_i = \beta \delta_{ij}\xi^j$.

Therefore, the most general solution reads
\begin{align}\label{general sol}
 \Phi^a_{\mathrm{in}}(r) + \Phi^a_{\mathrm{c}} (r)   \equiv J^a + \cdots + \frac{R^a}{r^{\delta_a}} + \cdots \,,
\end{align}
where we defined $J^a$ and $R^a$. For arbitrary sources $J^a$ we can always find $c^I$  
\begin{equation}
 c^I = {(\mathbb{S}^{-1})^{I}}_{a} J^a \,,
\end{equation}
where $I = 1, \dots, N$. 
The corresponding response $R^a$ is expressed as
\begin{equation} \label{response1}
R^a =  {\mathbb{O}^a}_{I}  c^{I} =  {\mathbb{O}^a}_{I} {(\mathbb{S}^{-1})^{{I}}}_{b} J^b \,.
\end{equation}

The general on-shell quadratic action in terms of the sources and the responses can be written as
\begin{equation} \label{sb}
S_{\mathrm{ren}}^{(2)}
= \frac{{\mathcal V}_2}{2} \int_0^\infty \frac{\dd \omega}{(2\pi)}  \left[ \bar J^a \mathbb{A}_{a b} J^b
+  \bar J^a \mathbb{B}_{a b} {R^b} \right]  ,
\end{equation}
where $\mathbb{A}$ and $\mathbb{B}$ are regular matrices of order $N$ and the argument of $\bar{J}^a$ is $-\omega$. For example, the action \eqref{onshell2} is the case with: 
\begin{equation}
J^a =
\begin{pmatrix}
    a_x^{(0)}  \\
     a_y^{(0)}  \\   
    h_{tx}^{(0)} \\
    h_{ty}^{(0)} \\
   \psi_x^{(0)} \\
    \psi_y^{(0)} \\
\end{pmatrix}\,, \quad
R^a =
\begin{pmatrix}
    a_x^{(1)}  \\
        a_y^{(1)}  \\
    h_{tx}^{(3)} \\
        h_{ty}^{(3)} \\
   \psi_x^{(3)} \\
      \psi_y^{(3)} \\
\end{pmatrix}\,, \quad
\mathbb{A} = \begin{pmatrix}
    0 & -\mu& 0  \\
   0 & -2m_0& 0  \\
   0 & 0 & 0  \\
\end{pmatrix} \otimes \mathbf{1}_{2}  \,, \quad
   \mathbb{B}= \begin{pmatrix}
 1 & 0 & 0 \\
 0 & -3 & 0 \\
 0 & 0 & 3 \\
\end{pmatrix} \otimes \mathbf{1}_{2}  \,,
\end{equation}
where the index $\omega$ is suppressed and $\mathbf{1}_2$ is the $2\times 2$ unit matrix.  Plugging the relation  \eqref{response1}  into the action  \eqref{sb}   we have
\begin{equation} \label{Gab}
S_{\mathrm{ren}}^{(2)}
 = \frac{{\mathcal V}_2}{2}  \int_0^\infty \frac{\dd \omega}{(2\pi)}   {\bar J}^a \left[\mathbb{A}_{a b} + \mathbb{B}_{ac}{\mathbb{O}^c}_{I} {(\mathbb{S}^{-1})^{I}}_{b}\right]J^b    \,,
\end{equation}
which yields the retarded Green's function
\begin{align}
G_{ab}  =  \mathbb{A}_{a b} + \mathbb{B}_{ac}{\mathbb{O}^c}_{I} {(\mathbb{S}^{-1})^{I}}_{b}~.
\end{align}

In summary, to compute the retarded Green's function, we need four square $N \times N$ matrices, $\mathbb{A}, \mathbb{B}, \mathbb{S}$ and  $\mathbb{O}$. The matrices $\mathbb{A}$ and $\mathbb{B}$ can be read off from the action \eqref{sb},  which is given by the on-shell expansion near the boundary. The matrices $\mathbb{S}$ and $\mathbb{O}$ are obtained by solving a set of the differential equations. Part of them  comes from the solutions with incoming boundary conditions and the others may be related to the constraints. Notice that the Green's functions do not depend on the choice of initial conditions \eqref{init}. 

In the case of the dyonic black branes in section \ref{sec41},  we may construct a $6 \times 6$ matrix of the retarded Green's function. We will focus on the $4 \times 4$ submatrix corresponding to $a_{i}^{(0)}$ and $h_{t i}^{(0)}$ in \eqref{nearb},
\begin{equation}
\left(\begin{array}{cc}   G_{JJ}^{ij} & G_{JT}^{ij} \\   G_{TJ}^{ij}  &  G_{TT}^{ij} \end{array}\right) \,,
\end{equation}
where every $G^{ij}_{\alpha\beta}$ is a $2\times 2$ retarded Green's function with $i=x,y$ for given $\alpha$ and $\beta$. The sub-induces $\alpha,\beta$ denote the operators corresponding to the sources. i.e. The $a_{i}^{(0)}$ is dual to the electric current $J^{i}$ and the $h_{t i}^{(0)}$ is dual to the energy-momentum tensor $T^{t i}$.
From the linear response theory,  we have the following relation between the response functions and the sources:
\begin{equation}
\label{theo}
\left(\begin{array}{c}\langle J^{i} \rangle \\ \langle T^{t i} \rangle \end{array}\right)=
\left(\begin{array}{cc}   G_{JJ}^{ij} & G_{JT}^{ij} \\   G_{TJ}^{ij}  &  G_{TT}^{ij} \end{array}\right)
\left(\begin{array}{c}  a_{j}^{(0)} \\  h_{t j}^{(0)}\end{array}\right)\,,
\end{equation}
where $\langle J^{i} \rangle,\langle T^{t i} \rangle, a_{j}^{(0)} $ and $h_{t j}^{(0)}$ are  understood as $2 \times 1$ column matrices, with $i=x,y$. 
We want to relate these Green's functions  to the electric($\hat{\sigma}$), thermal($\hat{\bar{\kappa}}$), and thermoelectric($\hat{\alpha}, \hat{\bar{\alpha}}$)  conductivities defined as
\begin{equation}
\label{pheno}
\left(\begin{array}{c}\langle J^{i} \rangle \\ \langle Q^{i} \rangle \end{array}\right)
=
\left(\begin{array}{cc}   \hat{\sigma}^{ij} & \hat{\alpha}^{ij} T \\  \hat{ \bar{\alpha}}^{ij} T & \hat{\bar{\kappa}}^{ij} T \end{array}\right)
\left(\begin{array}{c} E_{j} \\ - (\nabla_{j} T)/T\end{array}\right)~,
\end{equation}
where  $Q^{i}$ is a heat current, $E_{i}$ is an electric field and $\nabla_{i} T$ is a temperature gradient along the $i$ direction. Notice that the electric and heat current here contain the contribution of magnetization, so we use the conductivities with hat(\eqref{DCtrans01}-\eqref{DCtrans04}). 
By taking into account diffeomorphism invariance \cite{Hartnoll:2009sz,Herzog:2009xv,Kim:2014bza}, \eqref{pheno} can be expressed as
\begin{equation}
\label{pheno1}
\left(\begin{array}{c}\langle J^{i} \rangle \\ \langle T^{t i} \rangle -\mu \langle J^{i} \rangle \end{array}\right)
=
\left(\begin{array}{cc}   \hat{\sigma}^{ij} & \hat{\alpha}^{ij} T \\  \hat{ \bar{\alpha}}^{ij} T & \hat{\bar{\kappa}}^{ij} T \end{array}\right)
\left(\begin{array}{c} i\omega  ( a_{j}^{(0)} + \mu  h_{t j}^{(0)})   \\   i \omega  h_{t j}^{(0)} \end{array}\right).
\end{equation}
From \eqref{theo} and \eqref{pheno1} with the magnetization subtraction \eqref{DCtrans1}-\eqref{DCtrans3},   the conductivities are expressed in terms of the retarded Green's functions as follows
\begin{equation} \label{pheno2}
\left(\begin{array}{cc}   \sigma^{ij} & \alpha^{ij} T \\   \bar{\alpha}^{ij} T & \bar{\kappa}^{ij} T \end{array}\right) =
\left(
\begin{array}{cc}
 -\frac{i G_{JJ}^{ij}}{\omega }  & \frac{i (\mu G_{JJ}^{ij} -G_{JT}^{ij})}{\omega } \\
 \frac{i (\mu G_{JJ}^{ij}  -G_{TJ}^{ij})}{\omega } & -\frac{i (G_{TT}^{ij} -G_{TT}^{ij}(\omega=0)-\mu  (G_{JT}^{ij}+ G_{TJ}^{ij}-\mu G_{JJ}^{ij}  ))}{\omega } \\
\end{array}
\right)
-
\frac{B}{T}\left(
\begin{array}{cc}
0 & \Sigma_1 \epsilon^{ij}\\
 \Sigma_1 \epsilon^{ij} &  \Sigma_2 \epsilon^{ij} \\
\end{array}
\right).
\end{equation}

\subsection{AC conductivities and the cyclotron poles}
In this section we present our numerical results. Some examples of the AC electric conductivity are shown in Figure \ref{fig:Hall1} and \ref{fig:Hall2}; the thermoelectric conductivity is in Figure  \ref{fig:Thermo}; and the thermal conductivity is in Figure \ref{fig:Thermal}.
\begin{figure}[]
\centering
  \subfigure[Re ${\sigma}^{xx}$ ] 
   {\includegraphics[width=4cm]{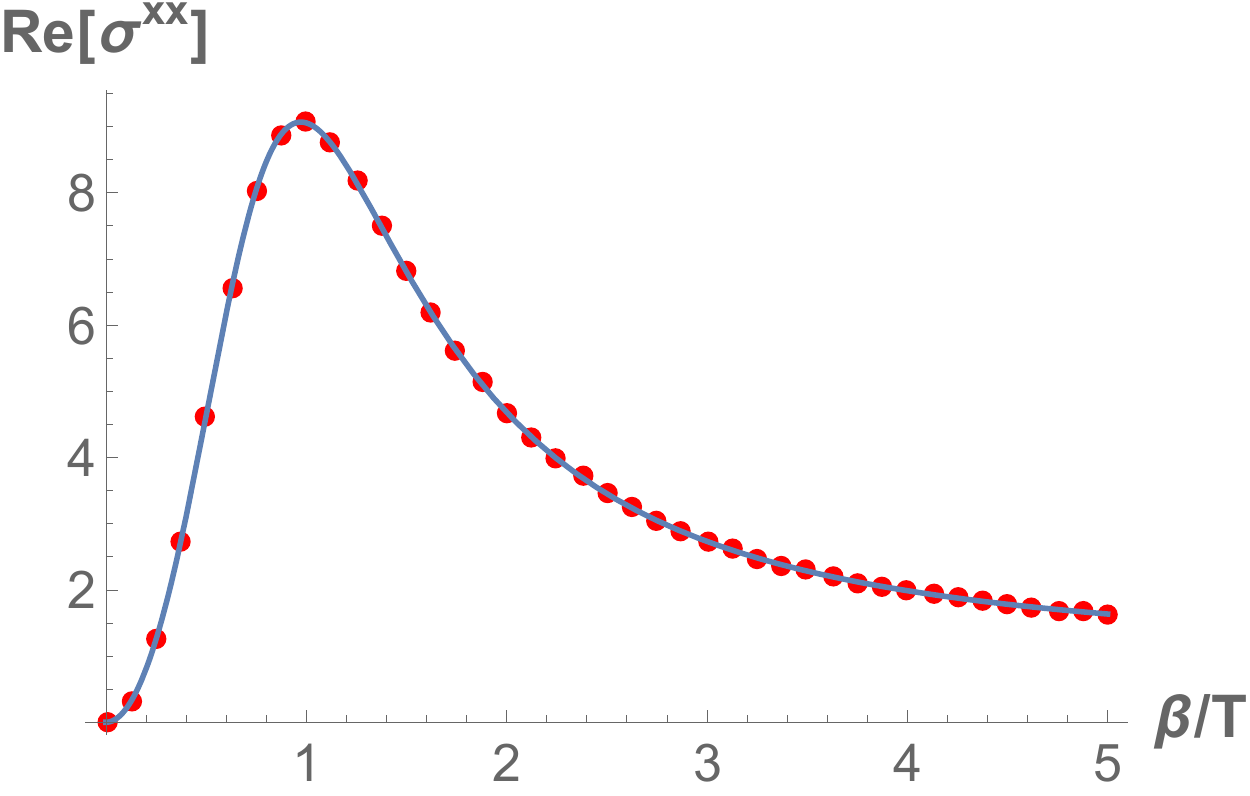} \label{}} \ \
     \subfigure[Re ${\alpha}^{xx}$ ]
   {\includegraphics[width=4cm]{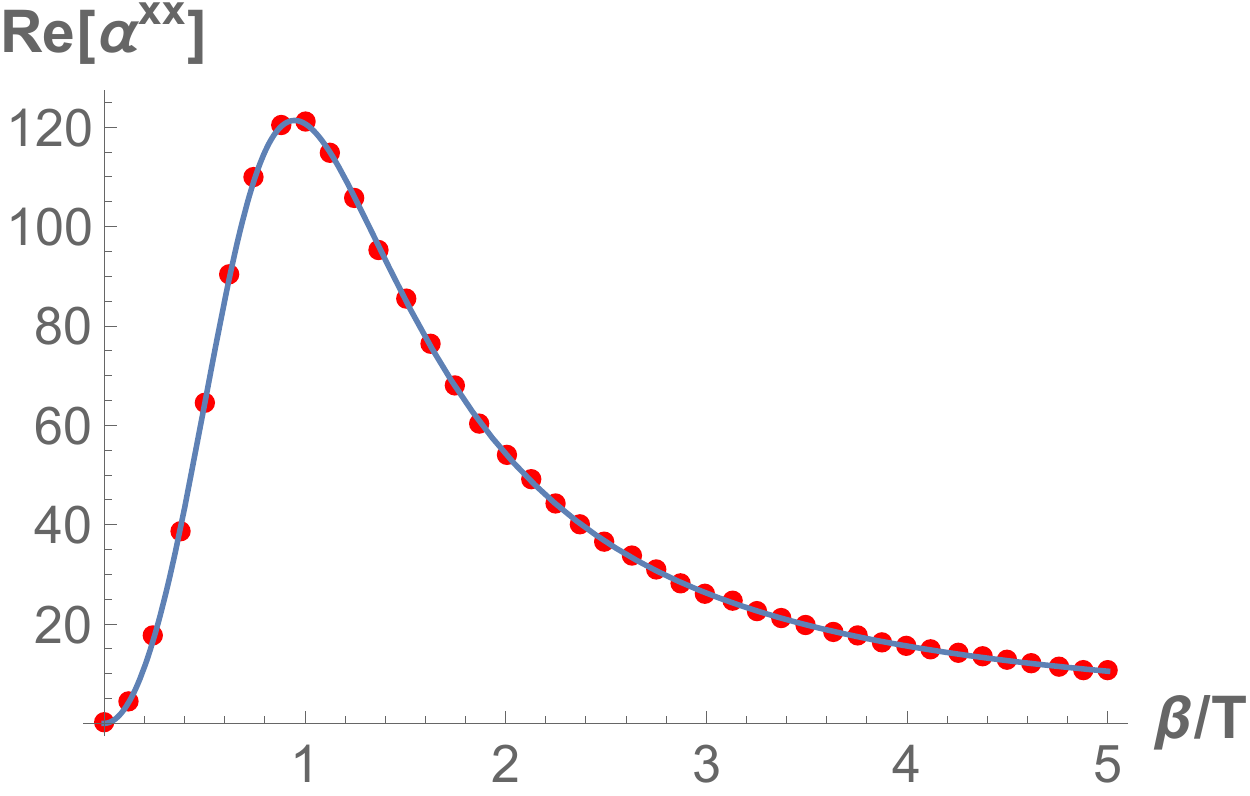} \label{}} \ \
        \subfigure[Re ${\bar{\kappa}}^{xx}$]
   {\includegraphics[width=4cm]{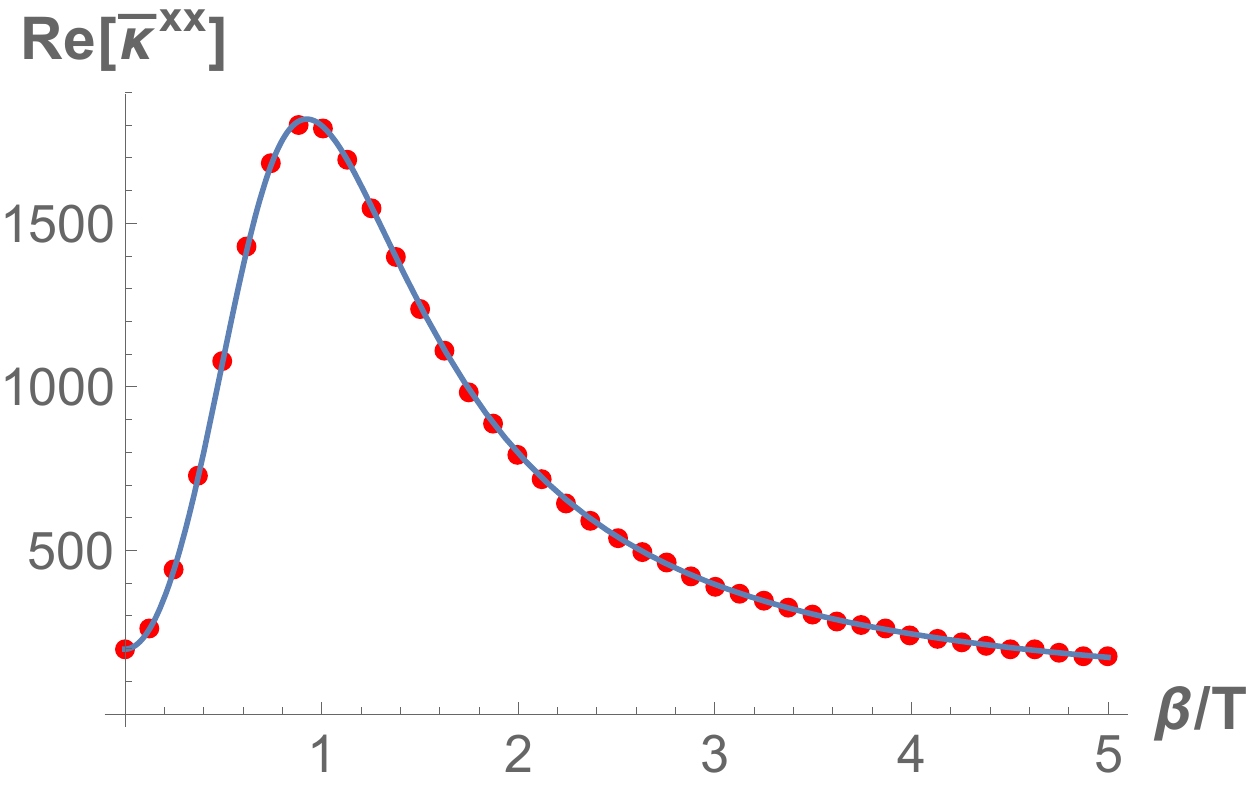} \label{} }
        \subfigure[Re ${\sigma}^{xy}$]
   {\includegraphics[width=4cm]{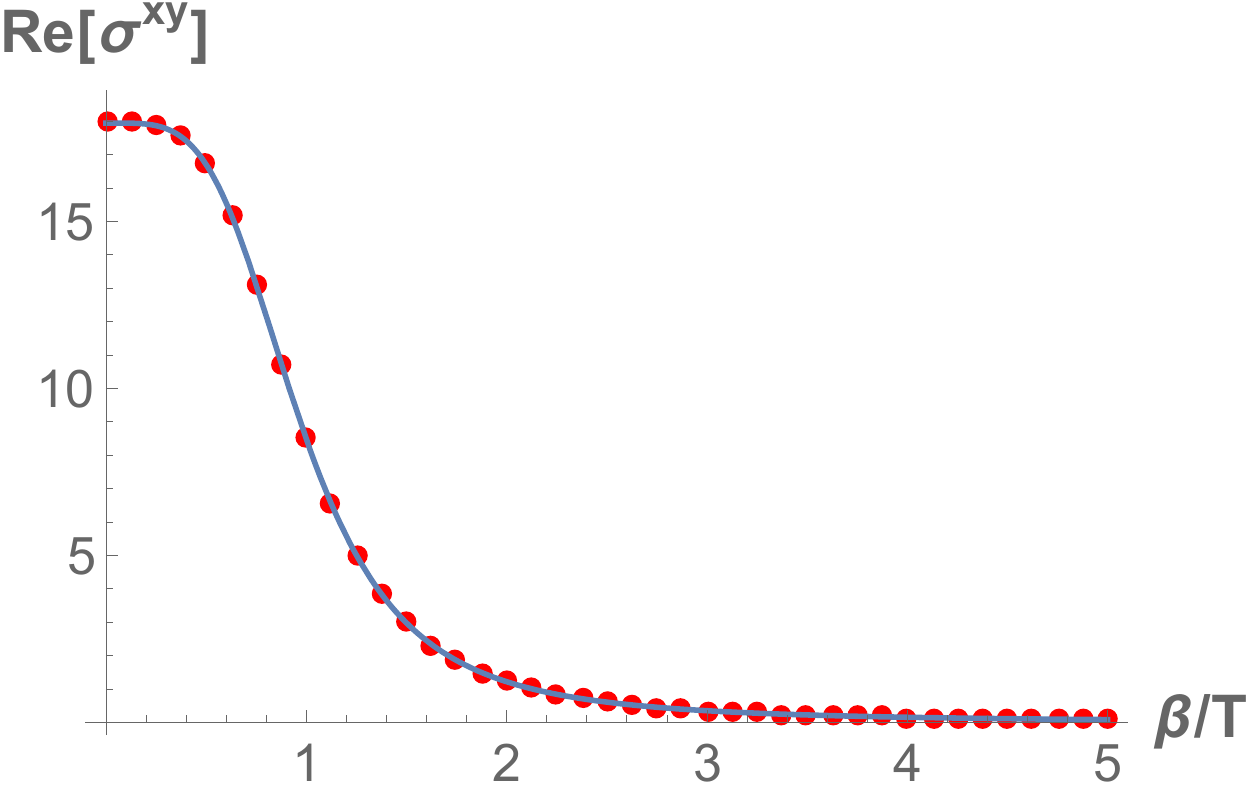} \label{} }\ \  
     \subfigure[Re ${\alpha}^{xy}$]
   {\includegraphics[width=4cm]{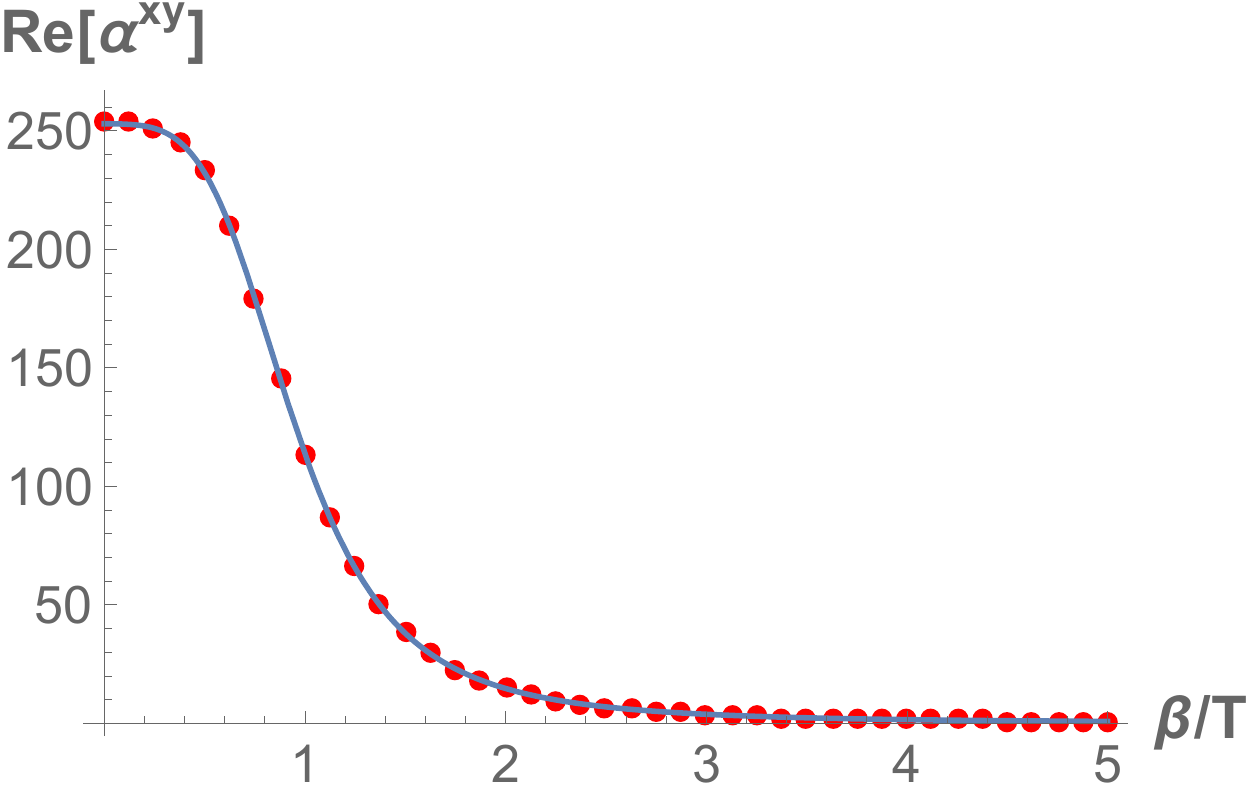} \label{} }\ \
     \subfigure[Re ${\bar{\kappa}}^{xy}$]
   {\includegraphics[width=4cm]{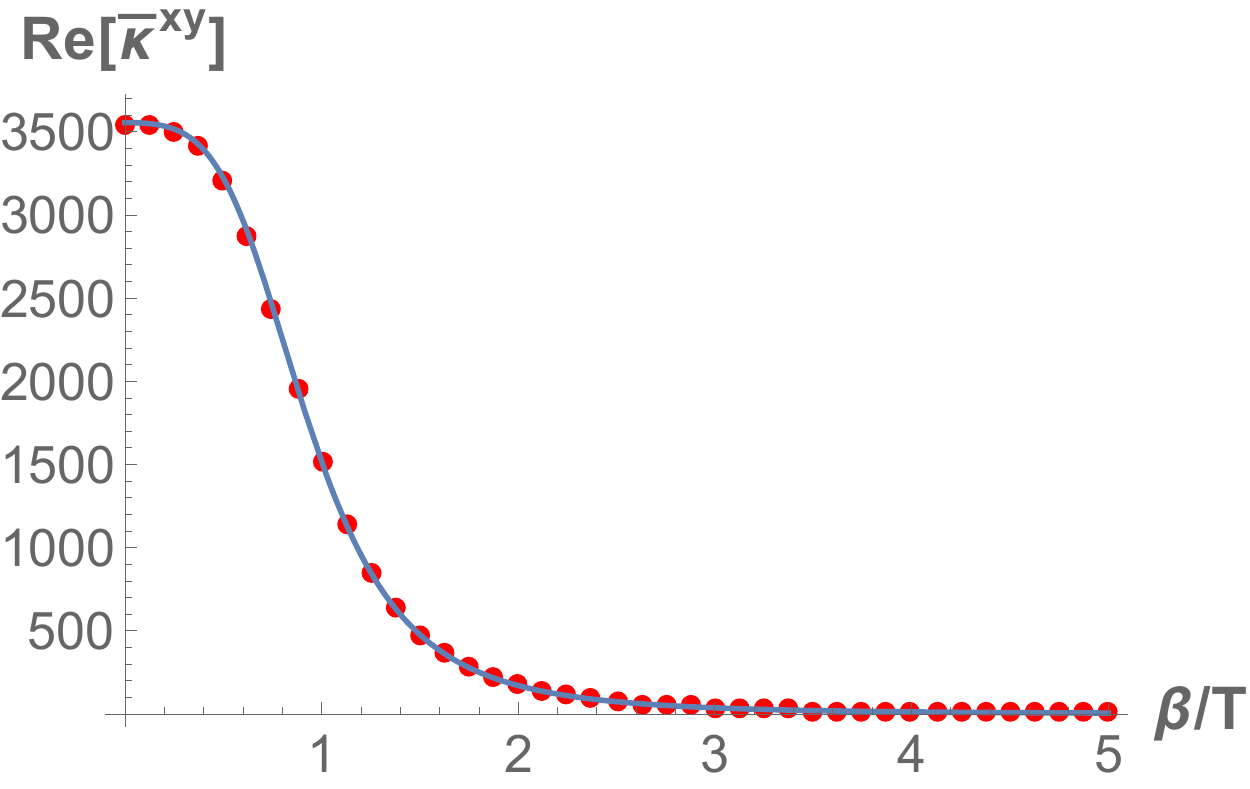} \label{} }
 \caption{ Agreement of DC analytic formulas(solid curves) obtained in section \ref{sec:DCconduct} with numerical results(red dots): DC conductivities vs $\beta/T$ with $\mu/T=4$, $B/T^2 =1$  
           } \label{fig:numericalDC}
\end{figure}

Before discussing the AC nature of the conductivities, we first examine the DC limit($\omega \rightarrow 0$) of the AC conductivities and compare them with the analytic expressions derived in  section \ref{sec:DCconduct}. The comparisons are shown in Figure \ref{fig:numericalDC}, where all conductivities are plotted as a function of $\beta/T$ with the other parameters fixed; $\mu/T=4$ and $B/T^2=1$. The solid lines were drawn by the analytic expressions,  \eqref{dyonicSxx}, \eqref{dyonicSxy}, \eqref{t1}-\eqref{t4}  and the red dots were read off from the AC numerical results in the limit $\omega \rightarrow 0$.   Both results agree, which serves as a supporting evidence for the validity of our analytic and numerical methods. Indeed this agreement is not so trivial in technical perspective. In the DC computation we turned on $h_{ri}$ and read off the physics from the horizon data while in the AC computation we worked in the gauge $h_{ri}=0$ and considered full evolution in the $r$-direction.

Let us turn to the AC properties of the conductivities. Figure \ref{fig:Hall1}, \ref{fig:Thermo}, and  \ref{fig:Thermal}  show the $\beta$ dependence of the electric, thermoelectric, and thermal conductivity  respectively.  The dotted lines are the cases for $\beta=0$ and the red, green, blue curves are for $\beta/T=0.5,1,1.5$. The red dots at $\omega=0$ in the real part of the conductivities are the values given by  the analytic formulas  \eqref{dyonicSxx}, \eqref{dyonicSxy}, \eqref{t1}-\eqref{t4}. They agree to the numerical AC conductivities in the limit $\omega \rightarrow 0$.   As $\beta$ increases, when all other scales are fixed,  the  curves become flatter, which is expected from stronger momentum relaxation.  Notice that there is no $1/\omega$ pole in the imaginary part of the conductivities even when $\beta =0$.  It is because the gauge field for $B$ \eqref{ansatz2}  breaks translation invariance in the same way as the axion fields do.

\begin{figure}[]
\centering
  \subfigure[Re ${\sigma}^{xx}$ ]
   {\includegraphics[width=3.5cm]{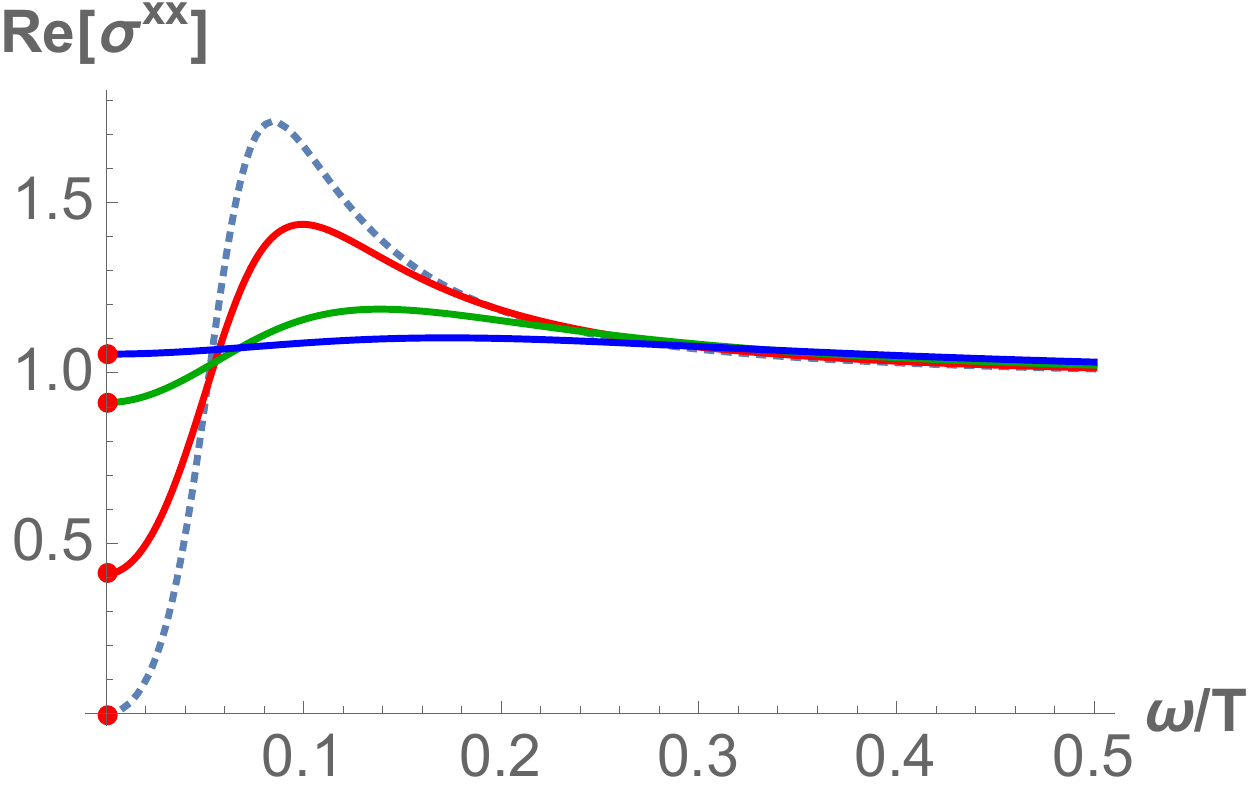} \label{}} 
     \subfigure[ Im ${\sigma}^{xx}$ ]
   {\includegraphics[width=3.5cm]{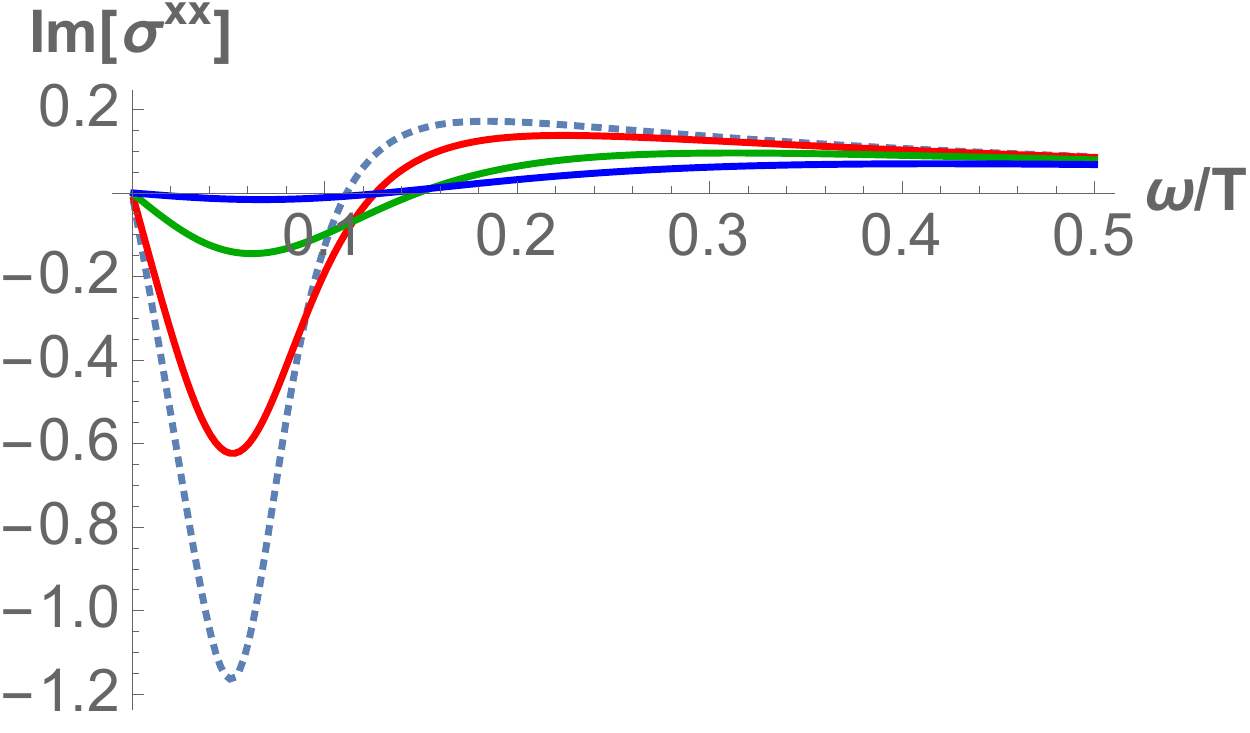} \label{}} \  \  
      \subfigure[Re ${\sigma}^{xy}$]
   {\includegraphics[width=3.5cm]{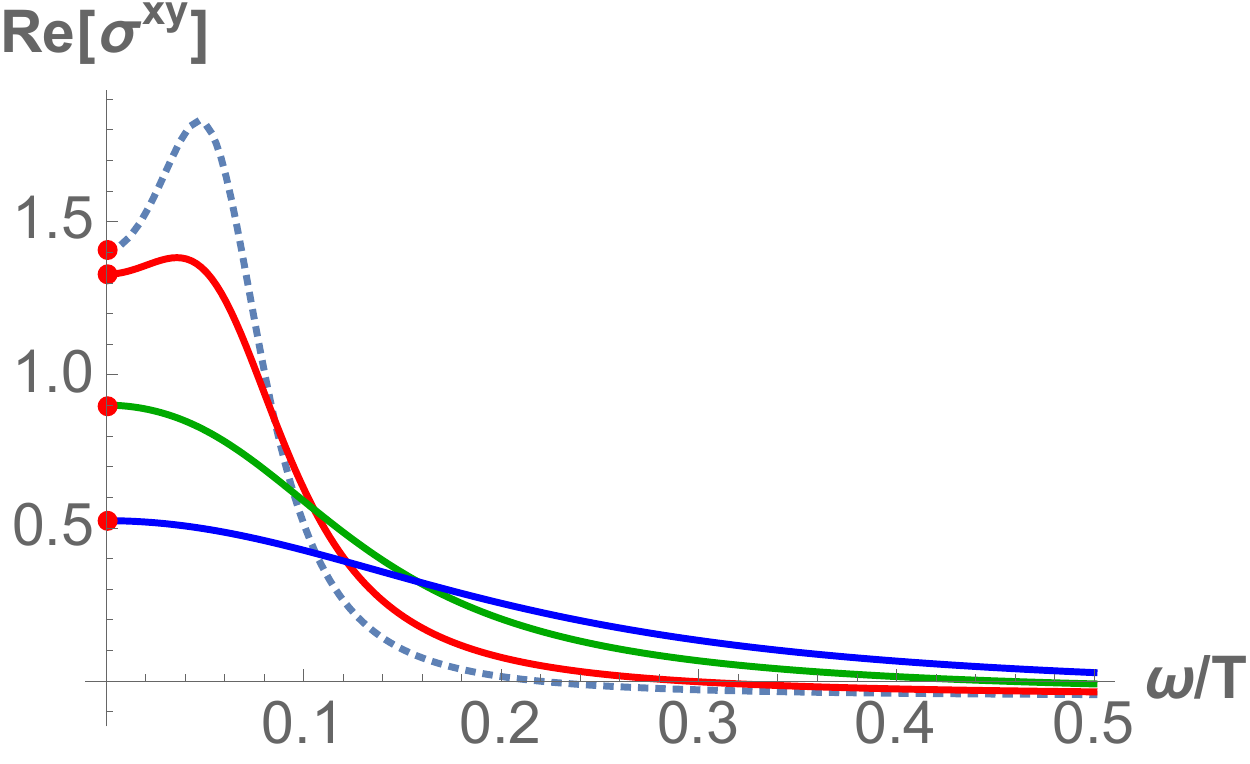} \label{} }
     \subfigure[Im $Ò{\sigma}^{xy}$]
   {\includegraphics[width=3.5cm]{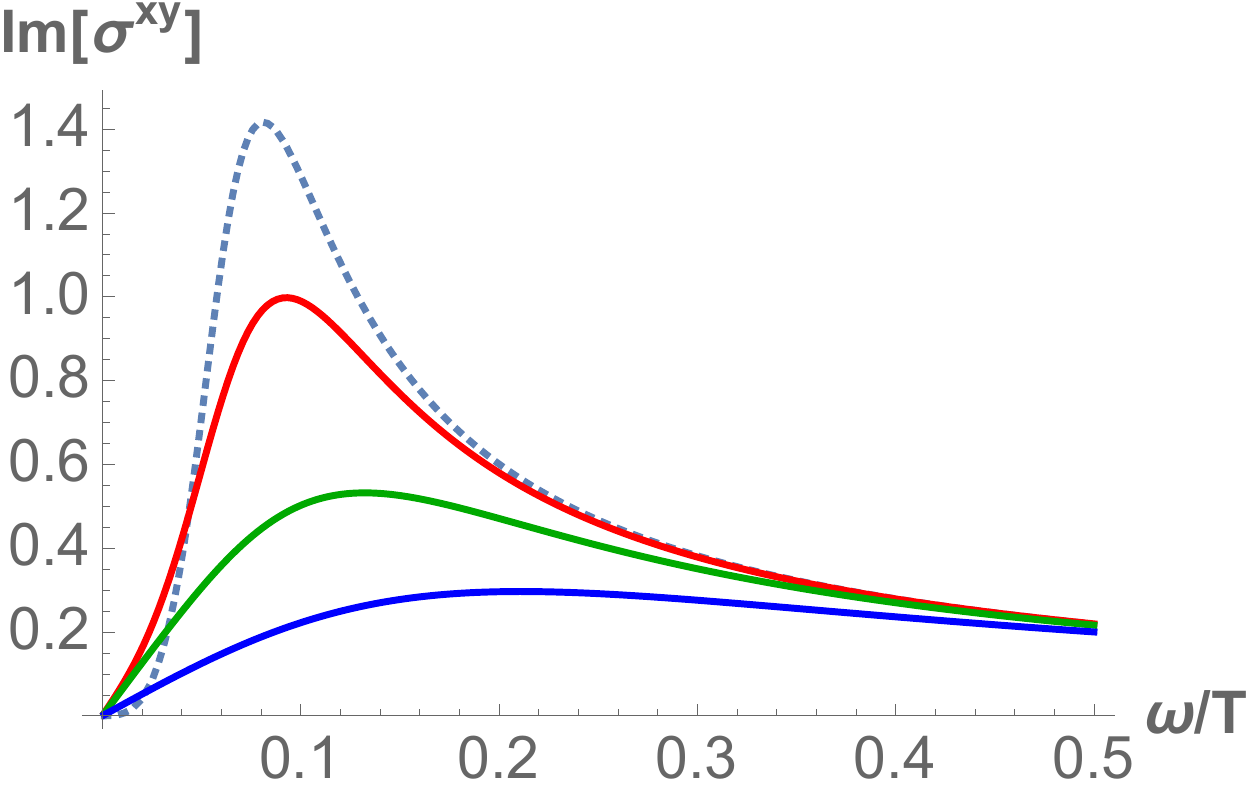} \label{} }
 \caption{$\beta$ dependence of electric conductivities: $\beta /T= 0, \ 0.5, \ 1, \ 1.5$(dotted, red, green, blue) with $\mu/T = 1, \ B/T^2=3$} \label{fig:Hall1}
\end{figure}
\begin{figure}[]
\centering
  \subfigure[Re ${\alpha}^{xx}$ ]
   {\includegraphics[width=3.5cm]{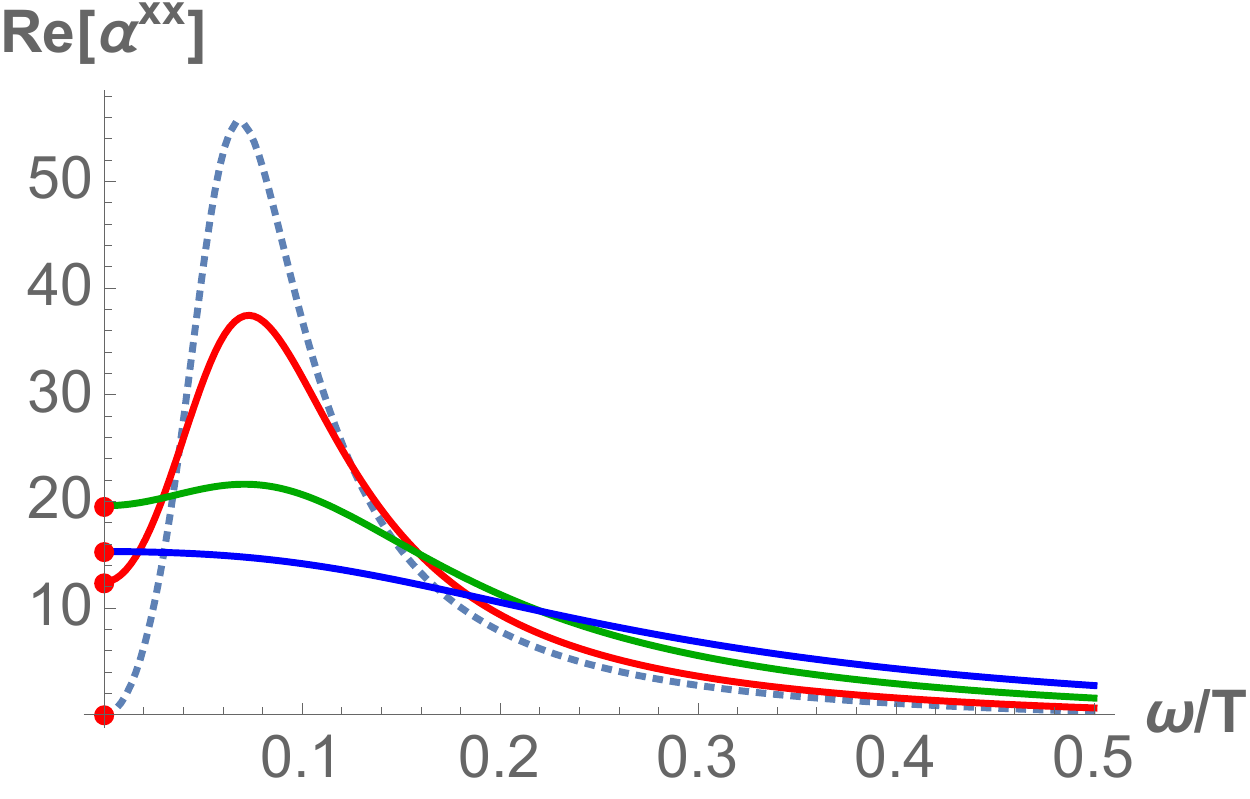} \label{}}
        \subfigure[ Im ${\alpha}^{xx}$ ]
   {\includegraphics[width=3.5cm]{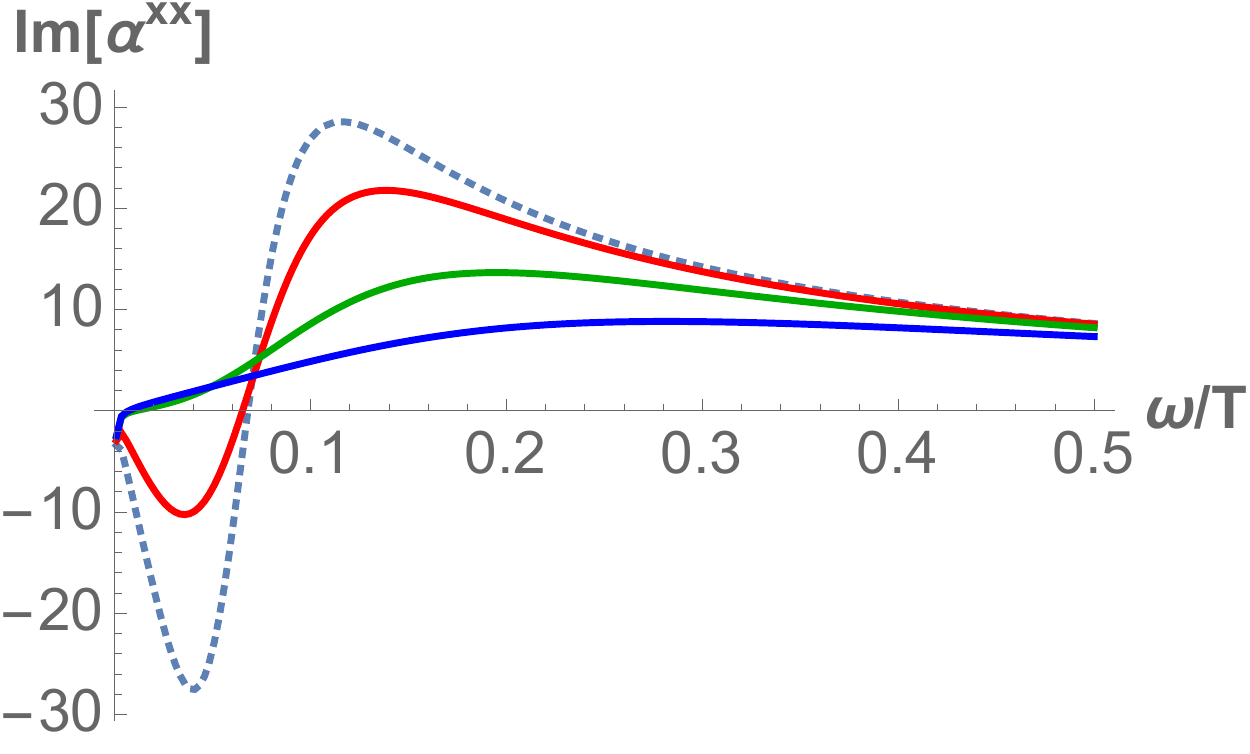} \label{}} \ \
     \subfigure[Re ${\alpha}^{xy}$]
   {\includegraphics[width=3.5cm]{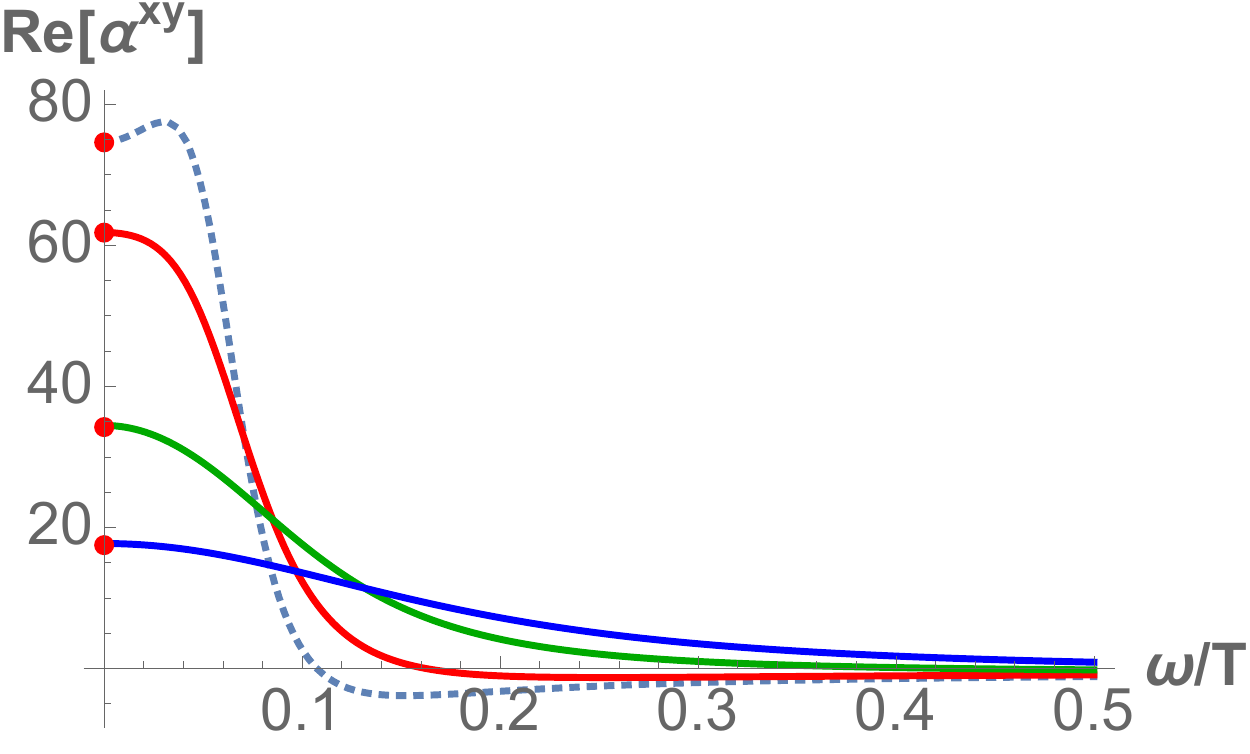} \label{} } 
     \subfigure[Im ${\alpha}^{xy}$]
   {\includegraphics[width=3.5cm]{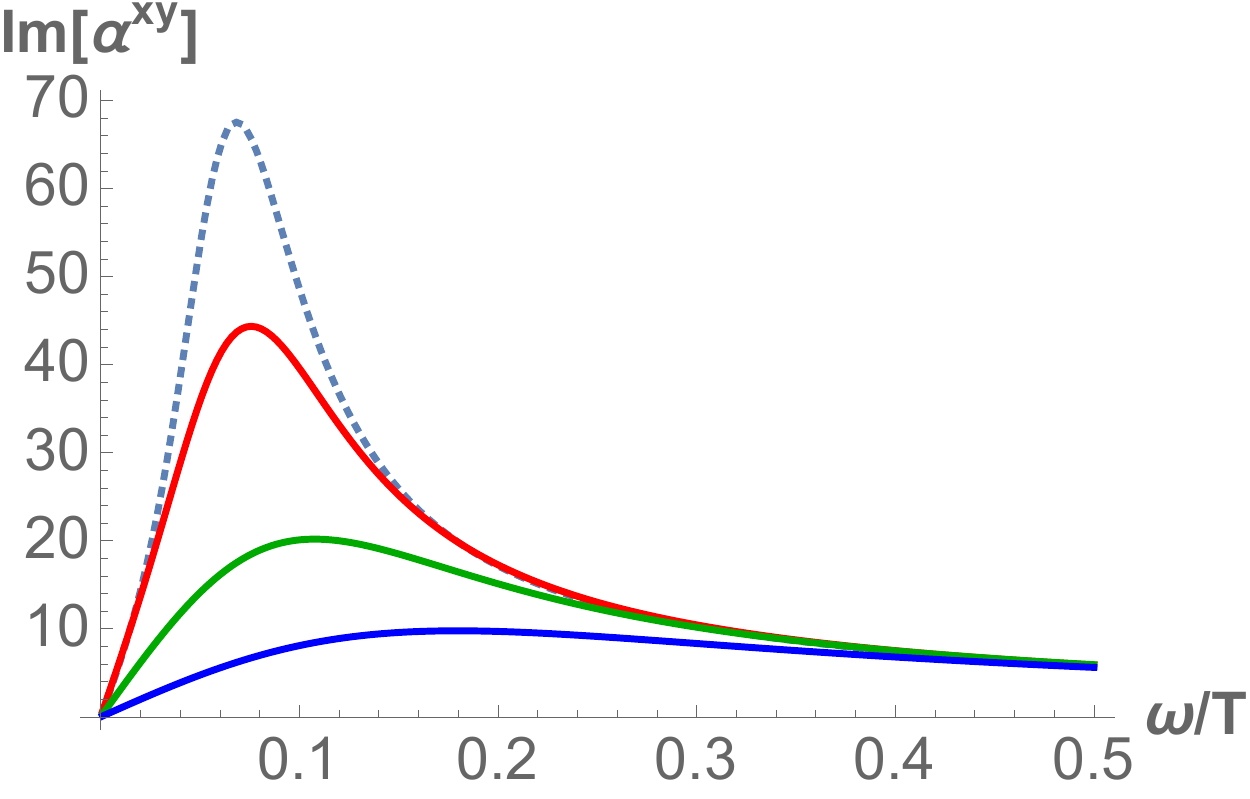} \label{} }
 \caption{ $\beta$ dependence of thermoelectric conductivities: $\beta /T= 0, \ 0.5, \ 1, \ 1.5$(dotted, red, green, blue) with $\mu/T = 1, \ B/T^2=3$} \label{fig:Thermo}
\end{figure}
\begin{figure}[]
\centering
  \subfigure[Re ${\bar{\kappa}}^{xx}$ ]
   {\includegraphics[width=3.5cm]{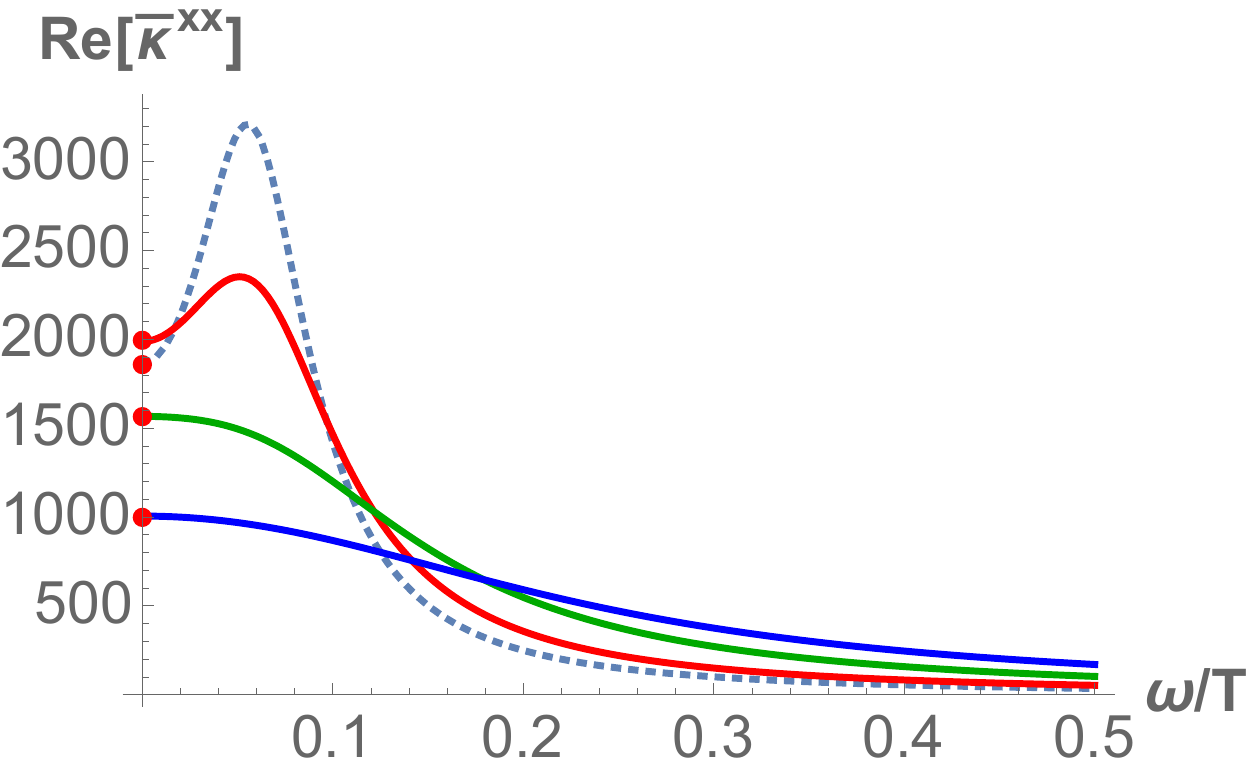} \label{}}
        \subfigure[ Im ${\bar{\kappa}}^{xx}$ ]
   {\includegraphics[width=3.5cm]{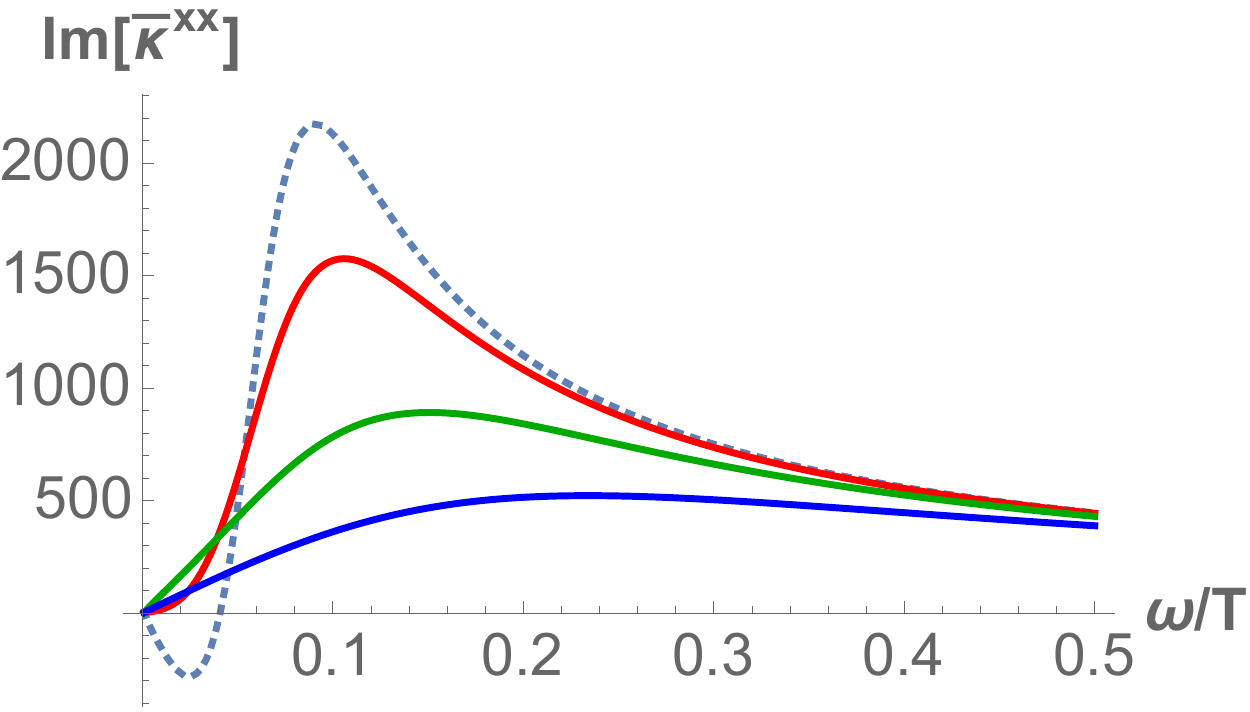} \label{}} \  \
     \subfigure[Re ${\bar{\kappa}}^{xy}$]
   {\includegraphics[width=3.5cm]{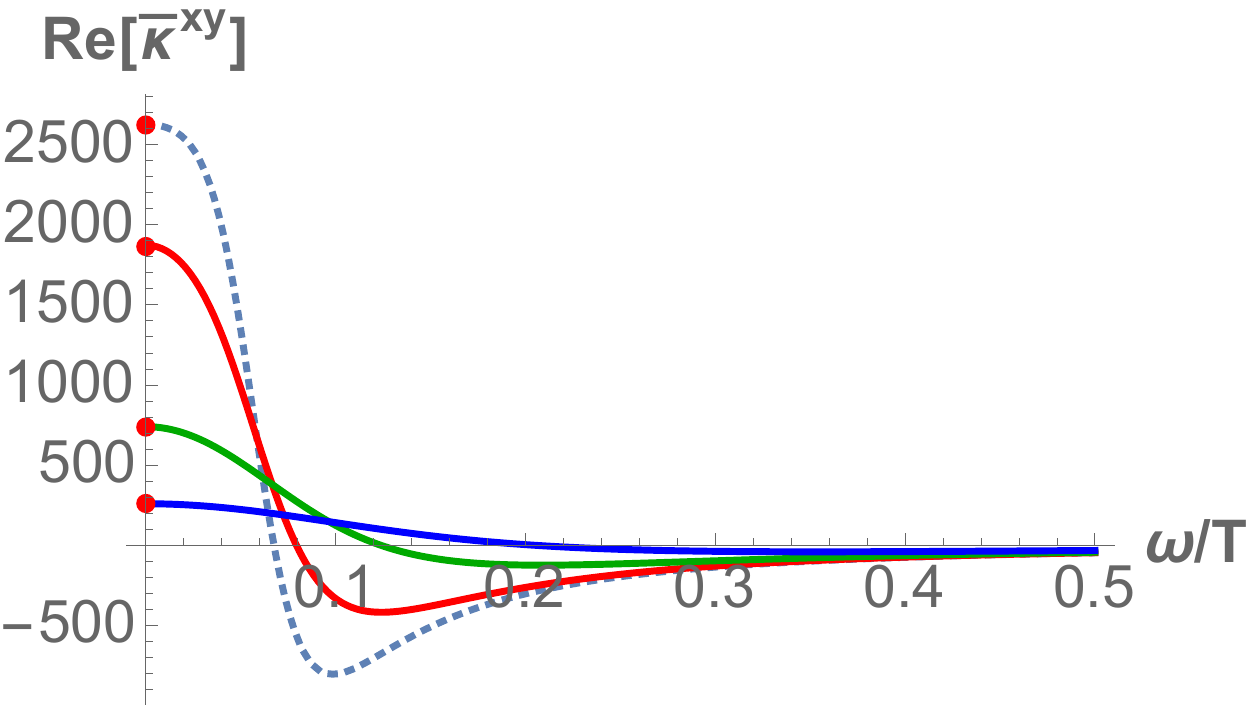} \label{} }
     \subfigure[Im ${\bar{\kappa}}^{xy}$]
   {\includegraphics[width=3.5cm]{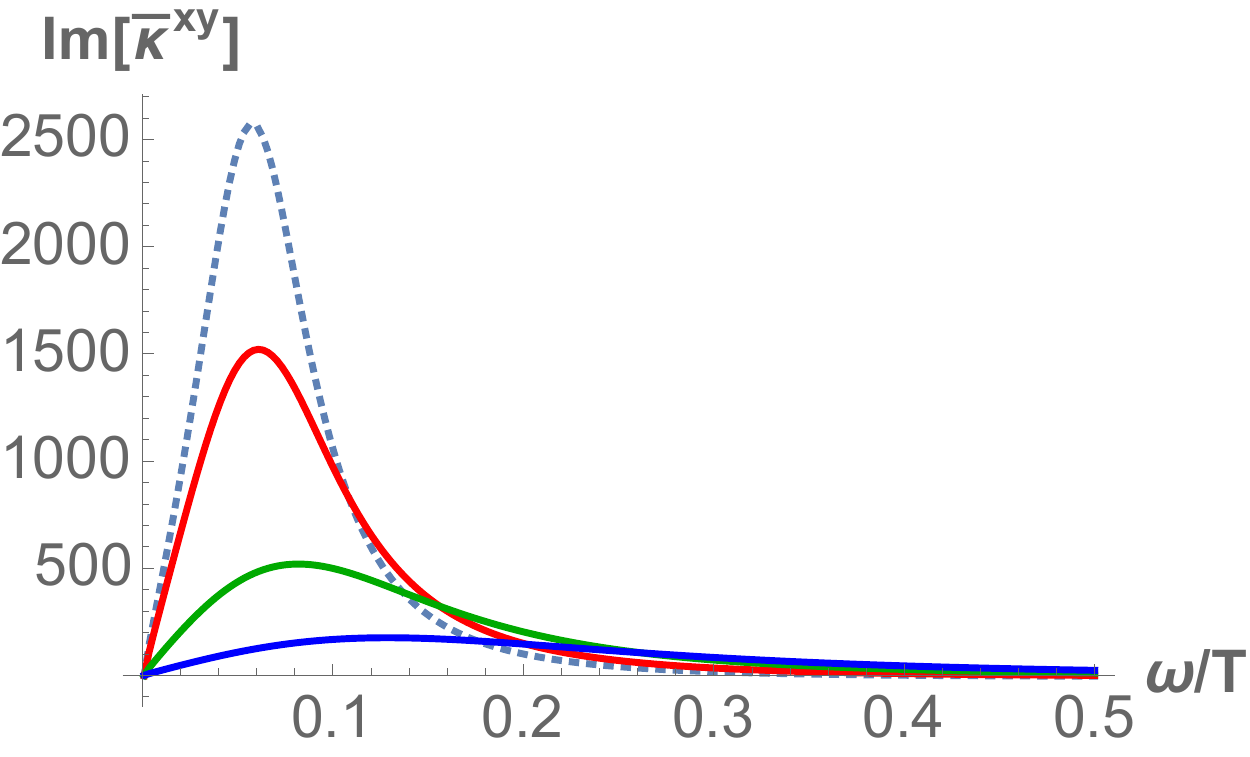} \label{} }
 \caption{ $\beta$ dependence of thermal conductivities: $\beta /T= 0, \ 0.5, \ 1, \ 1.5$(dotted, red, green, blue) with $\mu/T = 1, \ B/T^2=3$ } \label{fig:Thermal}
\end{figure}
\begin{figure}[]
\centering
  \subfigure[Re $\sigma^{xx}$ ]
   {\includegraphics[width=3.5cm]{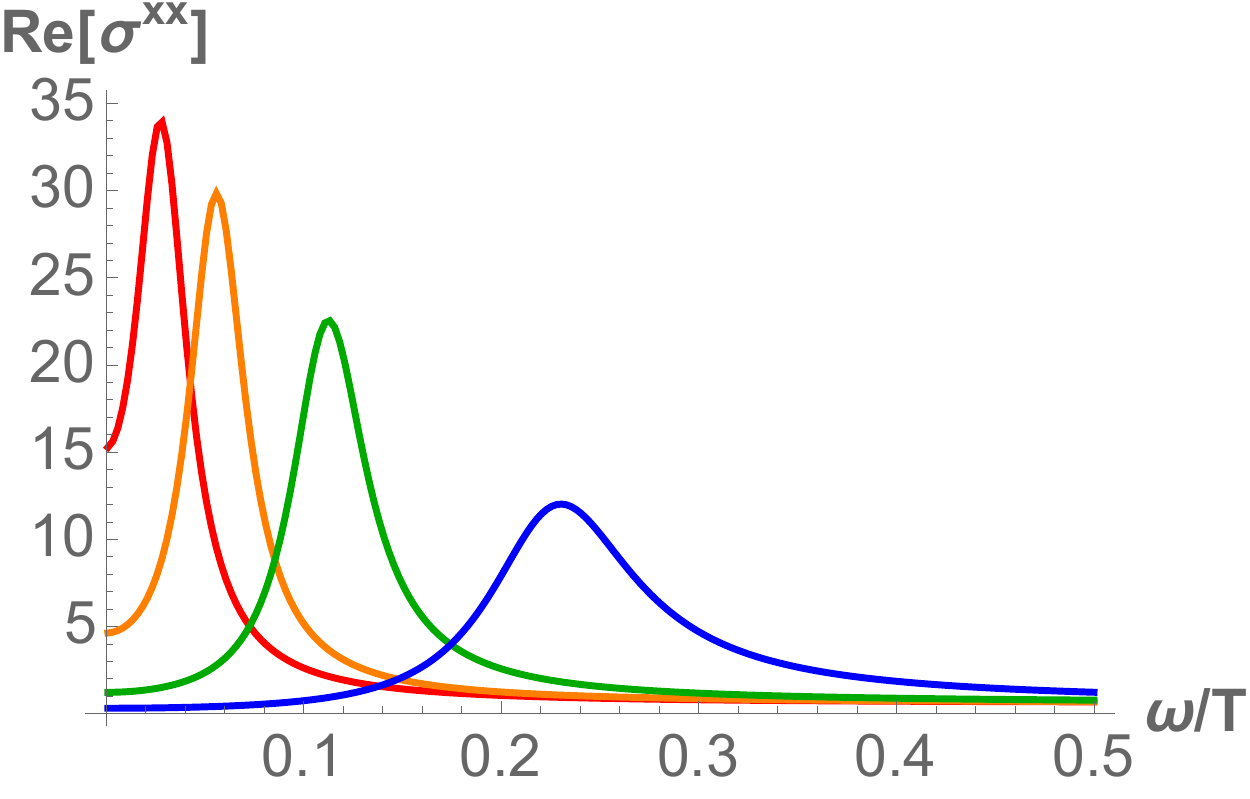} \label{}}
      \subfigure[Im $\sigma^{xx}$ ]
   {\includegraphics[width=3.5cm]{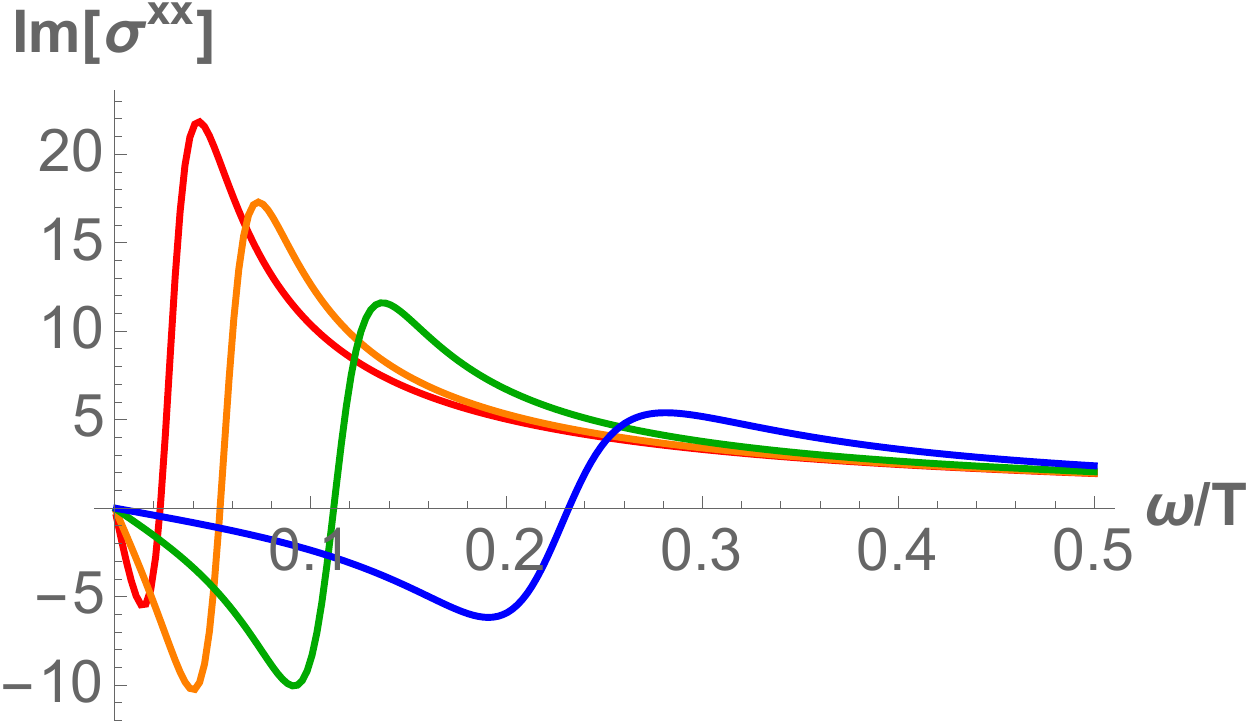} \label{}} \  \ 
     \subfigure[Re $\sigma^{xy}$]
   {\includegraphics[width=3.5cm]{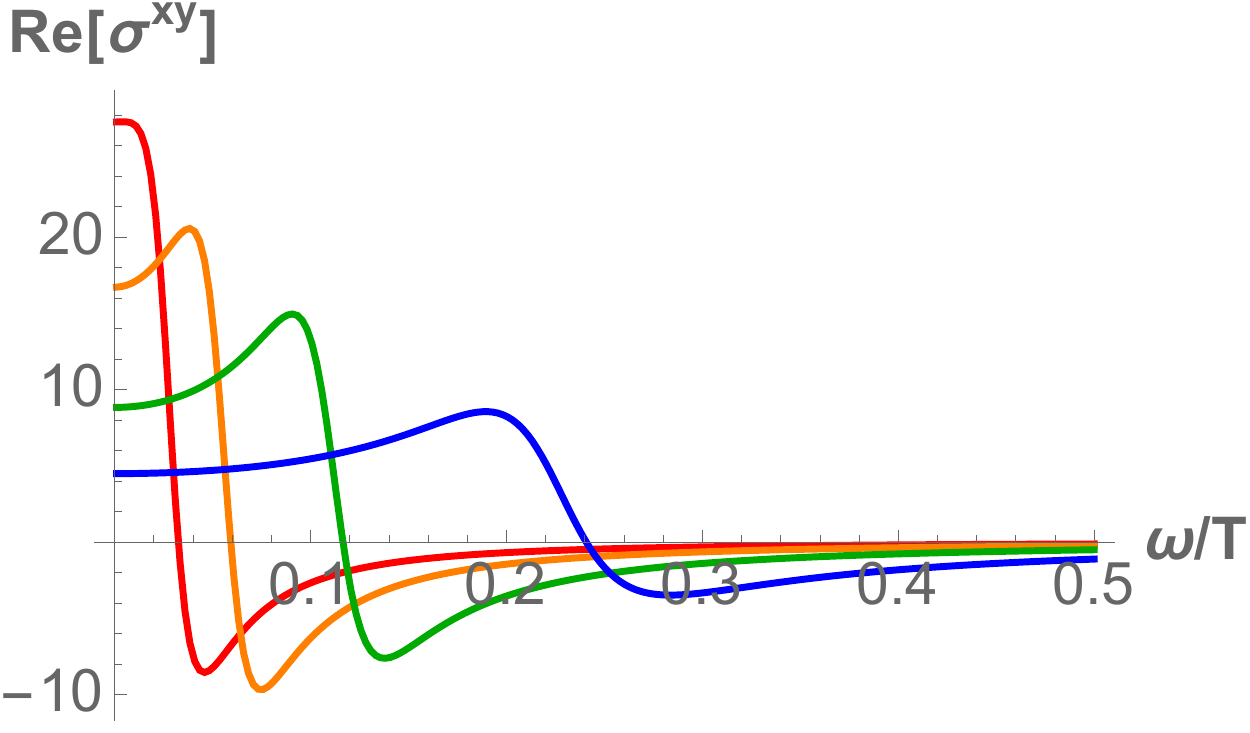} \label{} } 
     \subfigure[Im $\sigma^{xy}$]
   {\includegraphics[width=3.5cm]{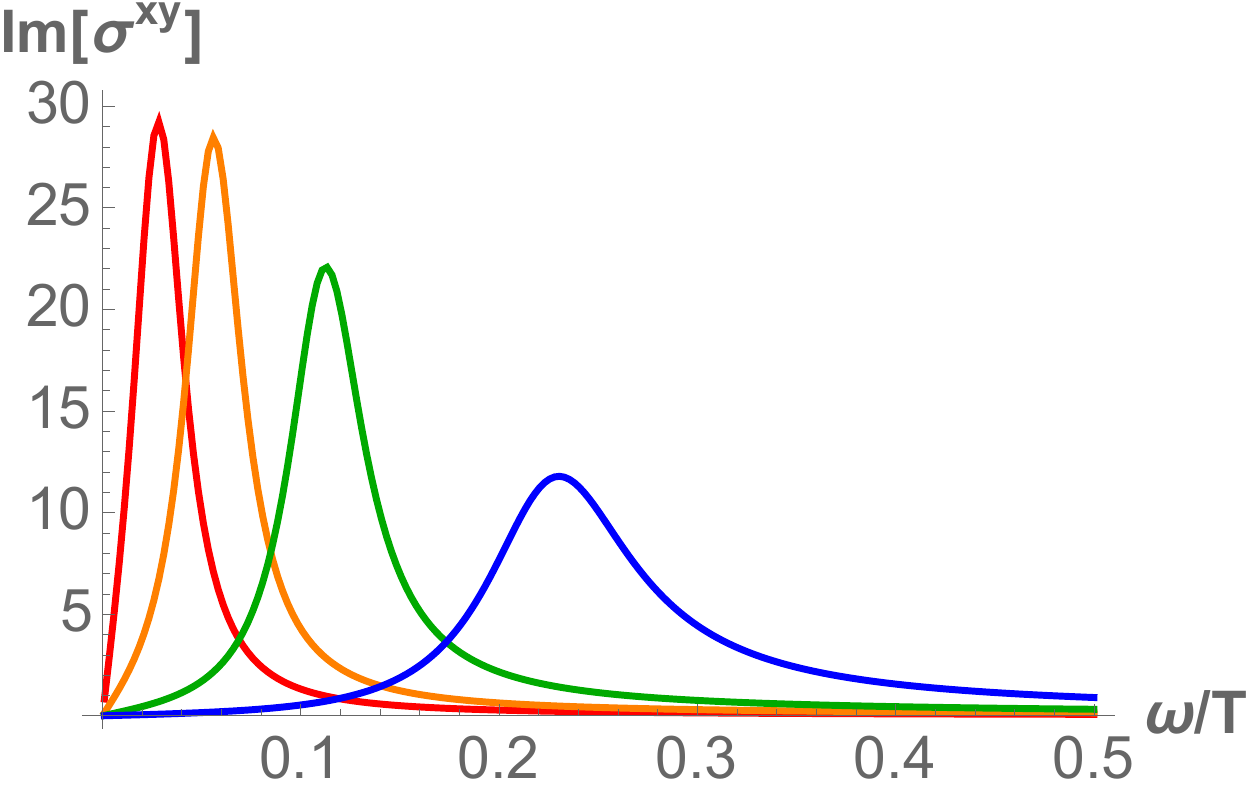} \label{} }
 \caption{$B$ dependence of electric conductivities: $B/T^2=  0.5,\ 1, \ 2, \ 4$(red,orange,green,blue) with $\beta/T = 1/2, \ \mu/T = 4$
           } \label{fig:Hall2}
\end{figure}

There is a peak in the curves in Figures  \ref{fig:Hall1}-\ref{fig:Thermal}, which is related to the cyclotron resonance pole,  the position of the pole of the conductivity in complex  $\omega$  plane~\cite{Hartnoll:2007ih,Hartnoll:2007ip}
\be
\omega_* \equiv \omega_c -i\gamma  \,,
\ee
where the cyclotron frequency $\omega_c$ is the relativistic hydrodynamic analog of the free particle case, $\omega_f = eB/mc$. However, the resonance due to $\omega_*$ here should be understood to be due to a collective fluid motion rather than to free particles. A damping $\gamma$ could be thought of as arising from interactions between the positively charged current and the negatively charged current of the fluid, which are counter-circulating.
In the hydrodynamic regime for small $B$ at $\beta=0$, the $\omega_*$ was computed analytically in \cite{Hartnoll:2007ip} as follows:
\be\label{omegaC}
\omega_c= \frac{\rho B}{{\cal E}+{\cal P}}  \equiv  \omega_c^0 \,,  \qquad  \gamma  =  \frac{B^2}{g^2 ({\cal E}+{\cal P})}  \equiv \gamma^0 \,.  
\ee

 Because $\beta$ is related to momentum relaxation, it is expected that $\gamma$ will increase as $\beta$ increases.  It is indirectly shown in the plots since all curves become flat, which may reflect the fact the pole goes away from the real $\omega$ axis. It turns out that the $\omega_c$ tends to increase as $\beta$ increase, although it is not so clear in the plots. It will be discussed later in Figure \ref{fig:cyclo} and the equation \eqref{omegaC2}.  
While in Figure  \ref{fig:Hall1}-\ref{fig:Thermal} we focused on the effect of $\beta$ at fixed $B$, in Figure  \ref{fig:Hall2} we investigated the effect of $B$ at fixed $\beta$; the red, orange, green, and blue curves are for $B/T^2 = 0.5, 1, 2, 4$ respectively. A relatively big $\mu/\beta = 8$ has been chosen  since the peak is shaper when $\mu/\beta$ is bigger. 
We find the peaks of the curves shift towards higher frequencies as $B$ increases, which is consistent with the hydrodynamic analysis \eqref{omegaC}.  

\begin{figure}[]
\centering
  \subfigure[$q_m/T=0,\ \mu/T=12.5 $ ]
   {\includegraphics[width=4cm]{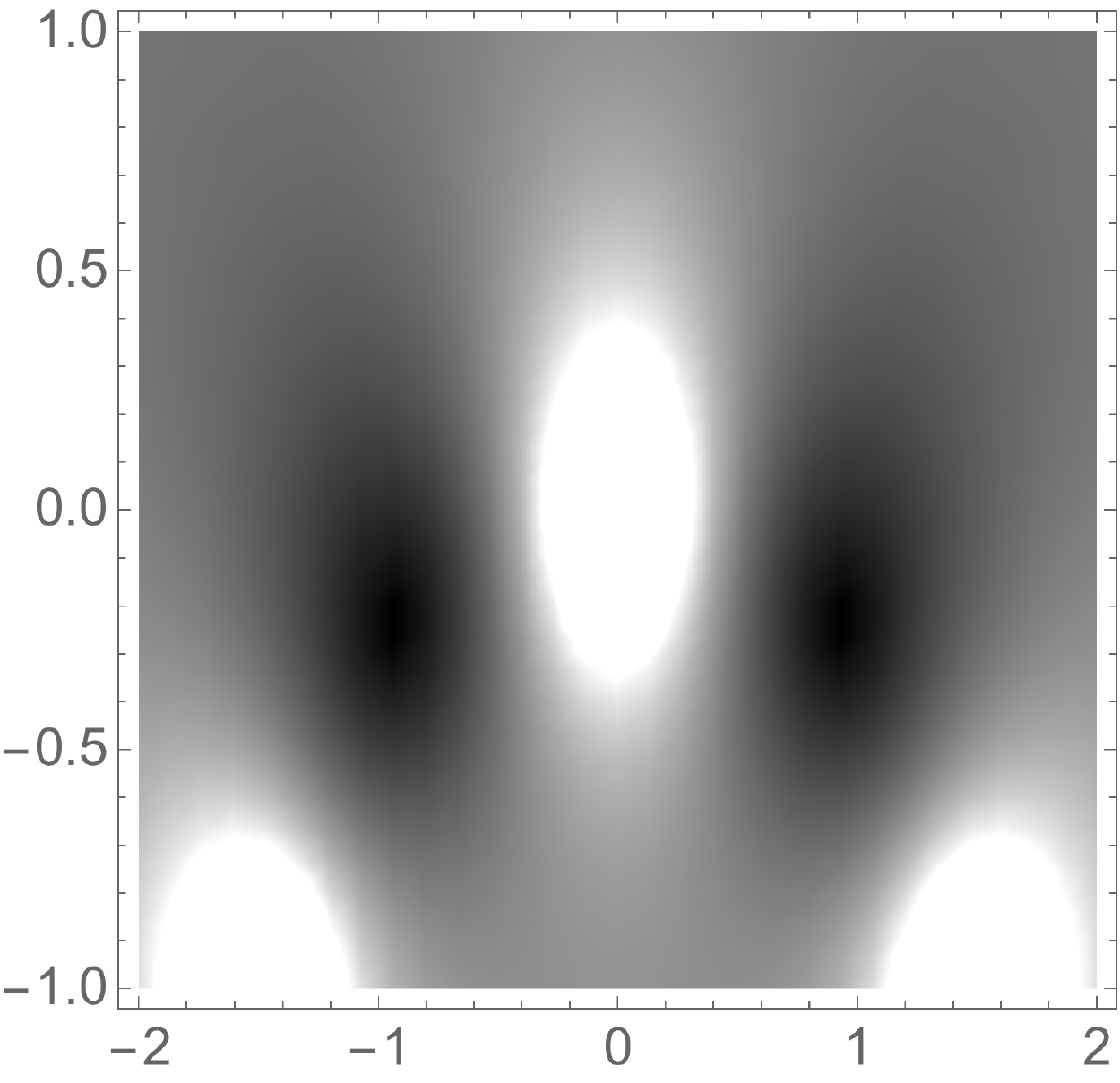} \label{}}
     \subfigure[$q_m/T=8.9,\ \mu/T=8.9 $]
   {\includegraphics[width=4.cm]{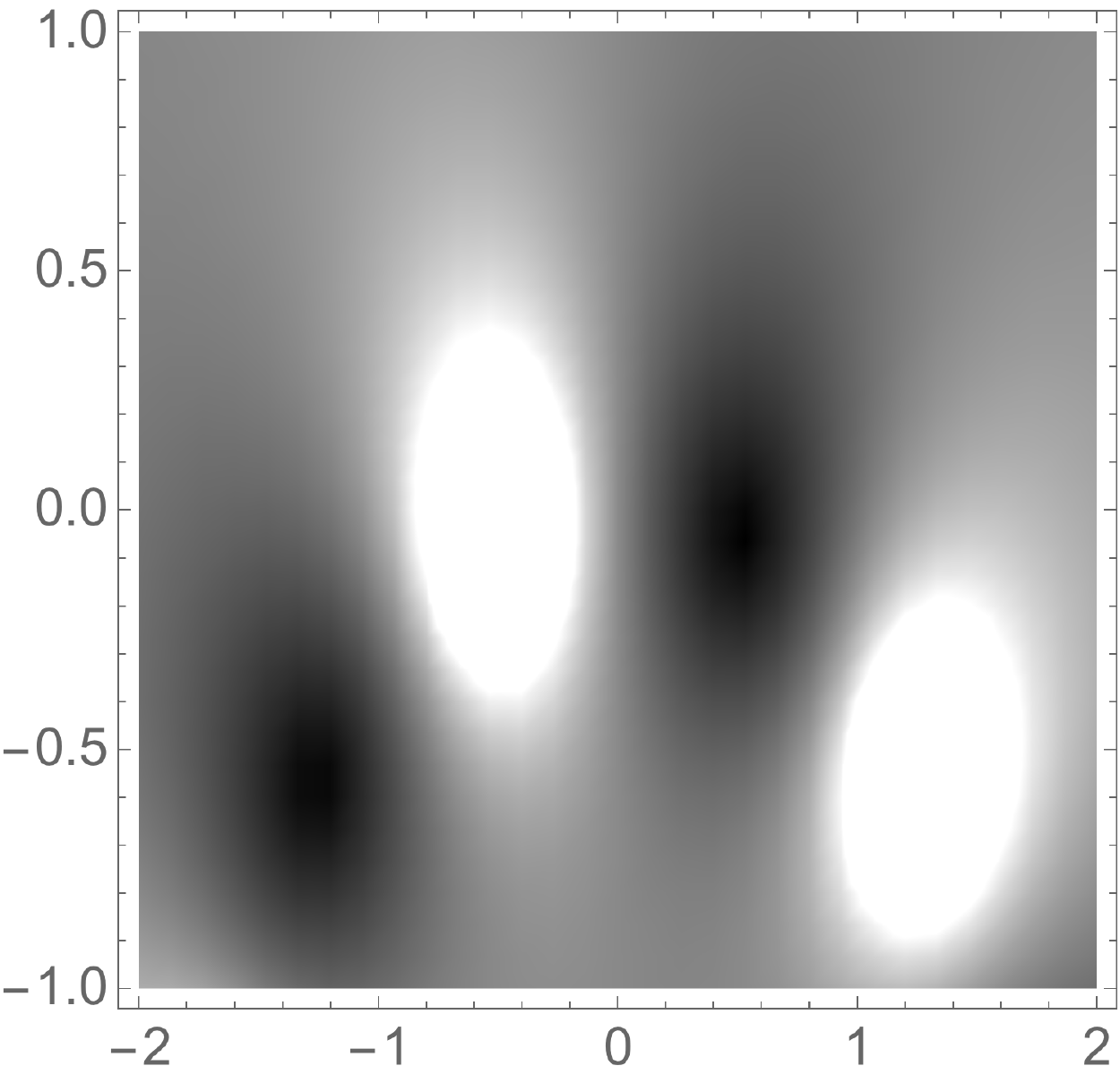} \label{} }
     \subfigure[$q_m/T=12.5,\ \mu/T=0 $]
   {\includegraphics[width=4cm]{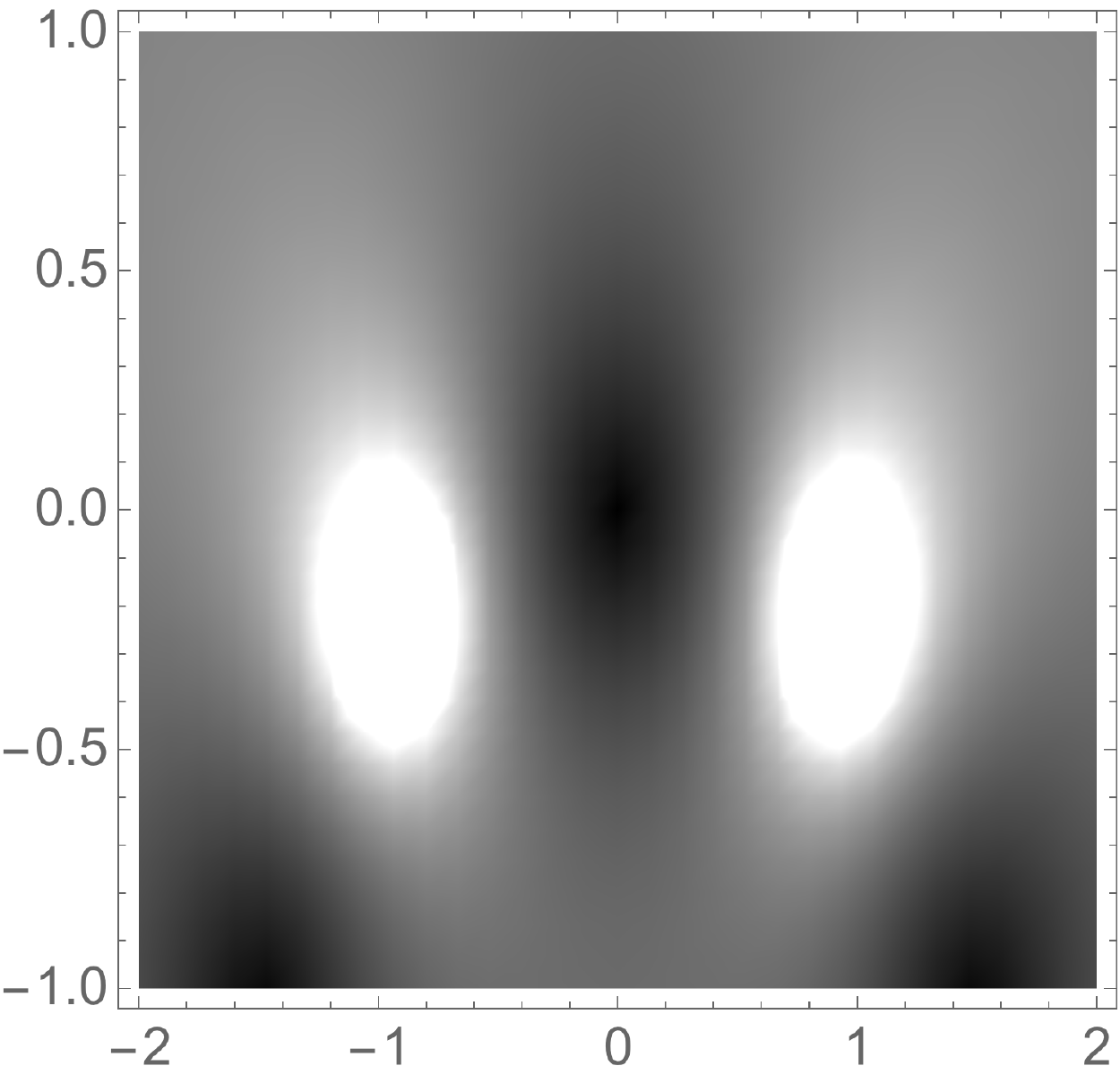} \label{}}
     \subfigure[$q_m/T=0,\ \mu/T=12.5 $]
   {\includegraphics[width=4cm]{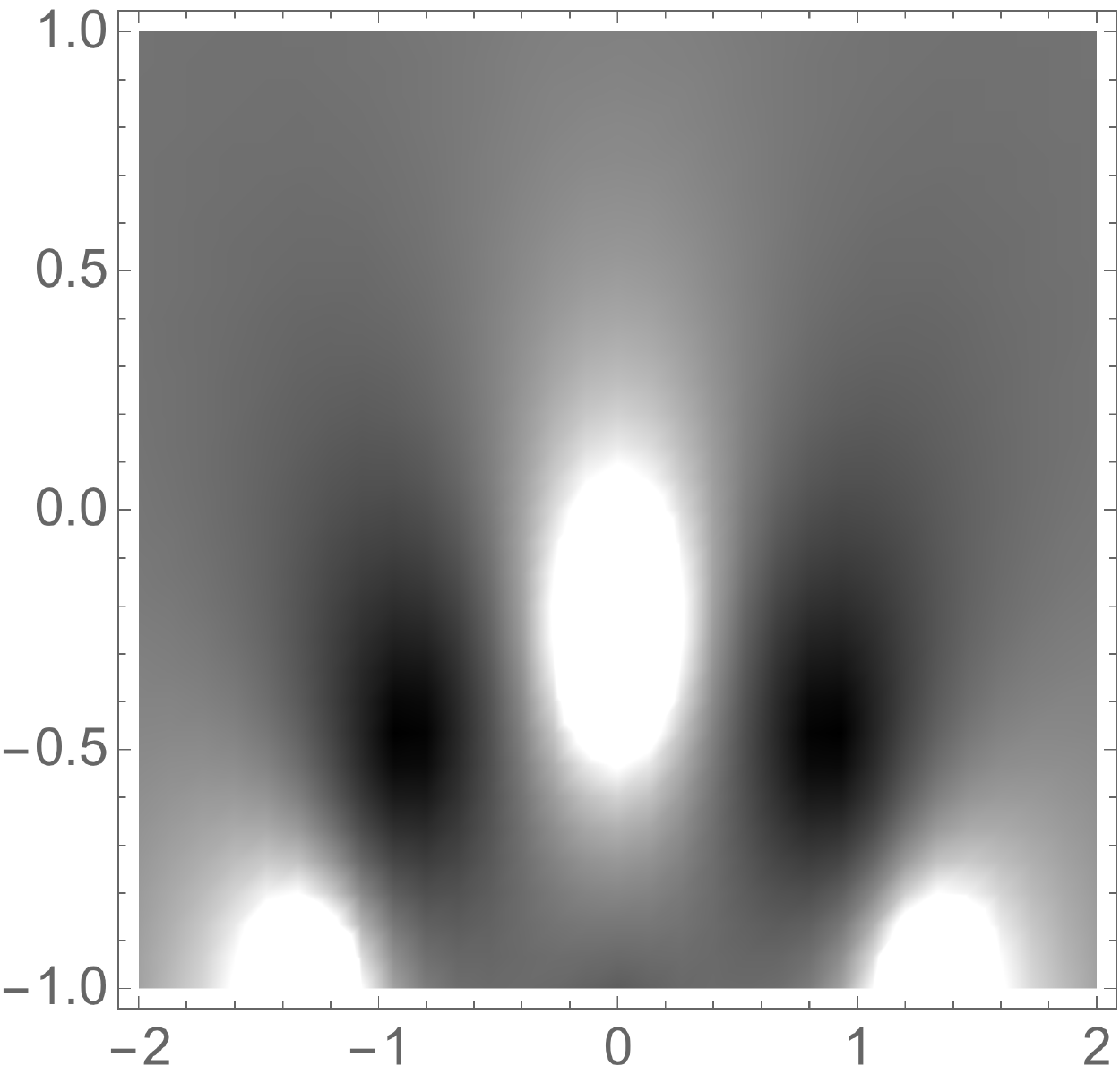} \label{} }
     \subfigure[$q_m/T=8.9,\ \mu/T=8.9 $]
   {\includegraphics[width=4cm]{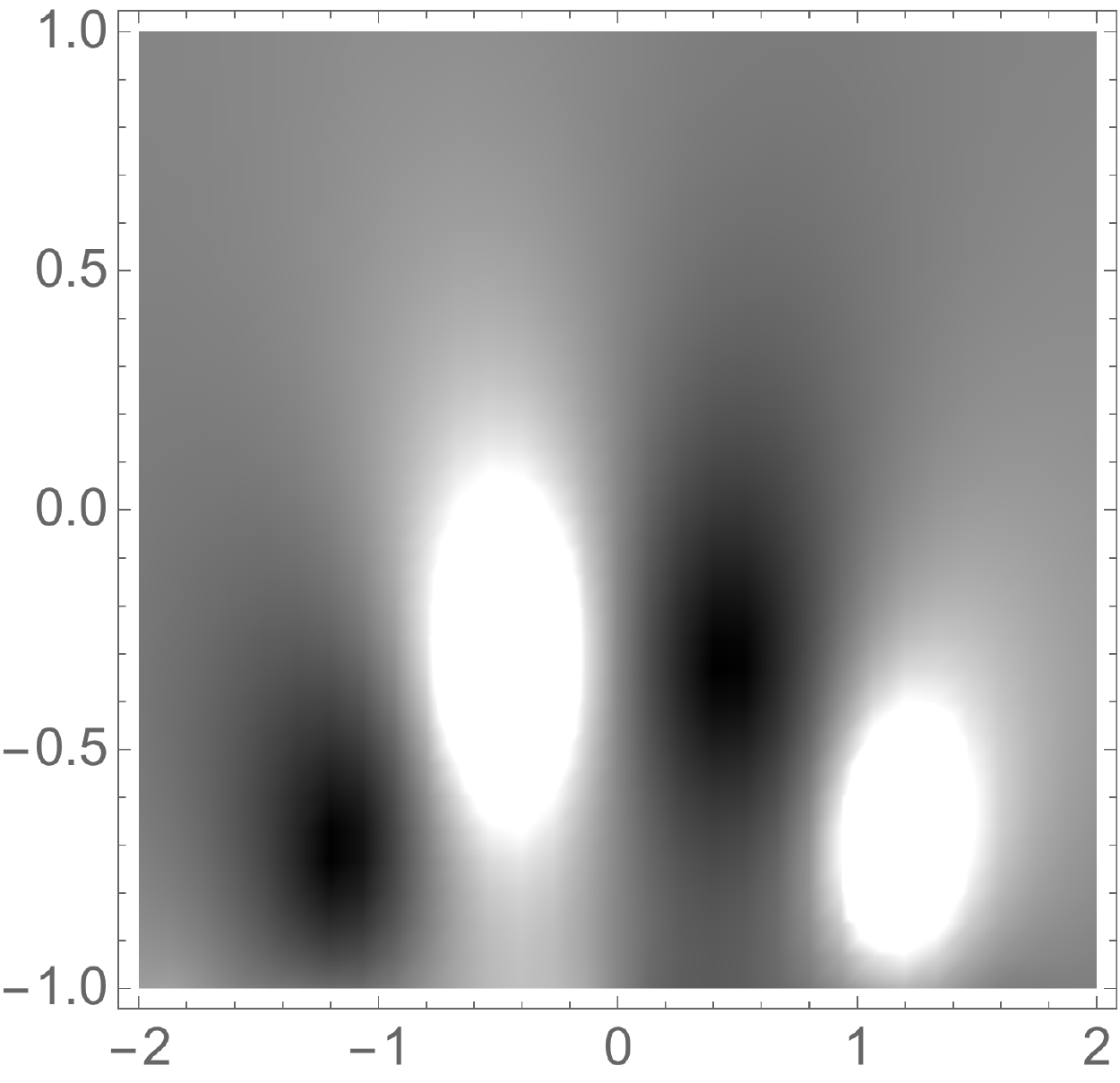} \label{}}
     \subfigure[$q_m/T=12.5,\ \mu/T=0 $]
   {\includegraphics[width=4cm]{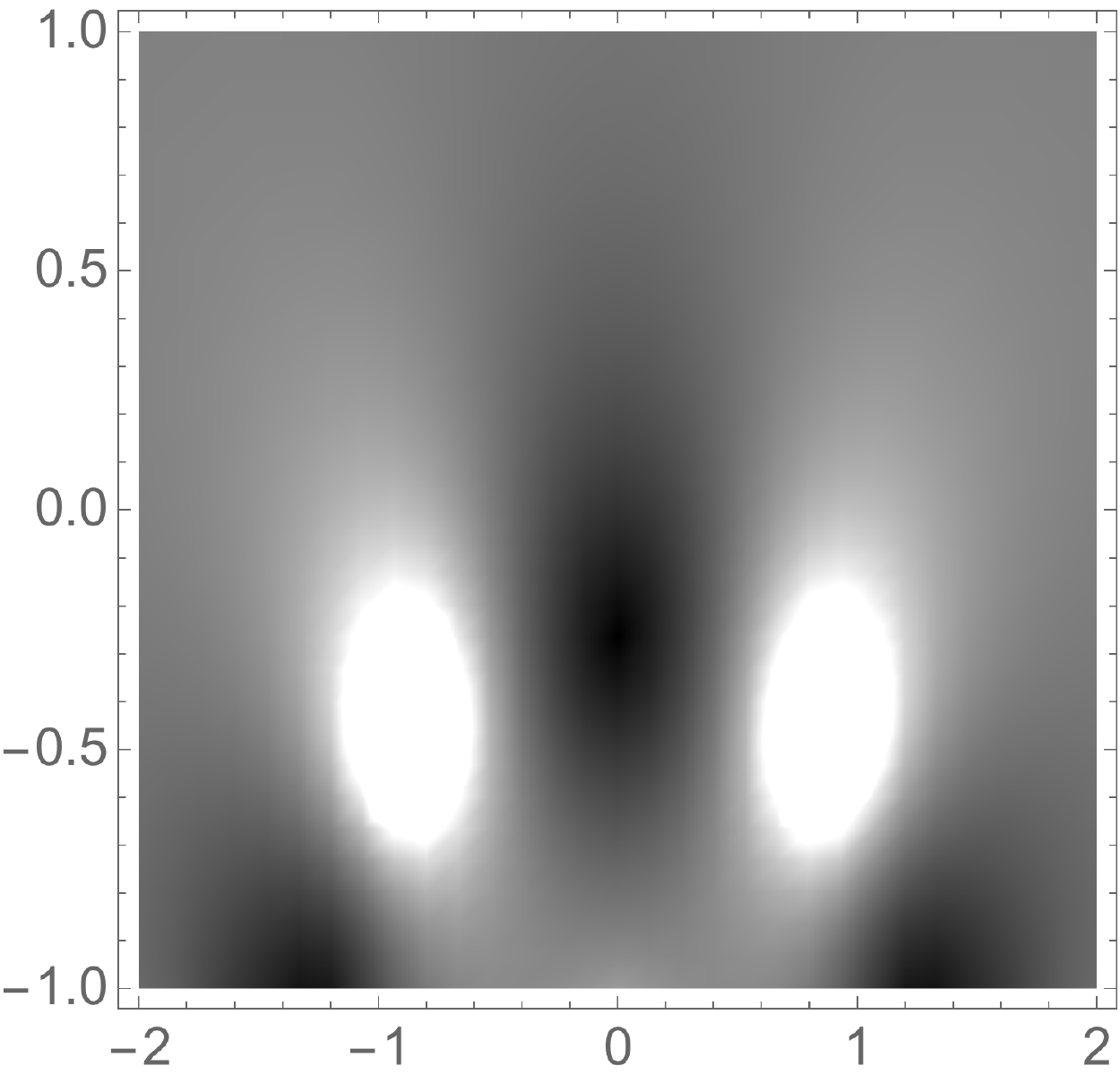} \label{} }
 \caption{A density plot of $\abs{\sigma^+}$ in complex $\omega$ plane: White areas correspond to poles and dark areas are zeroes of $\sigma^+$. $\beta = 0$ for  (a)(b)(c)  and $\beta/T= 10 $ for (d)(e)(f).
           } \label{fig:cyclo01}
\end{figure}

\begin{figure}[]
\centering
  \subfigure[Re $\omega_*$(=$\omega_c$) ]
   {\includegraphics[width=6cm]{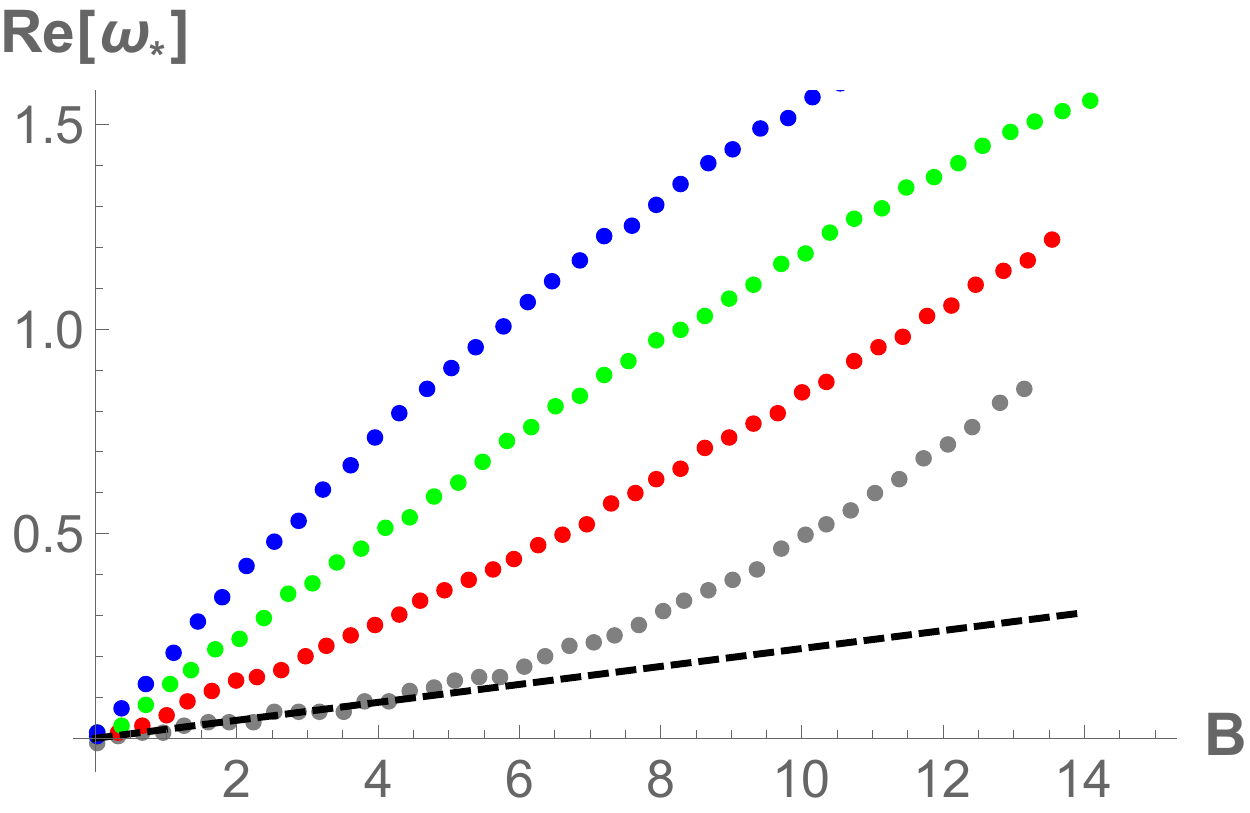} \label{}} \ \  \ \ \  \  \ \ 
     \subfigure[Im $\omega_*$(= $-\gamma$)]
   {\includegraphics[width=6cm]{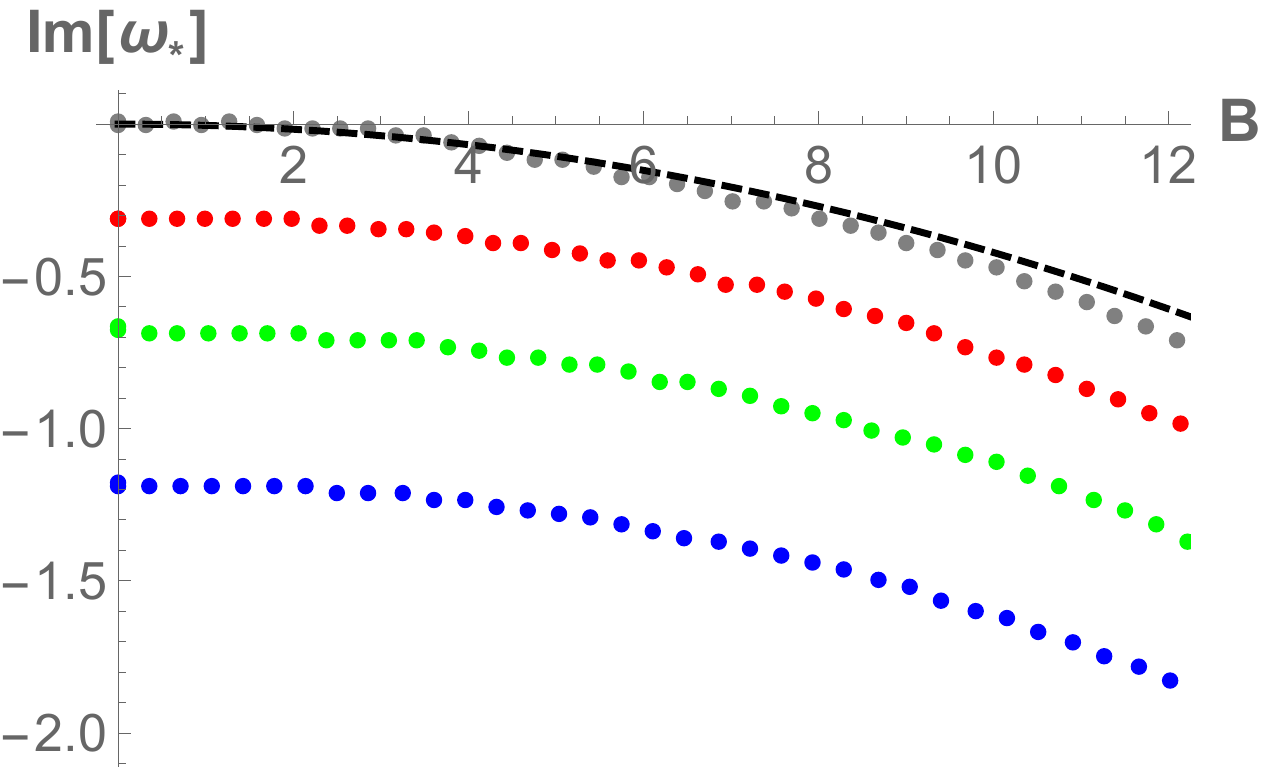} \label{} }
 \caption{The magnetic field dependence of the position of the cyclotron resonance pole: $\beta = 0, 2, 3, 4$(gray, red, green blue). 
  The dotted lines are the results by the hydrodynamic analysis (\ref{omegaC}). 
           } \label{fig:cyclo}
\end{figure}

To see the cyclotron pole directly in complex  $\omega$  plane, we introduce the following combination 
\be
\sigma^{\pm} = \sigma^{xy} \pm i \sigma^{xx} \,,
\ee
for easy comparison with \cite{Hartnoll:2007ih,Hartnoll:2007ip}. The density plots of $\abs{\sigma_+}$ in complex $\omega$ plane are shown in Figure \ref{fig:cyclo01}. Figure \ref{fig:cyclo01}(a),(b),(c) are the cases with $\beta=0$ and \ref{fig:cyclo01}(d),(e),(f) are the cases with $\beta/T = 10$. We choose $\mu$ and $q_m$ such that $q_m^2 + \mu^2$ to be constant. First, at $\beta=0$ we recover the result of \cite{Hartnoll:2007ip}\footnote{Our numerical values of $\mu$ and $q_m$ are different from \cite{Hartnoll:2007ip} due to convention difference.}. White areas correspond to poles and dark areas are zeroes of $\sigma^+$.  
There is a symmetry: $\mu \rightarrow q_m$ and $q_m \rightarrow -\mu$ at $\beta=0$ inherited from the electromagnetic duality of the bulk theory~\cite{Hartnoll:2007ip}.  Since this bulk duality holds at finite $\beta$ we expect the same symmetry is preserved. It is demonstrated in Figure \ref{fig:cyclo01}(d)(e)(f); the figure (d) and (f) are symmetric under the exchange of the white and dark region. 
The finite $\beta$ shifts the position of the poles to the negative imaginary direction.  This implies the width of the peak increases at real $\omega$ axis as discussed previously.


The magnetic field dependence of the cyclotron poles at different values of $\beta$ is shown in Figure \ref{fig:cyclo}, where (a) and (b) show the real part($\omega_c$) and the imaginary part($-\gamma$) respectively. 
The gray, red, green,  and blue dots are the numerical results for $\beta=0,2,3,4$ respectively, while the black dashed line is the analytic result at $\beta=0$ for small $B$ from the hydrodynamic analysis \eqref{omegaC}. 
At  $\beta =0$, our numerical result(gray dots) agrees to \cite{Hartnoll:2007ip} and also fits well to the hydrodynamic analysis(black dashed line) for small magnetic field.  We investigated how $\omega_*$ changes as $\beta$ is introduced. Based on the red, green, and blue dots in Figure \ref{fig:cyclo} and additional similar numerical data for different parameters, we found the following relation at small $B$
\begin{equation}\label{omegaC2}
\omega_* =  \omega_c -i\gamma = \omega_c^0 + c_1 \beta^2 B - i (\gamma^0 + c_2 \beta^2),
\end{equation}
where $\omega_c^0$ and $\gamma^0$ are defined in \eqref{omegaC}. It seems that $c_1$ and $c_2$ are dimensionful constants independent of $\beta$ and $B$. For this formula we focused on small $B$ region where the dots in Figure \ref{fig:cyclo}(a) are linear to $B$. However, at large magnetic field $B$ we numerically found the tendency that $\omega_c \sim c_3 B$, where $c_3$ seems independent of $\beta$.  
 
In the presence of dissipation,  the cyclotron frequency was shown \cite{Hartnoll:2007ih} to be changed as
\be
\gamma \rightarrow \gamma +\frac{1}{\tau_{imp}} \,,
\ee
while $\omega_c$ is intact. In our case $\tau_{imp}$ is proportional to $1/\beta^2$ for small $\beta$ so the shift of the imaginary part in (\ref{omegaC2}) is consistent with the hydrodynamic calculation \cite{Hartnoll:2007ih}. However, our result implies that the cyclotron frequency($\omega_c$) is also shifted by $c_1 \beta^2 B \sim B/\tau_{imp}$. We suspect that the analysis in \cite{Hartnoll:2007ih}, where $B/T^2 \ll 1$ is assumed, is valid in the limit $c_1 B$ is small.  We leave this issue and the analytic justification of the specific form \eqref{omegaC2} for a future project. 

\section{Conclusions} \label{seccon}

In this paper, we have computed the electric, thermoelectric, and thermal conductivity at finite magnetic field by means of the gauge/gravity duality. First, by considering  a general class of  Einstein-Maxwell-Dilaton theories with axion fields imposing  momentum relaxation, we have derived the analytic DC conductivities, which are expressed in terms of the black hole horizon data. As an explicit model we have studied the dyonic black hole modified by a momentum relaxation effect. The background solution is analytically obtained and the AC electric, thermoelectric, and thermal conductivity were numerically computed. The zero frequency limit of the numerical AC conductivities agree to the DC formulas. This is a non-trivial consistency check of our analytic and numerical methods to compute conductivities. Our numerical method can be applied to other cases in which multiple transport coefficients need to be computed at the same time. 

The Nernst signal,  the Hall angle, and the cyclotron resonance pole were discussed following \cite{Hartnoll:2007ai,Hartnoll:2007ih,Hartnoll:2007ip}. Our general analytic formulas of the Nernst signal can be used to build a realistic model and to investigate the universal properties of the model. In particular, in the dyonic black hole case, the Nernst signal is a bell-shaped curve as a function of the magnetic field, if momentum relaxation is small. It is similar to an experimental result in the normal state of cuprates~\cite{Wang:2006fk}, which was speculatively explained by a  vortex-liquid effect~\cite{Anderson:2006, Anderson:2007aa}.  For large momentum relaxation the Nernst signal is proportional to the magnetic field, which is a typical property of conventional metals. The Hall angle for the dyonic black hole was computed explicitly.  The Hall angle ranges between $1/T^0$ and $1/T^1$ and scales as $1/T$ for large $T$. However, in the strange metal phase, it was known that the Hall angle is proportional to $1/T^2$.   The cyclotron poles($\omega_*$) we found are consistent with the hydrodynamic results at $\beta=0$~\cite{Hartnoll:2007ih,Hartnoll:2007ip}. They are shifted by momentum relaxation($\beta \ne 0$) and our numerical analysis suggests the specific dependence of the cyclotron poles on $B$ and $\beta$, \eqref{omegaC2}, when $B$ is small. If $B$ is large it was proposed that  $\omega_c \sim c_3 B$ with $c_3$ independent of $\beta$.  We plan to investigate the properties of the cyclotron poles in more detail both numerically and analytically. 

It is important to compare our AC conductivities with the general expressions based on the memory matrix formalism~\cite{Lucas:2015vna,Lucas:2015pxa}.
It would be also interesting to compare our AC conductivity results with \cite{Alanen:2009cn}, where the AC electric conductivities have been studied at finite magnetic field in the probe brane set up, focusing on the transport at quantum Hall critical points~\cite{Sachdev:ab}. The metal phase of our model at $B=0$ does not have the property of linear-$T$ resistivity, so it is worthwhile to start with the models having that property and then investigate the Hall angle and the Nernst effect in those models. 

\acknowledgments

The work of KYK and KKK was supported by Basic Science Research Program through the National Research Foundation of Korea(NRF) funded by the Ministry of Science, ICT \& Future Planning(NRF-2014R1A1A1003220). 
The work of SS and YS was supported by Mid-career Researcher Program through the National Research Foundation of Korea (NRF) grant No. NRF-2013R1A2A2A05004846. YS was also supported in part by Basic Science Research Program through NRF grant No. NRF-2012R1A1A2040881.
We acknowledge the hospitality at APCTP(``Aspects of Holography'', Jul. 2014) and Orthodox Academy of Crete(``Quantum field theory, string theory and condensed matter physics'', Sep. 2014), where part of this work was done.

\bibliographystyle{JHEP}


\providecommand{\href}[2]{#2}\begingroup\raggedright\endgroup

\end{document}